\documentclass[aps, english, prl, reprint, superscriptaddress, secnumarabic, showkeys, showpacs, raggedbottom, longbibliography]{revtex4-1}
\usepackage[T1]{fontenc}
\usepackage[utf8]{inputenc}
\usepackage{xcolor}
\usepackage{pdfcolmk}
\usepackage{babel}
\usepackage{refstyle}
\usepackage{bm}
\usepackage{amsmath}
\usepackage{amssymb}
\usepackage{stackrel}
\usepackage{graphicx}
\PassOptionsToPackage{normalem}{ulem}
\usepackage{ulem}
\usepackage[unicode=true]{hyperref}
\usepackage{makecell}
\usepackage[export]{adjustbox}

\makeatletter

\AtBeginDocument{\providecommand\secref[1]{\ref{sec:#1}}}
\RS@ifundefined{subsecref}
  {\newref{subsec}{name = \RSsectxt}}
  {}
\RS@ifundefined{thmref}
  {\def\RSthmtxt{theorem~}\newref{thm}{name = \RSthmtxt}}
  {}
\RS@ifundefined{lemref}
  {\def\RSlemtxt{lemma~}\newref{lem}{name = \RSlemtxt}}
  {}

\hypersetup{
	colorlinks,
	breaklinks,
	linkcolor={blue},
	citecolor={blue},
	urlcolor={blue}
}

\makeatother

\usepackage{listings}

\begin{document}

\author{Stefan Krastanov}
\author{Sisi Zhou}
\affiliation{Yale Quantum Institute, Yale University, New Haven, Connecticut 06520, USA}

\author{Steven T. Flammia}
\affiliation{Centre for Engineered Quantum Systems, School of Physics,
	University of Sydney, Sydney NSW, Australia}
\affiliation{Yale Quantum Institute, Yale University, New Haven, Connecticut 06520, USA}

\author{Liang Jiang}
\affiliation{Yale Quantum Institute, Yale University, New Haven, Connecticut 06520, USA}

\title{Stochastic Estimation of Dynamical Variables}

\date{\today}

\begin{abstract}
Estimating the parameters governing the dynamics of a system is a prerequisite for its optimal control. 
We present a simple but powerful method that we call STEADY, for STochastic Estimation Algorithm for DYnamical variables, to estimate the Hamiltonian (or Lindbladian) governing a quantum system of a few qubits. 
STEADY makes efficient use of all measurements and its performance scales as the information-theoretic limits for such an estimator. 
Importantly, it is inherently robust to state preparation and measurement errors. 
It is not limited to evaluating only a fixed set of possible gates, rather it estimates the complete Hamiltonian of the system. 
The estimator is applicable to any Hamiltonian that can be written as a piecewise-differentiable function and it can easily include estimators for the non-unitary parameters as well. 
At the heart of our approach is a stochastic gradient descent over the difference between experimental measurement and model prediction.
\end{abstract}

\maketitle

A common task in physics and engineering is the control of a system, where the control pulses sent to the system pass through a complex transfer function before they effect a useful change to the state of the system. 
There are two overarching prerequisites for good control: learning the dynamical law that governs the system (the goal of disciplines like experimental design and parameter estimation) and, consecutively, the derivation of control pulses for the given system (broadly covered by optimal control theory). 
Advances in these areas are crucial for applications in quantum information science, where the precise control of well-characterized quantum systems will form the basis for quantum computers. 

Here we present STEADY, a conceptually simple but performant method for approaching the parameter estimation problem for dynamical variables. 
We can model a piece of quantum hardware with a Hamiltonian (or Lindbladian) $\tilde{H}(\bm{\omega};\bm{d})$ which depends on the parameters to be estimated $\bm{\omega}$ and on the control pulses $\bm{d}(t)$. 
Our goal becomes finding the value for $\bm{\omega}$ that leads to an $\tilde{H}$ that (for any value of the control pulses $\bm{d}$) most closely mimics the dynamical law $H$ governing the real hardware. 
As is commonly done in parameter estimation, we do this by searching for a value of $\bm{\omega}$ that minimizes some measure of distance between $\tilde{H}$ and $H$. 

Our contribution follows in the rich traditions of stochastic methods and compressed sensing: instead of performing full process tomography on the hardware which would be extremely time consuming, we run a relatively small number of random control pulses on it and study its response. 
For each control pulse we sample the final state of the system (for instance by projectively measuring the qubits in the computational basis). 
We then estimate the difference between this experimental measurement and the prediction based on the $\tilde{H}({\bm{\omega}})$ model. 
This measure of ``difference'' is stochastic, as it uses only a small finite sample of possible control drives. 

This leads to a number of properties that make STEADY perform particularly well. 
First, the distance measure that we use is differentiable with respect to $\omega$ which lets us use efficient (stochastic) gradient descent to rapidly find optimal values for the parameters being estimated. 
Moreover, the stochastic nature of our estimator leads to much lower resource requirements. 
We avoid doing full tomography, which greatly reduces the number of necessary measurements, while only modestly increasing the number of steps required by the (now stochastic) gradient descent. 
Simultaneously, the stochasticity lets us surpass the error floor otherwise imposed by the finite number of measurements performed when sampling the final states of the system. 
By using pulse sequences of varying lengths, our method becomes inherently insensitive to state preparation and measurement (SPAM) errors. 
The fact that dynamical variables---Hamiltonians and Lindbladians---are local quantities means that our estimates are generally \emph{sparse} descriptions of the noise in a system, in contrast to an estimate that reconstructs a finite-time evolution of the noise. 
This allows regularized estimators to be used that avoid overfitting and lead to good estimates with surprisingly few data. 
In fact, our estimator can be restated as a maximum likelihood estimator which
naturally approaches the information-theoretic limit of the Cram\'{e}r-Rao bound.

\subsection*{Background}

\begin{figure}
\includegraphics[width=9cm]{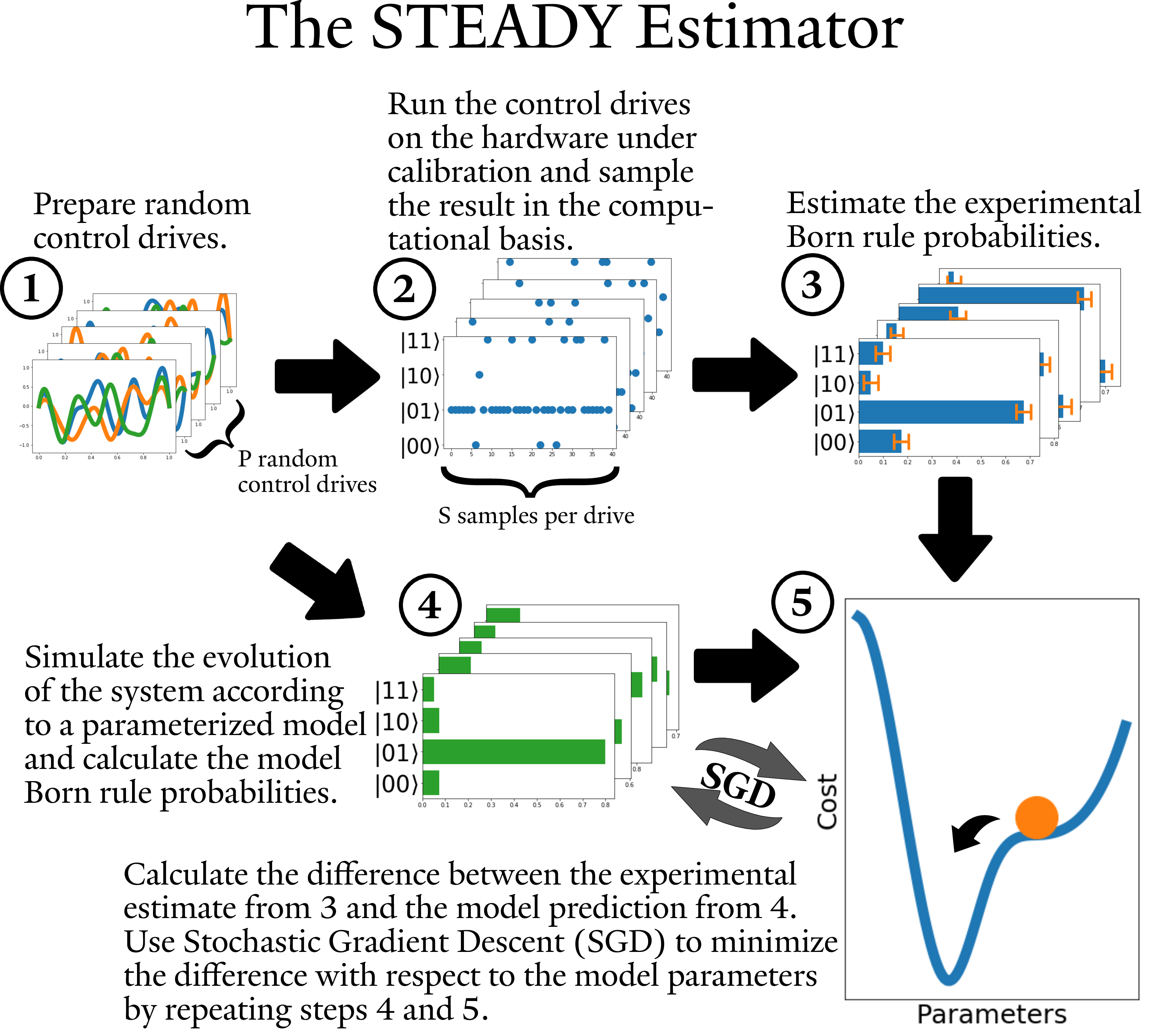}
\caption{\label{fig:blockdiagram} A pictorial representation of the STEADY
parameter estimation protocol. We sample the behavior of the hardware under
calibration for a set of randomly chosen contron drives and compare it to
the prediction of a parameterized model. Through a stochastic gradient
descent we find the parameter values that minimize the difference between
model prediction and experimental result, repeating steps 4 and 5 at each
iteration of the gradient descent.}
\end{figure}

In the field of quantum computing there is a long history of achievements in both
parameter estimation and control theory: the design of precise control schemes~\cite{controlable_peirce1988optimal, compilation_dawson2005solovay, krotov_krotov1983iterative, grape_khaneja2005optimal, crab_caneva2011chopped}
for the preparation of states~\cite{chemcontrol_tannor1985control, stateprep_law1996arbitrary},
unitary operations~\cite{ucontrol_unanyan1998robust, ucontrol_mischuck2013qudit, ucontrol_heeres2015cavity, ucontrol_heeres2017implementing, ucontrol_krastanov2015universal},
and even complete quantum channels~\cite{ccontrol_bacon2001universal, ccontrol_shen2017quantum}
has permitted advances in metrology, chemistry, communication, and---more
recently---quantum computation. These control schemes are,
in return, informed by the precise description of the system, obtained
through various tomographic measurements~\cite{tomography_leibfried1996experimental, tomography_chuang1997prescription, tomography_poyatos1997complete}.

One can simplify the control problem by considering only a discrete
set of gates instead of studying the continuous control that a complete
knowledge of the Hamiltonian would provide. As long as the available
set of basic gates generates the unitary group, there is a known efficient
compilation procedure~\cite{compilation_dawson2005solovay,reversiblecircuits2017svore,verifiedcircuits2017svore,automatedcircuits2017svore}. The performance
bottleneck is in estimating the exact behavior of the given gates.
Techniques like process tomography~\cite{tomography_leibfried1996experimental, tomography_chuang1997prescription,tomography_poyatos1997complete}
have existed for a while, but they are susceptible to state preparation
and measurement (SPAM) errors~\cite{tomography_merkel2013selfconsistent}. Gate set tomography~\cite{tomography_merkel2013selfconsistent, gstomography_blume2013robust, gstomography_greenbaum2015introduction}
mostly circumvents issues of SPAM, by requiring the preparation of
only one type of state (e.g.\ the ground state) and only one type of
measurement (e.g.\ in the computational basis). However gate set tomography
is still susceptible to what~\cite{gstomography_greenbaum2015introduction}
calls ``intrinsic SPAM errors'': state preparation errors for the
ground state (e.g.\ a finite temperature of the system) and any imperfections
in the projective measurement in the computational basis. Finally,
there are tools like randomized benchmarking~\cite{Emerson2005, benchmark_knill2008randomized}
that quantify average error rates of quantum processes (instead of the entire channel) without SPAM, and so can be used only as a benchmark, not as a tool for direct
calibration (though extensions of the idea do allow this~\cite{Kimmel2013}).

On the other end of the spectrum are control schemes that compute
new control pulses for every single unitary operation (instead of
compiling them out of the predetermined ``universal'' set of gates).
This type of ``continuous'' control provides for quantum circuits
with an order of magnitude smaller depth~\cite{heeres2017implementing, supremacy_boixo2018characterizing, supremacy_neill2018blueprint},
however, it is also computationally more difficult. A popular approach
to it is the use of gradient-based methods like GRAPE~\cite{grape_khaneja2005optimal}.
The fidelity of a given operation (the difference between the desired
operation and the operation actually implemented by the control pulse)
is computed as a function of the control pulse parameters. The fidelity
is a differentiable function and its gradient with respect to the
control pulse parameters is also computed, thus permitting efficient
gradient descent, leading to a locally optimal control pulse. A large
body of work is available discussing how to avoid getting stuck in
local optima~\cite{grapeimprov_brif2010control,grapeimprov_zahedinejad2015high, grapeimprov_palittapongarnpim2016learning}.
Recent development in the use of reinforcement learning has even provided
for gradient-free techniques robust to noise~\cite{rl_chen2014fidelity, rl_bukov2017machine, rl_niu2018universal}.

However, most of the gradient-based control techniques require precise
knowledge of the Hamiltonian in order to provide high-fidelity control
drives. Some adaptive techniques get around this problem by switching
to an in-situ method when they are near the optimal pulse. When the
imprecisions in the model Hamiltonian start dominating, they forsake
the model and start measuring the pulse fidelity experimentally through
process tomography~\cite{grapenm_egger2014adaptive,grapenm_wu2018data}
or randomized benchmarking~\cite{grapenm_kelly2014optimal}. However,
this makes the optimization much less rapid as the gradient is not
directly available anymore and techniques like downhill simplex become
necessary. Such optimization techniques are limited by the statistical
error in the fidelity estimation. Recently, an elegant workaround
based on ``simultaneous perturbation stochastic approximation''
was suggested in the ACRONYM method~\cite{grapespsa_ferrie2015robust},
breaking through this statistical error floor. Impressive
improvements have been seen in experiments following variations of
these methods~\cite{expcalib_rol2017restless}. In either case, however,
many additional experimental samples are required, but are then discarded
after the current iteration of the optimizer. Moreover, all of these
additional measurements are done for the sake of designing one specific
gate with exquisite precision, but they do not contribute to estimating
the Hamiltonian of the system and are forsaken when later on one tries
to design another gate. 

Here we focus on the precise estimation of the Hamiltonian itself, which can later
be used in any control scheme. We suggest STEADY, a simple, but
powerful approach that exploits the entirety of the measurement data
in the Hamiltonian estimation, reaching the fidelity limits imposed
by information theory. Similarly to ACRONYM, we use stochastic techniques
to surpass the statistical error floor. Similarly to gate set tomography
we are inherently insensitive to SPAM errors. Moreover, borrowing
ideas from randomized benchmarking, our method can circumvent even
the intrinsic SPAM errors (like finite temperature in the preparation
of the ground state). 
The random pulses, informationally incomplete measurements, sparse models, and regularized estimators that we use take advantage of compressed sensing methods for state and process tomography~\cite{Gross2010, Flammia2012} to improve accuracy.
Lastly, STEADY estimates the complete
Hamiltonian or Lindbladian (or other dynamical models, e.g. a stochastic master equation) of a system, not just one gate or a set of gates.

A recent independent preprint~\citep{flurin2018rnn} by Flurin et al.
presents techniques similar to STEADY's with significant differences in the
design choices. While the prescription for gathering experimental data and the
comparison of the model to that data are very similar in both approaches,
Flurin et al. use a recurrent neural net (RNN) to model the dynamics, while we use
physical models with very general parametrization. Flurin's black box approach
is promising for models that are particularly difficult to differentiate, like
stochastic master equations, however modern autodifferentiation
frameworks~\citep{innes2018blackmagic}
enable the use of STEADY as well. Furthermore, the RNN size is expected
to grow exponentially with the number of qubits in order to simulate the
quantum dynamics. That exponential cost is explicitly present in STEADY. It
would be interesting to examine whether the RRN, after hyperparameter optimization,
would find a sparse representation of the dynamics, similarly to STEADY's
use of Hamiltonian and Lindbladian generators which are explicitly a sparser representation of
the otherwise dense superoperator.

In the following we specify the formalism we use to describe the protocol
and demonstrate that it reaches the information theoretical limit in estimation
fidelity. We discuss the effect of the intrinsic SPAM errors and how
to circumvent them. Experimental design techniques that further improve
the fidelity of our estimator are described. We briefly discuss the
effects of parameter drift, non unitary errors, and nonlinearities
in the Hamiltonian (as a function of the control pulses). Together
with this manuscript we also provide a software package based on a
popular differentiable programming framework~\cite{diffprog_tensorflow2015-whitepaper}
that implements our techniques for various models including unitary
or non-unitary evolution.


\subsection*{Problem statement}

A system of $Q$ qubits (a $2^{Q}$-dimensional Hilbert space) is
controlled by a Hamiltonian $H(\bm{d})$ (that is itself a function
of time-dependent control pulse $\bm{d}(t)$ set by the experimentalist).
$\bm{d}$ is a $D$-dimensional real vector, where $D$ is the number
of control parameters available to the experimentalist. The evolution
of an initial state $|\psi\rangle$ will then be expressed as 
\begin{equation}
|\dot{\psi}(t)\rangle=-iH\left(\bm{d}(t)\right)|\psi(t)\rangle.
\end{equation}

To accurately predict this dynamics we must learn the Hamiltonian $H$ as a map $H:\bm{d}\rightarrow H(\bm{d})$ in order to be able to control the quantum hardware using a control pulse $\bm{d}$. 
We introduce a parameterized model for the Hamiltonian, $\tilde{H}(\bm{\omega};\bm{d})$, in which case the problem becomes finding the values of all parameters in the array $\bm{\omega_{0}}$ for which $H(\bm{d})=\tilde{H}(\bm{\omega_{0}};\bm{d})$ for all $\bm{d}$.
We will also discuss the case where the model $\tilde{H}$ cannot exactly represent the reality of $H$, as well as cases where non-unitary evolution is non-negligible.

With some \textit{a priori} knowledge of the physical system, an experimentalist
might be able to deduce an approximation of $\bm{\omega_{0}}$, however
the experimentalist could also run experiments on the hardware to
learn successively better approximations of $\bm{\omega_{0}}$. 
An experimentalist can run a control pulse $\bm{d}(t)$ and then project
and measure the final state of the system in the computational basis.
The details of how the pulses are chosen and how the measurement
data is used distinguishes the various approaches to Hamiltonian estimation
and calibration like process tomography~\cite{tomography_leibfried1996experimental,tomography_poyatos1997complete,tomography_chuang1997prescription},
gate set tomography~\cite{gstomography_blume2013robust,gstomography_greenbaum2015introduction},
and randomized benchmarking~\cite{Emerson2005, benchmark_knill2008randomized,Kimmel2013}.

Unlike most estimation techniques, we work at the level of control
pulses, without hiding them behind a set of precompiled gates. Moreover,
our protocol is inherently untroubled by SPAM errors, as no special
states or measurements are necessary, besides preparing the ground
state and performing measurements in the computational basis, just
like in gate set tomography. We improve even further by partially
circumventing the intrinsic SPAM errors found in gate set tomography~\cite{gstomography_greenbaum2015introduction}.

\section*{Methods}

\begin{figure}
\includegraphics[width=9cm,valign=T]{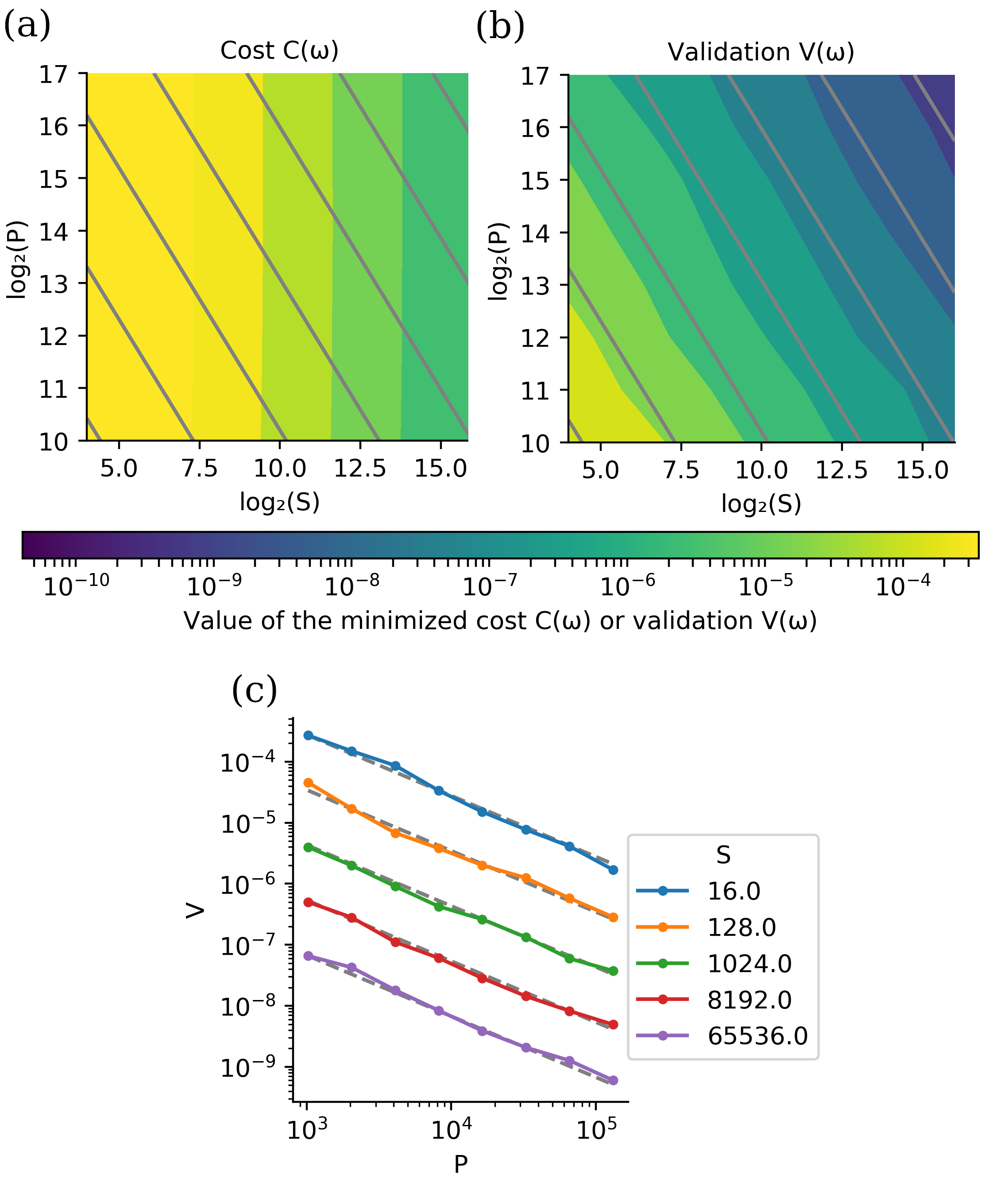}
\caption{\label{fig:cost_and_validation}The fidelity of our estimator increases
steadily with more measurement data, scaling as the Cram\'{e}r--Rao
bound. The performance is not limited by the uncertainty (due to finite
sampling) in the population
measurements.  (a) The minimized cost function $C(\bm{\omega})$
and (b) the minimized validation function $V(\bm{\omega})$. The horizontal
axis gives the number of samples $S$ taken per pulse, while the vertical
axis gives the number of unique random pulses $P$ used. The scales
are binary logarithmic (i.e.\ $10$ corresponds to $2^{10}$ samples
or pulses). \emph{The gray lines show constant $P\times S$ (and
they are logarithmically spaced).} The color map is logarithmic as
well. In (c) we explicitly plot all data points used for the construction
of the contour plots in order to more clearly show the $V(\bm{\omega})\propto\frac{1}{P\times S}$ \emph{power law} (represented exactly by the gray dashed lines).}
\end{figure}

To describe our protocol, we will first consider only Hamiltonians
that are linear in the control pulse. 
This is an appropriate description when some prior knowledge of the relationship between the control fields and the Hamiltonian is known, but more general mappings could be addressed as well, as we discuss below. 
The model for such a Hamiltonian in its
most general form would be (in index notation for $i,j\in[1..2^{Q}]$)
\begin{equation}
\tilde{H}_{ij}(\bm{\sigma},\bm{h};\bm{d})=h_{ij}+\stackrel[k=1]{D}{\sum}\sigma_{ijk}d_{k},
\end{equation}
where $h_{ij}$ and $\sigma_{ijk}$ belong to arrays of complex numbers
representing the parameters $\bm{\omega}$ that need to be learned.
The numbers $h_{ij}$ form the elements of a $2^{Q}\times2^{Q}$ matrix and can be interpreted as the drift Hamiltonian of the system, and the numbers $\sigma_{ijk}$
(forming a $2^{Q}\times2^{Q}\times D$ array) can be interpreted as the list of control
Hamiltonians (one for each control parameter in $\bm{d}$, where $\bm{d}$ could be time dependent). Hermiticity
can be ensured if the parametrization is done in terms of pairs symmetric
and antisymmetric real matrices. There are $2^{2Q}\times(D+1)$ real
parameters to be learned in this case.

Such a high level of parametrization might be unnecessary for a well
studied system. In such a case one can list the fixed known Hermitian
operators that are part of the Hamiltonian in a large list $\{A_{1},A_{2},\dots,A_{M}\}$
(where we have $M$ such possible operators) and only parameterize
how these operators are summed together in 
\begin{equation}
\begin{array}{c}
\tilde{H}(\bm{\alpha},\bm{\beta};\bm{d})=\stackrel[k=1]{M}{\sum}a_{k}A_{k},\\
\text{where }a_{k}=\stackrel[l=1]{D}{\sum}\alpha_{kl}d_{l}+\beta_{k}.
\end{array}
\end{equation}
 Here $\alpha_{kl}$ and $\beta_{k}$ are arrays of real numbers representing
the parameters $\bm{\omega}$ that need to be learned. The $M$-dimensional
real vector $\bm{\beta}$ represents all the drift components of the Hamiltonian,
while the $M\times D$ real matrix $\bm{\alpha}$ represents the linear
mixing/crosstalk between drives. There are $M\times(D+1)$ real parameters
to be learned in this case. For the next few paragraphs we will consider
only this low-degree-of-freedom parametrization, and later on we will
discuss when one might want to use the more general approach. 
We stress that our approach is not limited to Hamiltonians linear in the control pulse
parameters---any functional dependence can be used in
place of the two discussed above. In particular, we discuss the estimation
of a Lindbladian later in this manuscript, and even the measurement backaction parameters of a stochastic master equation can be studied using STEADY.

Experiments in this setup proceed by performing some form of regression on data that has been gathered from the hardware in order to find an approximation of $\bm{\omega_0}$. 
To gather the data we suggest the following approach. 
Begin by generating a large number $P$ of random control pulses on the classical computer controlling the hardware. 
For simplicity we will initially consider only constant pulses of fixed duration $T$ where each random pulse is taken from a normal distribution of unit variance centered on zero, but our approach is equally easy to apply to random time-dependent pulses sampled from other distributions with fixed variance. 
This provides us with a list $\{\bm{d}_{1},\dots\bm{d}_{P}\}$ of control pulses. 
Each control pulse is run on the hardware, initialized to the ground state, resulting in a final state 
\begin{equation}
|\psi_{i}\rangle=\mathrm{e}^{-iH(\bm{d}_{i})T}|0\rangle.
\end{equation}
Experimentalists cannot exactly read the components of $|\psi_{i}\rangle$
in a given basis (e.g.\ the computational basis), rather they can only
estimate them through projective measurements. 
Specifically, our protocol requires the experimentalist to run each pulse $S$ times in order
to repeatedly sample through projective measurements. 
For each $\bm{d}_{i}$ (and corresponding $|\psi_{i}\rangle$) this provides a vector of
estimated Born rule probabilities $\hat{\bm{p}}_{i}$. With an infinite
number of samples $S$, and the assumption that the unknown Hamiltonian
parameters do not drift, the estimate would converge to be exactly the Born rule
probability vector $\bm{p}_{i}:=\lim_{S\to\infty}\hat{\bm{p}}_{i}$ (the $k$-th
component of $\bm{p}_{i}$ is $\left|\langle k|\psi_{i}\rangle\right|^{2}$,
where $|k\rangle$ enumerates the computational basis). Similarly, the Born
rule probability for a given drive $\bm{d}_{i}$ predicted by the model
Hamiltonian $\tilde{H}(\bm{\omega};\bm{d}_{i})$ will be denoted
$\tilde{\bm{p}}_{i}(\bm{\omega})$ (its $k$-th component is $\left|\langle k|\mathrm{e}^{-i\tilde{H}(\bm{\omega};\bm{d}_{i})T}|0\rangle\right|^{2}$).
We note that when the model contains the true Hamiltonian then $\bm{p}_{i} = \tilde{\bm{p}}_{i}(\bm{\omega_0})$.

We can define the ``distance'' between the measured estimate for the population
and the predicted population, averaged over the $P$ random pulses:
\begin{equation}
\label{eq:cost}
    C(\bm{\omega})=\frac{1}{P}\sum_{i=1}^{P}\text{dist}\left(\hat{\bm{p}}_{i},\tilde{\bm{p}}_{i}(\bm{\omega})\right).
\end{equation}
Our estimator $\hat{\bm{\omega}}$ for $\bm{\omega_{0}}$ is the minimum of this
distance measure (called the ``cost function'' from here on):
\begin{equation}
    \hat{\bm{\omega}}=\arg\min_{\bm{\omega}} C(\bm{\omega}).
\end{equation}

If the distance function $\text{dist}()$ is the cross entropy, i.e.\ $\text{dist}(\bm{a},\bm{b})=-\bm{a}.\log(\bm{b})$,
then our estimator is a maximum-likelihood estimator. For most of
the numerical examples, the distance function we use is the mean squared
error $\text{dist}(\bm{a},\bm{b})=\left(\bm{a}-\bm{b}\right)^{2}$,
which is simpler, but in practice leads to the same estimate. 
The cost function is differentiable, which permits us to run automated
stochastic gradient descent optimizers in the search for $\bm{\omega_{0}}$. 
In practice, we also augment this cost function with a regularization cost for $\bm{\omega}$ to avoid overfitting and enhance convergence, as discussed in the supplement. 
Ideally, we would have the sum run over all possible control pulses,
but this is unfeasible in finite time. 
Stochastic gradient descent, where only a small number $P$ of random pulses is used, is what enables our method thanks to its strong guarantees of convergence to the same minimum.

\section*{Results}

In the next few paragraphs we study the performance of this method.
In order to check our susceptibility to over-fitting or convergence
failures, we also introduce a validation cost function (which would
be unavailable to the experimentalist, but is available on our simulated
``mock'' hardware) 
\begin{equation}
V(\bm{\omega}) = \frac{1}{P_{v}}\sum_{i=1}^{P_{v}}\text{dist}\left(
    \bm{p}_{i},
    \tilde{\bm{p}}_{i}\left(\bm{\omega}\right)
	\right),
\end{equation}
 where $P_{v}$ is the size of $\{\bm{d}_{1}^{v},\dots\bm{d}^{v}_{P_{v}}\}$,
a new set of random ``validation'' control pulses sampled from a unit-variance
distribution. This validation function does not suffer from the statistical
noise inherent to finite $S$: it is a sample estimate of the cost function over pulses, and it is the expected value over measurements. 
Moreover $V$ is non-negative and $V(\bm{\omega_{0}})=0$,
hence $V(\bm{\omega})=\mathcal{O}\left(\left(\bm{\omega}-\bm{\omega_{0}}\right)^{2}\right)$,
given our choice of distance function (see supplementary materials).
We will keep the validation set $\{\bm{d}_{i}^{v}\}$ the same in
all comparisons, even if we change the size, variance, or anything
else related to the training set $\{\bm{d}_{1},\dots\bm{d}_{P}\}$.
Similarly, for the validation function we always use a pulse of unit
duration, even if we decide to use different duration pulses for the
cost function.

The validation function is defined so that small values of $V(\bm{\hat{\omega}})$ imply good predictive power of the empirically reconstructed model. 
Our validation function was chosen for this reason and for its close connection to the
cost function, but in the supplementary materials we demonstrate that
other more common measures of fidelity have the same scaling.

For most of this manuscript the distance function in $C$ and $V$
is the mean squared error and other choices are discussed in the appendix.
Unless specified otherwise, numerical results are given for $T=1$,
and a simulated system of $Q=3$ qubits, driven by Pauli drives and
$\sigma_{i}^{+}\sigma_{i+1}^{-}$ nearest neighbor exchange interactions,
where $\bm{\omega}$ gives the relative strengths of each drive. Details are provided in the supplementary materials.

\subsection*{Statistical Sampling Errors}

\begin{figure*}
\includegraphics[width=18.5cm]{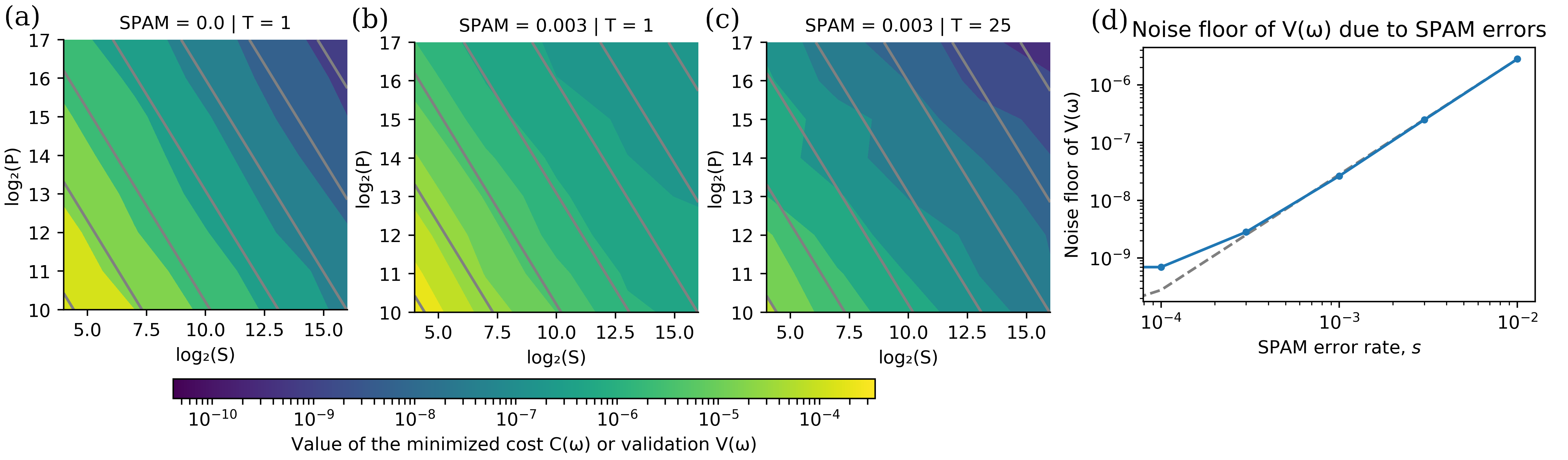} 
\caption{\label{fig:spam}Intrinsic
SPAM errors, e.g.\ in the
preparation of the ground state, create an error floor. (a) Similarly to Fig.~\ref{fig:cost_and_validation},
this is the minimized validation function $V(\bm{\omega})$ in the absence of SPAM errors.
(b) The same plot in the presence of intrinsic SPAM errors on the order of 0.3\% shows much worse values for the validation function.
(c) By using longer control drives ($T=25$) the estimator becomes more sensitive to deviations in parameter values, hence surpassing the error floor that
the intrinsic SPAM errors have imposed in (b).
(d) The error floor (i.e.\
minimal value of $V(\bm{\omega})$) versus the intrinsic SPAM error rate $s$ for short pulses ($T=1$). 
This is plotted using the values $S, P = 2^{16}$, which are chosen to be so large that statistical errors are negligible compared to the bias. 
\emph{The dashed
line shows the $V\propto s^{2}$ power law.} 
In subfigures (b) and (d) one
can see the detrimental effects of intrinsic SPAM errors to the performance
of the parameter estimator while (c) shows it is feasible to surpass that error floor.}
\end{figure*}

If we could obtain a perfect estimate of the populations $\{\bm{p_{i}}\}_{i=1..P}$
(i.e.\ if we could have $S=\infty$), then even a small data set (a
small $P$) would be sufficient to perfectly estimate the Hamiltonian
parameters. The only issue would be ensuring our system is not under-constrained
by having $P\gtrsim M\times(D+1)$, and regularizing the gradient
descent procedure to ensure we are not stuck in a valley of the cost
function. Indeed, when $S=\infty$, we rapidly converge to $C(\bm{\omega})\approx V(\bm{\omega})\approx10^{-16}$,
a floor imposed by the floating point precision.

However, in a realistic case we could run the quantum hardware only
a finite number of times $P\times S$, where a finite $S$ will incur
a statistical error on $\{\hat{\bm{p}}_{i}\}_{i=1..P}$. Given that
obtaining $\hat{\bm{p}}_{i}$ is a multinomial sampling procedure from
the distribution $\bm{p}_{i}$, we can expect an error
$\propto\frac{1}{\sqrt{S}}$, which would cause an error floor of
$C(\bm{\omega})\propto\frac{1}{S}$. We would need to increase $S$
in order to get a better estimate of $\bm{\omega_{0}}$, as can be
seen from the behavior of $C(\bm{\omega})$ in Fig.~\ref{fig:cost_and_validation}a.
From that figure one could think that increasing $S$ is important
while increasing $P$ is a waste of resources, however, the minimized
value of $C(\bm{\omega})$ is a bad proxy of the quality of our parameter
estimation given that it is inherently plagued by the $\frac{1}{S}$
statistical error. This is why we have introduced the validation function
$V(\bm{\omega})=\mathcal{O}\bigl((\bm{\omega}-\bm{\omega_{0}})^{2}\bigr)$,
which in Fig.~\ref{fig:cost_and_validation}b. shows that the total
amount of information $P\times S$ is the important resource expended
in parameter estimation. The precision of our estimate scales as $\frac{1}{P\times S}$,
i.e.\ inversely with the total amount of measurements we take from
the hardware. It is inconsequential how we group the data (more pulses
or better estimation of the result from fewer pulses) as long as $P\times S$
is kept constant and $P$ is sufficiently large to constrain the system. In fact, as long as we have good regularization
that ensures convergence of the gradient descent procedure, even $S=1$
(where $\hat{\bm{p}}_{i}$ becomes a binary vector) performs just as
well.

On first sight it can be counter intuitive that the validation function
continues to improve even when the actual cost function reaches a
floor, but this is similar to the difference between the standard
deviation of a distribution (taking the role of $C$) and the possibly
much lower standard error in the estimator of the mean of that distribution
(taking the role of $V$). Such behavior is typical of stochastic
optimizers and can be seen in ACRONYM~\cite{grapespsa_ferrie2015robust}
as well. A linearized example of this can be seen in the supplementary material.
A more rigorous understanding of this effect can be presented in terms
of the Cram\'{e}r--Rao bound~\cite{cramer2016bound}. The variance of each component
$\omega_{l}$ of our estimator for $\bm{\omega_{0}}$ is bounded by
$\frac{1}{\mathcal{I}_{l}(\omega_{l})}$, where $\mathcal{I}_{l}$
is the corresponding component of the diagonal of the Fisher information
matrix for our measurements (without loss of generality, we assume the parameter vector $\bm{\omega}$
is chosen so that no two distinct components have nonzero correlation). Moreover, $\mathcal{I}_{l}=P\times S\times\langle\mathcal{I}_{l}^\text{samp}\rangle$,
where $\mathcal{I}_{l}^\text{samp}$ is the Fisher information for the
estimator of a single projective measurement for a given pulse and
$\langle\dots\rangle$ denotes an average over all control pulses
$\{\bm{d_{1}},\dots\bm{d_{P}}\}$. The projective measurement is equivalent
to sampling once the multinomial distribution described by $\bm{p}_{i}$,
hence $\mathcal{I}_{l}^\text{samp}=\stackrel[k=1]{2^{Q}}{\sum}p_{k}\left(\frac{\partial p_{k}}{\partial\omega_{l}}\right)^{2}$,
where $p_{k}=\left|\langle k|\psi_{i}\rangle\right|^{2}$. For the
purposes of statistical errors we can already observe that for every
component $\omega_{l}$ of $\bm{\omega}$ we have 
\begin{equation}
\text{var}(\omega_{l})\ge\frac{1}{P\times S\times\langle\mathcal{I}_{l}^\text{samp}\rangle}.
\end{equation}
 This confirms our observations from Fig.~\ref{fig:cost_and_validation},
and proves the efficiency and unbiasedness of our estimator.

\subsection*{Intrinsic State Preparation and Measurement Errors}

The discussion from the preceding paragraphs did not consider the
effects of imperfect state preparation and measurements (SPAM). Unlike
with process tomography, we do not need to prepare initial states
spanning the whole Hilbert space, nor do we need to make projections
in anything but the computational basis. As such, STEADY is not susceptible
to the usual SPAM errors, an advantage we share with gate set tomography.
However, both gate set tomography and STEADY suffer if the
ground state is not properly cooled or if the measurement in the computational
basis is not perfect (effects called ``intrinsic SPAM'' by~\cite{gstomography_greenbaum2015introduction}).
Below we study the error floor caused by the intrinsic SPAM and describe
how our protocol is able to deal with it. 
Our model for intrinsic SPAM in the numerical examples is parameterized by a single parameter $s$, which is defined as the probability that any given qubit might have flipped from $|0\rangle$ to $|1\rangle$ during state preparation. 
We also choose the probability that a projective measurement is incorrectly reported as its opposite to be this same value $s$. 
For numerical simplicity second order
effects are neglected, i.e.\ two qubits cannot flip at the same time
(during preparation or measurement). The detailed exact expression
for the intrinsic SPAM model we use is also given in the supplementary
materials. 

As one can see from Fig.~\ref{fig:spam}(b), increasing the intrinsic
SPAM probability (the per-qubit error rate for initial state preparation
or final projective measurement) leads to a breakdown of the $\frac{1}{P\times S}$
scaling and an intrinsic error floor that cannot be surpassed by
simply increasing the available data. The Cram\'{e}r-Rao bound
is not immediately useful in explaining this effect, because we have
initially phrased it in terms of an unbiased estimator for $\bm{\omega_{0}}$.
However, the intrinsic SPAM errors contribute to a bias in our estimator,
in which case the bound becomes 
\begin{equation}
\mathbb{E}\left((\omega_{l}-\omega_{l,0})^{2}\right)\ge\frac{\left(1-\frac{\partial b_{l}}{\partial\omega_{l}}\right)^{2}}{P\times S\times\langle\mathcal{I}_{l}^\text{samp}\rangle}+b_{l}^{2},
\end{equation}
 where $\mathbb{E}(\dots)$ is an expectation value over all possible
sets of measurements, and $\bm{b}$ is the bias of the estimator of
$\bm{\omega_{0}}$.

In Fig.~\ref{fig:spam}(d) we see the appearance of an error floor in
the fidelity/validation function. We observe $V\propto s^{2}$, which
unsurprisingly points to a bias that grows with the intrinsic SPAM
error $\left|\bm{b}\right|\propto s$ (behavior that can be observed
by leading order expansions of the validation function, as done in
the supplementary materials).

An inspiration from randomized benchmarking leads to a way to surpass
this error floor, namely that longer control pulses would suffer from
fixed SPAM errors, but greater and greater coherent bias-induced errors.
Anywhere that $\bm{b}$ appears, it is multiplied by $T$, leading
to $\mathrm{e}^{-i\tilde{H}(\bm{\omega}+\bm{b};\bm{d})T}=\mathrm{e}^{-i\tilde{H}(\bm{\omega};\bm{d})T}+\mathcal{O}\left(\left|\bm{b}\right|T\right)$.
Doing the same Taylor expansion as described above leads us to $\left|\bm{b}\right|\propto\frac{s}{T}$.
We do observe in Fig.~\ref{fig:spam}(c) that we rapidly return
to the much lower statistical error by using longer pulses. Equivalently,
instead of using longer pulses, we can use pulses of greater average
power. Thanks to this, STEADY provides for addressing not only
statistical errors (by taking more samples), but also intrinsic SPAM
errors (by using longer/stronger pulses).

A figure with a more exhaustive numerical report over various pulse
lengths and SPAM error rates is presented in the supplementary materials.

Another way to fight the effects of SPAM is to include them in the
model of the dynamics, and estimate the parameters governing SPAM
together with the unitary parameters. Later in the manuscript we discuss
such approaches of extending the model to describe more general dynamics
like non-linear drives and non-unitary effects.

\subsection*{Optimal Control and Experimental Design}

\begin{figure}
\includegraphics[width=8.9cm]{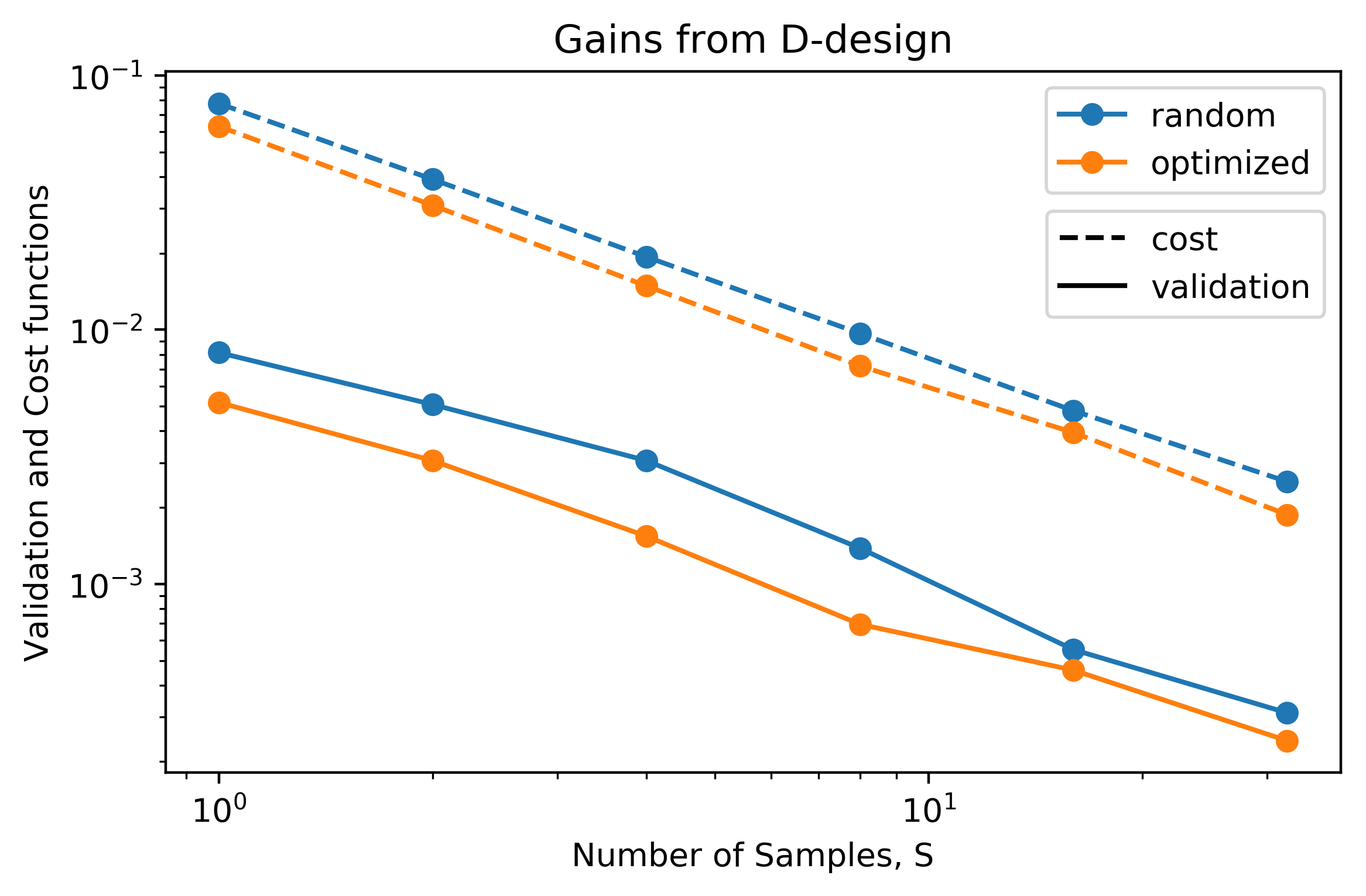} 

\caption{\label{fig:ddesign}Comparison between random (blue) and optimized (orange) control pulses for Hamiltonian estimation. The 
optimized cost function is the dashed line while the validation function is the
solid line (the statistical error in the validation is smaller than the
size of the data point markers).
$P=512$ pulses are used in the cost function. $S$ goes from $1$ to $64$.
At still larger values of $S$ (and much smaller values of $V$) annealing to the
minimum cost becomes difficult and imperfect convergence hides any possible
gains stemming from the optimized pulses.}
\end{figure}

The usual goal of estimating the parameters describing one's hardware
would be to permit high-fidelity open-loop control. There is a rich
history of methods designed for that purpose~\cite{stateprep_law1996arbitrary,ucontrol_mischuck2013qudit,ucontrol_krastanov2015universal,ucontrol_heeres2017implementing,grape_leung2017speedup,mavadia2017prediction}.
Lately, optimal control techniques have been used with great success.
The ``differentiable programming'' toolkits~\cite{diffprog_tensorflow2015-whitepaper}
that have enabled the rapid development of STEADY is readily applicable to the reverse problem of optimal
control~\cite{grape_leung2017speedup}: we ``freeze'' the model parameters
and now optimize with respect to the control drives, while the minimization
target is not the distance between a measured and predicted state,
but rather between the desired and predicted state.

However, in the context of our work, there is a more exciting application
of optimal control, that would permit parameter estimation at much
lower resource/time cost.
As we have described in previous sections,
the fidelity of any estimator is limited by the Fisher information
contained in our measurements.
Here we suggest a relatively straightforward optimization procedure to increase
the available Fisher information. This procedure will perform well in practice,
and as we have shown, it continues to saturate the scaling behavior of the Cram\'{e}r--Rao Bound, but it is
not globally optimal. As our task is a multiparameter estimation problem and
parameters that couple to noncommuting observables cannot be measured
simultaneously without disturbance, the optimal estimation scheme is unlikely
to be as simple as the use of random control pulses in STEADY (we refer the reader to Refs. \citep{expdesign_pezze2017optimal,crbnoncom_matsumoto2002new,crbnoncom_yang2018optimal} for an overview
of the literature on optimal multiparameter estimation).
Instead of using random control pulses
we can also run gradient descent to find
the control pulses that maximize the determinant of the Fisher information of the measurements
(also known as D-optimal design in the field of experimental design).
This optimization procedure does not need to converge with great precision
to the true maximum, as we are interested
in the gross gain obtained from switching from random pulses to optimized
pulses (the minor gain from precisely finding the maximum is negligible
in this context). Fig~\ref{fig:ddesign} demonstrates how STEADY can
halve the required number of measurements while obtaining the same
fidelity, thanks to this careful design of control pulses. We have
kept the average power of the control pulses fixed to a single unit,
both in the case of random and optimized pulses. In our mock tests this
technique has been able to provide similar gains in the fidelity of the
estimator over a large range of values for $P$ or $S$.


\begin{figure}
\includegraphics[width=9.2cm]{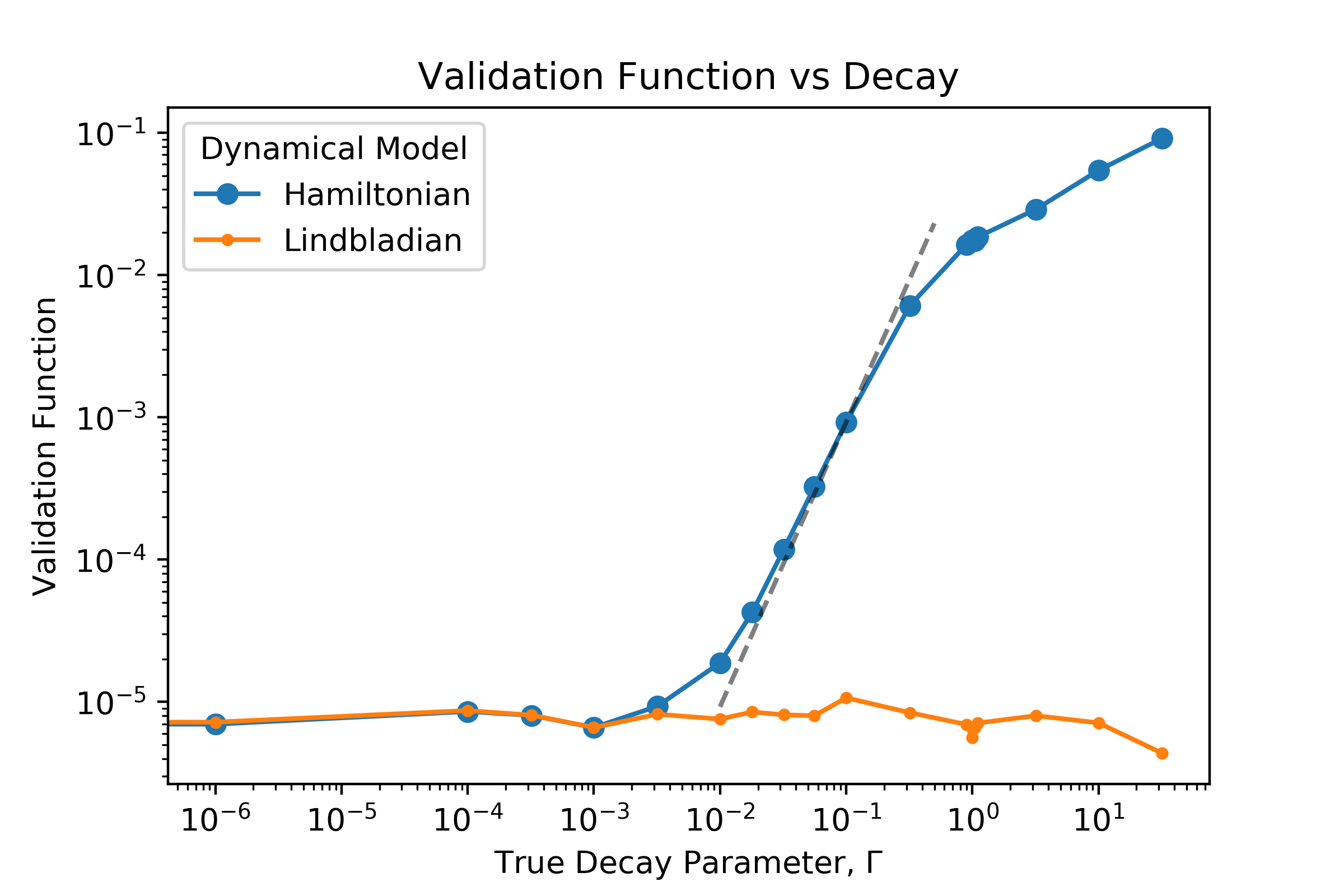} 

\caption{\label{fig:lindblad}If the model of the dynamics we are using in
our estimator is not capable of representing the actual dynamics,
the quality of the estimate will suffer. Above we compare the performance
of a Hamiltonian model (as described in the main text) to a Lindbladian
model (where each qubit suffers from a separate $T_{1}$ decay and
the decay parameter $\Gamma=\frac{1}{T_{1}}$ is estimated in the
fit). The fit is done at $S=1024$ and $P=512$, without intrinsic
SPAM in the data. Independently of the value of $\Gamma$, the Lindbladian
model reaches the statistical error floor. The Hamiltonian model,
which is not rich enough to represent the dynamics of the system,
meets an error floor imposed by $\Gamma$. The dashed line corresponds
to a $y\propto x^{2}$ power law.}
\end{figure}

\section*{Discussion}

There are a number of technical details and caveats that need to be
addressed.

First, the stochastic gradient descent could have trouble reaching
the optimum due to difficult-to-traverse valleys in the cost landscape.
In practice, annealing the learning rate in an adaptive Nesterov-momentum
Adam optimizer~\cite{opt_nesterov1983method,opt_kingma2014adam}, together
with annealing of a 1-norm regularizer (following MacKay's ``empirical
Bayes'' trick~\cite{optbayes_mackay1992practical}) was sufficient
for robust performance of our estimator. These choices were informed by a
hyper-parameter optimization study. 
We provide more details about this in the supplement. 

Many hardware systems could be plagued by slow drifts in the Hamiltonian.
Approaches to solving this issue span from estimating the drift (i.e.\
just making it one of the parameters describing the dynamics) to repeating
the estimation procedure at regular intervals. Both techniques are
readily applicable to our method, additionally opening the possibility
to use efficient transfer learning~\cite{transfer_pratt1993discriminability}
if necessary.

It is important to also note that our approach for circumventing
the intrinsic SPAM errors assumes that our model is capable of exactly describing the actual dynamics of the system. 
There are two cases in which this might not be true.
On one hand, we might have coherent errors, due to small corrections
to $H$ that are not described by the map $\bm{d}\mapsto\tilde{H}(\bm{d})$.
Such problems can be straightforwardly addressed by parameterizations
of $\tilde{H}$ that include quadratic and higher terms. 

A special type of error that can never be captured by the Hamiltonian formulation above 
would be an inherently non-unitary error such as decay or dephasing. 
The only way to fully capture the effects of such non-unitary dynamics is to include them
in the model used by the estimator, and the most natural approach is to simulate the evolution of the system with a master equation. 
This leads naturally to the \emph{Lindbladian} version of STEADY, which follows the same principles as the Hamiltonian version detailed above. 
Fig.~\ref{fig:lindblad} demonstrates that including the non-unitary dynamics in our model permits us to reach the statistical error floor that was unreachable with an incomplete Hamiltonian model. 
A more detailed discussion in terms of Fisher Information content for this Lindbladian case is provided in the appendix.
Even more general dynamical laws can be implemented in STEADY as well: whether
for use in classical mechanics, other dynamical systems, or in the rich field
of continuous weak measurements of quantum systems
\citep{flurin2018rnn,weak_tsang2009time,weak_cortez2017rapid}.

\subsection*{Conclusion}

We have introduced STEADY, a method for Hamiltonian or Lindbladian estimation
that reaches the information-theoretic performance limits. The method
is inherently insensitive to general SPAM errors plaguing approaches
like process tomography and can even circumvent the intrinsic SPAM
errors (e.g.\ errors in the preparation of the ground state). Working
at the Hamiltonian/Lindbladian level gives us greater control than
what methods restricted to sets of pre-compiled gates provide, letting
us use optimal control techniques when manipulating the system. This
versatility permits us to use well known techniques like D-optimal
experimental design to further improve the fidelity of our estimator.

There are many avenues that remain open for further exploration and extension. 
Because STEADY uses an inherently sparse description of the noise in the system, it is conceivable that the method could be made scalable. 
At the moment, this is precluded by the need to simulate the full quantum dynamics, but methods to speed this up could be employed such as using non-universal quantum circuits instead of random pulses, or using a well-calibrated quantum device together with STEADY to calibrate a new device. 
There is much room for additional theoretical work on the performance of STEADY, especially in regards to the family of models that we consider. 
Here we mainly considered a linear coupling to drives, but it should be possible to give a complete theoretical analysis of performance for more general classes of functions. 
It would also be interesting to include weak measurements in a stochastic master equation formalism and attempt to infer the measurement parameters as well. 
One additional open problem is to connect the performance of STEADY to other popular error metrics for quantum gates so as to facilitate an analysis of fault tolerant capabilities for quantum computation~\cite{Kueng2016}.
Lastly, experimental implementation of these ideas will undoubtedly lead to further improvements in the method.

\begin{acknowledgments}
This work would not have been possible without the contributions of
the Python, Jupyter, Matplotlib, Numpy, Qutip, and Julia open source
projects and the Yale HPC team. 
We acknowledge support from the ARL-CDQI (W911NF-15-2-0067, W911NF-18-2-0237), ARO (W911NF-18-1-0020, W911NF-18-1-0212), ARO MURI (W911NF-16-1-0349), AFOSR MURI (FA9550-14-1-0052, FA9550-15-1-0015), DOE (DE-SC0019406), NSF (EFMA-1640959), and the Packard Foundation (2013-39273).
This work was also supported by the Australian Research Council through the Centre of Excellence in Engineered Quantum Systems CE170100009, and by the US Army Research Office grant numbers W911NF-14-1-0098 and W911NF-14-1-0103.
\end{acknowledgments}

SK performed the majority of the software development and numerical tests. The
design, evaluation, and repeated iterative improvements of the method were
jointly done between all four authors.

\bibliography{hamiltonian_estimation_manuscript}

\begin{thebibliography}{60}%
\makeatletter
\providecommand \@ifxundefined [1]{%
 \@ifx{#1\undefined}
}%
\providecommand \@ifnum [1]{%
 \ifnum #1\expandafter \@firstoftwo
 \else \expandafter \@secondoftwo
 \fi
}%
\providecommand \@ifx [1]{%
 \ifx #1\expandafter \@firstoftwo
 \else \expandafter \@secondoftwo
 \fi
}%
\providecommand \natexlab [1]{#1}%
\providecommand \enquote  [1]{``#1''}%
\providecommand \bibnamefont  [1]{#1}%
\providecommand \bibfnamefont [1]{#1}%
\providecommand \citenamefont [1]{#1}%
\providecommand \href@noop [0]{\@secondoftwo}%
\providecommand \href [0]{\begingroup \@sanitize@url \@href}%
\providecommand \@href[1]{\@@startlink{#1}\@@href}%
\providecommand \@@href[1]{\endgroup#1\@@endlink}%
\providecommand \@sanitize@url [0]{\catcode `\\12\catcode `\$12\catcode
  `\&12\catcode `\#12\catcode `\^12\catcode `\_12\catcode `\%12\relax}%
\providecommand \@@startlink[1]{}%
\providecommand \@@endlink[0]{}%
\providecommand \url  [0]{\begingroup\@sanitize@url \@url }%
\providecommand \@url [1]{\endgroup\@href {#1}{\urlprefix }}%
\providecommand \urlprefix  [0]{URL }%
\providecommand \Eprint [0]{\href }%
\providecommand \doibase [0]{http://dx.doi.org/}%
\providecommand \selectlanguage [0]{\@gobble}%
\providecommand \bibinfo  [0]{\@secondoftwo}%
\providecommand \bibfield  [0]{\@secondoftwo}%
\providecommand \translation [1]{[#1]}%
\providecommand \BibitemOpen [0]{}%
\providecommand \bibitemStop [0]{}%
\providecommand \bibitemNoStop [0]{.\EOS\space}%
\providecommand \EOS [0]{\spacefactor3000\relax}%
\providecommand \BibitemShut  [1]{\csname bibitem#1\endcsname}%
\let\auto@bib@innerbib\@empty
\bibitem [{\citenamefont {Peirce}\ \emph {et~al.}(1988)\citenamefont {Peirce},
  \citenamefont {Dahleh},\ and\ \citenamefont
  {Rabitz}}]{controlable_peirce1988optimal}%
  \BibitemOpen
  \bibfield  {author} {\bibinfo {author} {\bibfnamefont {Anthony~P}\
  \bibnamefont {Peirce}}, \bibinfo {author} {\bibfnamefont {Mohammed~A}\
  \bibnamefont {Dahleh}}, \ and\ \bibinfo {author} {\bibfnamefont {Herschel}\
  \bibnamefont {Rabitz}},\ }\bibfield  {title} {\enquote {\bibinfo {title}
  {Optimal control of quantum-mechanical systems: Existence, numerical
  approximation, and applications},}\ }\href {\doibase
  10.1103/physreva.37.4950} {\bibfield  {journal} {\bibinfo  {journal}
  {Physical Review A}\ }\textbf {\bibinfo {volume} {37}},\ \bibinfo {pages}
  {4950} (\bibinfo {year} {1988})}\BibitemShut {NoStop}%
\bibitem [{\citenamefont {Dawson}\ and\ \citenamefont
  {Nielsen}(2005)}]{compilation_dawson2005solovay}%
  \BibitemOpen
  \bibfield  {author} {\bibinfo {author} {\bibfnamefont {Christopher~M}\
  \bibnamefont {Dawson}}\ and\ \bibinfo {author} {\bibfnamefont {Michael~A}\
  \bibnamefont {Nielsen}},\ }\bibfield  {title} {\enquote {\bibinfo {title}
  {The solovay-kitaev algorithm},}\ }\href {\doibase
  10.1017/cbo9780511976667.019} {\bibfield  {journal} {\bibinfo  {journal}
  {arXiv preprint quant-ph/0505030}\ } (\bibinfo {year} {2005}),\
  10.1017/cbo9780511976667.019},\ \Eprint
  {http://arxiv.org/abs/quant-ph/0505030} {quant-ph/0505030} \BibitemShut
  {NoStop}%
\bibitem [{\citenamefont {Krotov}\ and\ \citenamefont
  {Feldman}(1983)}]{krotov_krotov1983iterative}%
  \BibitemOpen
  \bibfield  {author} {\bibinfo {author} {\bibfnamefont {VF}~\bibnamefont
  {Krotov}}\ and\ \bibinfo {author} {\bibfnamefont {IN}~\bibnamefont
  {Feldman}},\ }\bibfield  {title} {\enquote {\bibinfo {title} {An iterative
  method for solving optimal-control problems},}\ }\href {\doibase
  10.1134/s0005117913120035} {\bibfield  {journal} {\bibinfo  {journal}
  {Engineering cybernetics}\ }\textbf {\bibinfo {volume} {21}},\ \bibinfo
  {pages} {123--130} (\bibinfo {year} {1983})}\BibitemShut {NoStop}%
\bibitem [{\citenamefont {Khaneja}\ \emph {et~al.}(2005)\citenamefont
  {Khaneja}, \citenamefont {Reiss}, \citenamefont {Kehlet}, \citenamefont
  {Schulte-Herbr{\"u}ggen},\ and\ \citenamefont
  {Glaser}}]{grape_khaneja2005optimal}%
  \BibitemOpen
  \bibfield  {author} {\bibinfo {author} {\bibfnamefont {Navin}\ \bibnamefont
  {Khaneja}}, \bibinfo {author} {\bibfnamefont {Timo}\ \bibnamefont {Reiss}},
  \bibinfo {author} {\bibfnamefont {Cindie}\ \bibnamefont {Kehlet}}, \bibinfo
  {author} {\bibfnamefont {Thomas}\ \bibnamefont {Schulte-Herbr{\"u}ggen}}, \
  and\ \bibinfo {author} {\bibfnamefont {Steffen~J}\ \bibnamefont {Glaser}},\
  }\bibfield  {title} {\enquote {\bibinfo {title} {Optimal control of coupled
  spin dynamics: design of nmr pulse sequences by gradient ascent
  algorithms},}\ }\href {\doibase 10.1016/j.jmr.2004.11.004} {\bibfield
  {journal} {\bibinfo  {journal} {Journal of magnetic resonance}\ }\textbf
  {\bibinfo {volume} {172}},\ \bibinfo {pages} {296--305} (\bibinfo {year}
  {2005})}\BibitemShut {NoStop}%
\bibitem [{\citenamefont {Caneva}\ \emph {et~al.}(2011)\citenamefont {Caneva},
  \citenamefont {Calarco},\ and\ \citenamefont
  {Montangero}}]{crab_caneva2011chopped}%
  \BibitemOpen
  \bibfield  {author} {\bibinfo {author} {\bibfnamefont {Tommaso}\ \bibnamefont
  {Caneva}}, \bibinfo {author} {\bibfnamefont {Tommaso}\ \bibnamefont
  {Calarco}}, \ and\ \bibinfo {author} {\bibfnamefont {Simone}\ \bibnamefont
  {Montangero}},\ }\bibfield  {title} {\enquote {\bibinfo {title} {Chopped
  random-basis quantum optimization},}\ }\href {\doibase
  10.1103/physreva.84.022326} {\bibfield  {journal} {\bibinfo  {journal}
  {Physical Review A}\ }\textbf {\bibinfo {volume} {84}},\ \bibinfo {pages}
  {022326} (\bibinfo {year} {2011})},\ \Eprint {http://arxiv.org/abs/1103.0855}
  {1103.0855} \BibitemShut {NoStop}%
\bibitem [{\citenamefont {Tannor}\ and\ \citenamefont
  {Rice}(1985)}]{chemcontrol_tannor1985control}%
  \BibitemOpen
  \bibfield  {author} {\bibinfo {author} {\bibfnamefont {David~J}\ \bibnamefont
  {Tannor}}\ and\ \bibinfo {author} {\bibfnamefont {Stuart~A}\ \bibnamefont
  {Rice}},\ }\bibfield  {title} {\enquote {\bibinfo {title} {Control of
  selectivity of chemical reaction via control of wave packet evolution},}\
  }\href {\doibase 10.1063/1.449767} {\bibfield  {journal} {\bibinfo  {journal}
  {The Journal of chemical physics}\ }\textbf {\bibinfo {volume} {83}},\
  \bibinfo {pages} {5013--5018} (\bibinfo {year} {1985})}\BibitemShut {NoStop}%
\bibitem [{\citenamefont {Law}\ and\ \citenamefont
  {Eberly}(1996)}]{stateprep_law1996arbitrary}%
  \BibitemOpen
  \bibfield  {author} {\bibinfo {author} {\bibfnamefont {CK}~\bibnamefont
  {Law}}\ and\ \bibinfo {author} {\bibfnamefont {JH}~\bibnamefont {Eberly}},\
  }\bibfield  {title} {\enquote {\bibinfo {title} {Arbitrary control of a
  quantum electromagnetic field},}\ }\href {\doibase
  10.1103/physrevlett.76.1055} {\bibfield  {journal} {\bibinfo  {journal}
  {Physical review letters}\ }\textbf {\bibinfo {volume} {76}},\ \bibinfo
  {pages} {1055} (\bibinfo {year} {1996})}\BibitemShut {NoStop}%
\bibitem [{\citenamefont {Unanyan}\ \emph {et~al.}(1998)\citenamefont
  {Unanyan}, \citenamefont {Fleischhauer}, \citenamefont {Shore},\ and\
  \citenamefont {Bergmann}}]{ucontrol_unanyan1998robust}%
  \BibitemOpen
  \bibfield  {author} {\bibinfo {author} {\bibfnamefont {R}~\bibnamefont
  {Unanyan}}, \bibinfo {author} {\bibfnamefont {M}~\bibnamefont
  {Fleischhauer}}, \bibinfo {author} {\bibfnamefont {BW}~\bibnamefont {Shore}},
  \ and\ \bibinfo {author} {\bibfnamefont {K}~\bibnamefont {Bergmann}},\
  }\bibfield  {title} {\enquote {\bibinfo {title} {Robust creation and
  phase-sensitive probing of superposition states via stimulated raman
  adiabatic passage (stirap) with degenerate dark states},}\ }\href {\doibase
  10.1016/s0030-4018(98)00358-7} {\bibfield  {journal} {\bibinfo  {journal}
  {Optics Communications}\ }\textbf {\bibinfo {volume} {155}},\ \bibinfo
  {pages} {144--154} (\bibinfo {year} {1998})}\BibitemShut {NoStop}%
\bibitem [{\citenamefont {Mischuck}\ and\ \citenamefont
  {M{\o}lmer}(2013)}]{ucontrol_mischuck2013qudit}%
  \BibitemOpen
  \bibfield  {author} {\bibinfo {author} {\bibfnamefont {Brian}\ \bibnamefont
  {Mischuck}}\ and\ \bibinfo {author} {\bibfnamefont {Klaus}\ \bibnamefont
  {M{\o}lmer}},\ }\bibfield  {title} {\enquote {\bibinfo {title} {Qudit quantum
  computation in the jaynes-cummings model},}\ }\href {\doibase
  10.1103/physreva.87.022341} {\bibfield  {journal} {\bibinfo  {journal}
  {Physical Review A}\ }\textbf {\bibinfo {volume} {87}},\ \bibinfo {pages}
  {022341} (\bibinfo {year} {2013})},\ \Eprint {http://arxiv.org/abs/1210.4488}
  {1210.4488} \BibitemShut {NoStop}%
\bibitem [{\citenamefont {Heeres}\ \emph {et~al.}(2015)\citenamefont {Heeres},
  \citenamefont {Vlastakis}, \citenamefont {Holland}, \citenamefont
  {Krastanov}, \citenamefont {Albert}, \citenamefont {Frunzio}, \citenamefont
  {Jiang},\ and\ \citenamefont {Schoelkopf}}]{ucontrol_heeres2015cavity}%
  \BibitemOpen
  \bibfield  {author} {\bibinfo {author} {\bibfnamefont {Reinier~W}\
  \bibnamefont {Heeres}}, \bibinfo {author} {\bibfnamefont {Brian}\
  \bibnamefont {Vlastakis}}, \bibinfo {author} {\bibfnamefont {Eric}\
  \bibnamefont {Holland}}, \bibinfo {author} {\bibfnamefont {Stefan}\
  \bibnamefont {Krastanov}}, \bibinfo {author} {\bibfnamefont {Victor~V}\
  \bibnamefont {Albert}}, \bibinfo {author} {\bibfnamefont {Luigi}\
  \bibnamefont {Frunzio}}, \bibinfo {author} {\bibfnamefont {Liang}\
  \bibnamefont {Jiang}}, \ and\ \bibinfo {author} {\bibfnamefont {Robert~J}\
  \bibnamefont {Schoelkopf}},\ }\bibfield  {title} {\enquote {\bibinfo {title}
  {Cavity state manipulation using photon-number selective phase gates},}\
  }\href {\doibase 10.1103/physrevlett.115.137002} {\bibfield  {journal}
  {\bibinfo  {journal} {Physical review letters}\ }\textbf {\bibinfo {volume}
  {115}},\ \bibinfo {pages} {137002} (\bibinfo {year} {2015})},\ \Eprint
  {http://arxiv.org/abs/1503.01496} {1503.01496} \BibitemShut {NoStop}%
\bibitem [{\citenamefont {Heeres}\ \emph
  {et~al.}(2017{\natexlab{a}})\citenamefont {Heeres}, \citenamefont {Reinhold},
  \citenamefont {Ofek}, \citenamefont {Frunzio}, \citenamefont {Jiang},
  \citenamefont {Devoret},\ and\ \citenamefont
  {Schoelkopf}}]{ucontrol_heeres2017implementing}%
  \BibitemOpen
  \bibfield  {author} {\bibinfo {author} {\bibfnamefont {Reinier~W}\
  \bibnamefont {Heeres}}, \bibinfo {author} {\bibfnamefont {Philip}\
  \bibnamefont {Reinhold}}, \bibinfo {author} {\bibfnamefont {Nissim}\
  \bibnamefont {Ofek}}, \bibinfo {author} {\bibfnamefont {Luigi}\ \bibnamefont
  {Frunzio}}, \bibinfo {author} {\bibfnamefont {Liang}\ \bibnamefont {Jiang}},
  \bibinfo {author} {\bibfnamefont {Michel~H}\ \bibnamefont {Devoret}}, \ and\
  \bibinfo {author} {\bibfnamefont {Robert~J}\ \bibnamefont {Schoelkopf}},\
  }\bibfield  {title} {\enquote {\bibinfo {title} {Implementing a universal
  gate set on a logical qubit encoded in an oscillator},}\ }\href {\doibase
  10.1038/s41467-017-00045-1} {\bibfield  {journal} {\bibinfo  {journal}
  {Nature communications}\ }\textbf {\bibinfo {volume} {8}},\ \bibinfo {pages}
  {94} (\bibinfo {year} {2017}{\natexlab{a}})}\BibitemShut {NoStop}%
\bibitem [{\citenamefont {Krastanov}\ \emph {et~al.}(2015)\citenamefont
  {Krastanov}, \citenamefont {Albert}, \citenamefont {Shen}, \citenamefont
  {Zou}, \citenamefont {Heeres}, \citenamefont {Vlastakis}, \citenamefont
  {Schoelkopf},\ and\ \citenamefont {Jiang}}]{ucontrol_krastanov2015universal}%
  \BibitemOpen
  \bibfield  {author} {\bibinfo {author} {\bibfnamefont {Stefan}\ \bibnamefont
  {Krastanov}}, \bibinfo {author} {\bibfnamefont {Victor~V}\ \bibnamefont
  {Albert}}, \bibinfo {author} {\bibfnamefont {Chao}\ \bibnamefont {Shen}},
  \bibinfo {author} {\bibfnamefont {Chang-Ling}\ \bibnamefont {Zou}}, \bibinfo
  {author} {\bibfnamefont {Reinier~W}\ \bibnamefont {Heeres}}, \bibinfo
  {author} {\bibfnamefont {Brian}\ \bibnamefont {Vlastakis}}, \bibinfo {author}
  {\bibfnamefont {Robert~J}\ \bibnamefont {Schoelkopf}}, \ and\ \bibinfo
  {author} {\bibfnamefont {Liang}\ \bibnamefont {Jiang}},\ }\bibfield  {title}
  {\enquote {\bibinfo {title} {Universal control of an oscillator with
  dispersive coupling to a qubit},}\ }\href {\doibase
  10.1103/physreva.92.040303} {\bibfield  {journal} {\bibinfo  {journal}
  {Physical Review A}\ }\textbf {\bibinfo {volume} {92}},\ \bibinfo {pages}
  {040303} (\bibinfo {year} {2015})},\ \Eprint
  {http://arxiv.org/abs/1502.08015} {1502.08015} \BibitemShut {NoStop}%
\bibitem [{\citenamefont {Bacon}\ \emph {et~al.}(2001)\citenamefont {Bacon},
  \citenamefont {Childs}, \citenamefont {Chuang}, \citenamefont {Kempe},
  \citenamefont {Leung},\ and\ \citenamefont
  {Zhou}}]{ccontrol_bacon2001universal}%
  \BibitemOpen
  \bibfield  {author} {\bibinfo {author} {\bibfnamefont {Dave}\ \bibnamefont
  {Bacon}}, \bibinfo {author} {\bibfnamefont {Andrew~M}\ \bibnamefont
  {Childs}}, \bibinfo {author} {\bibfnamefont {Isaac~L}\ \bibnamefont
  {Chuang}}, \bibinfo {author} {\bibfnamefont {Julia}\ \bibnamefont {Kempe}},
  \bibinfo {author} {\bibfnamefont {Debbie~W}\ \bibnamefont {Leung}}, \ and\
  \bibinfo {author} {\bibfnamefont {Xinlan}\ \bibnamefont {Zhou}},\ }\bibfield
  {title} {\enquote {\bibinfo {title} {Universal simulation of markovian
  quantum dynamics},}\ }\href {\doibase 10.1103/physreva.64.062302} {\bibfield
  {journal} {\bibinfo  {journal} {Physical Review A}\ }\textbf {\bibinfo
  {volume} {64}},\ \bibinfo {pages} {062302} (\bibinfo {year} {2001})},\
  \Eprint {http://arxiv.org/abs/quant-ph/0008070} {quant-ph/0008070}
  \BibitemShut {NoStop}%
\bibitem [{\citenamefont {Shen}\ \emph {et~al.}(2017)\citenamefont {Shen},
  \citenamefont {Noh}, \citenamefont {Albert}, \citenamefont {Krastanov},
  \citenamefont {Devoret}, \citenamefont {Schoelkopf}, \citenamefont {Girvin},\
  and\ \citenamefont {Jiang}}]{ccontrol_shen2017quantum}%
  \BibitemOpen
  \bibfield  {author} {\bibinfo {author} {\bibfnamefont {Chao}\ \bibnamefont
  {Shen}}, \bibinfo {author} {\bibfnamefont {Kyungjoo}\ \bibnamefont {Noh}},
  \bibinfo {author} {\bibfnamefont {Victor~V}\ \bibnamefont {Albert}}, \bibinfo
  {author} {\bibfnamefont {Stefan}\ \bibnamefont {Krastanov}}, \bibinfo
  {author} {\bibfnamefont {Michel~H}\ \bibnamefont {Devoret}}, \bibinfo
  {author} {\bibfnamefont {Robert~J}\ \bibnamefont {Schoelkopf}}, \bibinfo
  {author} {\bibfnamefont {SM}~\bibnamefont {Girvin}}, \ and\ \bibinfo {author}
  {\bibfnamefont {Liang}\ \bibnamefont {Jiang}},\ }\bibfield  {title} {\enquote
  {\bibinfo {title} {Quantum channel construction with circuit quantum
  electrodynamics},}\ }\href {\doibase 10.1103/physrevb.95.134501} {\bibfield
  {journal} {\bibinfo  {journal} {Physical Review B}\ }\textbf {\bibinfo
  {volume} {95}},\ \bibinfo {pages} {134501} (\bibinfo {year} {2017})},\
  \Eprint {http://arxiv.org/abs/1611.03463} {1611.03463} \BibitemShut {NoStop}%
\bibitem [{\citenamefont {Leibfried}\ \emph {et~al.}(1996)\citenamefont
  {Leibfried}, \citenamefont {Meekhof}, \citenamefont {King}, \citenamefont
  {Monroe}, \citenamefont {Itano},\ and\ \citenamefont
  {Wineland}}]{tomography_leibfried1996experimental}%
  \BibitemOpen
  \bibfield  {author} {\bibinfo {author} {\bibfnamefont {D}~\bibnamefont
  {Leibfried}}, \bibinfo {author} {\bibfnamefont {DM}~\bibnamefont {Meekhof}},
  \bibinfo {author} {\bibfnamefont {BE}~\bibnamefont {King}}, \bibinfo {author}
  {\bibfnamefont {CH}~\bibnamefont {Monroe}}, \bibinfo {author} {\bibfnamefont
  {Wayne~M}\ \bibnamefont {Itano}}, \ and\ \bibinfo {author} {\bibfnamefont
  {David~J}\ \bibnamefont {Wineland}},\ }\bibfield  {title} {\enquote {\bibinfo
  {title} {Experimental determination of the motional quantum state of a
  trapped atom},}\ }\href {\doibase 10.1103/physrevlett.77.4281} {\bibfield
  {journal} {\bibinfo  {journal} {Physical Review Letters}\ }\textbf {\bibinfo
  {volume} {77}},\ \bibinfo {pages} {4281} (\bibinfo {year}
  {1996})}\BibitemShut {NoStop}%
\bibitem [{\citenamefont {Chuang}\ and\ \citenamefont
  {Nielsen}(1997)}]{tomography_chuang1997prescription}%
  \BibitemOpen
  \bibfield  {author} {\bibinfo {author} {\bibfnamefont {Isaac~L}\ \bibnamefont
  {Chuang}}\ and\ \bibinfo {author} {\bibfnamefont {Michael~A}\ \bibnamefont
  {Nielsen}},\ }\bibfield  {title} {\enquote {\bibinfo {title} {Prescription
  for experimental determination of the dynamics of a quantum black box},}\
  }\href {\doibase 10.1080/09500349708231894} {\bibfield  {journal} {\bibinfo
  {journal} {Journal of Modern Optics}\ }\textbf {\bibinfo {volume} {44}},\
  \bibinfo {pages} {2455--2467} (\bibinfo {year} {1997})},\ \Eprint
  {http://arxiv.org/abs/quant-ph/9610001} {quant-ph/9610001} \BibitemShut
  {NoStop}%
\bibitem [{\citenamefont {Poyatos}\ \emph {et~al.}(1997)\citenamefont
  {Poyatos}, \citenamefont {Cirac},\ and\ \citenamefont
  {Zoller}}]{tomography_poyatos1997complete}%
  \BibitemOpen
  \bibfield  {author} {\bibinfo {author} {\bibfnamefont {JF}~\bibnamefont
  {Poyatos}}, \bibinfo {author} {\bibfnamefont {J~Ignacio}\ \bibnamefont
  {Cirac}}, \ and\ \bibinfo {author} {\bibfnamefont {Peter}\ \bibnamefont
  {Zoller}},\ }\bibfield  {title} {\enquote {\bibinfo {title} {Complete
  characterization of a quantum process: the two-bit quantum gate},}\ }\href
  {\doibase 10.1103/physrevlett.78.390} {\bibfield  {journal} {\bibinfo
  {journal} {Physical Review Letters}\ }\textbf {\bibinfo {volume} {78}},\
  \bibinfo {pages} {390} (\bibinfo {year} {1997})}\BibitemShut {NoStop}%
\bibitem [{\citenamefont {Parent}\ \emph {et~al.}(2017)\citenamefont {Parent},
  \citenamefont {Roetteler}, \citenamefont {Svore},\ and\ \citenamefont
  {Svore}}]{reversiblecircuits2017svore}%
  \BibitemOpen
  \bibfield  {author} {\bibinfo {author} {\bibfnamefont {Alex}\ \bibnamefont
  {Parent}}, \bibinfo {author} {\bibfnamefont {Martin}\ \bibnamefont
  {Roetteler}}, \bibinfo {author} {\bibfnamefont {Krysta~M.}\ \bibnamefont
  {Svore}}, \ and\ \bibinfo {author} {\bibfnamefont {Krysta~M.}\ \bibnamefont
  {Svore}},\ }\bibfield  {title} {\enquote {\bibinfo {title} {Revs: A tool for
  space-optimized reversible synthesis},}\ }\bibfield  {booktitle} {\emph
  {\bibinfo {booktitle} {Proceedings of the 9th International Conference on
  Reversible Computation (RC 2017)}},\ }\href {\doibase
  10.1007/978-3-319-59936-6_7} {\ \textbf {\bibinfo {volume} {10301}},\
  \bibinfo {pages} {90--101} (\bibinfo {year} {2017})},\ \bibinfo {note}
  {arxiv.org preprint arxiv:1510.00377}\BibitemShut {NoStop}%
\bibitem [{\citenamefont {Amy}\ \emph {et~al.}(2017)\citenamefont {Amy},
  \citenamefont {Roetteler},\ and\ \citenamefont
  {Svore}}]{verifiedcircuits2017svore}%
  \BibitemOpen
  \bibfield  {author} {\bibinfo {author} {\bibfnamefont {Matthew}\ \bibnamefont
  {Amy}}, \bibinfo {author} {\bibfnamefont {Martin}\ \bibnamefont {Roetteler}},
  \ and\ \bibinfo {author} {\bibfnamefont {Krysta~M.}\ \bibnamefont {Svore}},\
  }\bibfield  {title} {\enquote {\bibinfo {title} {Verified compilation of
  space-efficient reversible circuits},}\ }\bibfield  {booktitle} {\emph
  {\bibinfo {booktitle} {Proceedings of the 28th International Conference on
  Computer Aided Verification (CAV 2017)}},\ }\href {\doibase
  10.1007/978-3-319-63390-9_1} {\ ,\ \bibinfo {pages} {3--12} (\bibinfo {year}
  {2017})},\ \Eprint {http://arxiv.org/abs/1603.01635} {1603.01635}
  \BibitemShut {NoStop}%
\bibitem [{\citenamefont {Svore}\ \emph {et~al.}(2017)\citenamefont {Svore},
  \citenamefont {Roetteler}, \citenamefont {Wiebe},\ and\ \citenamefont
  {Wecker}}]{automatedcircuits2017svore}%
  \BibitemOpen
  \bibfield  {author} {\bibinfo {author} {\bibfnamefont {Krysta~M.}\
  \bibnamefont {Svore}}, \bibinfo {author} {\bibfnamefont {Martin}\
  \bibnamefont {Roetteler}}, \bibinfo {author} {\bibfnamefont {Nathan}\
  \bibnamefont {Wiebe}}, \ and\ \bibinfo {author} {\bibfnamefont {Dave}\
  \bibnamefont {Wecker}},\ }\bibfield  {title} {\enquote {\bibinfo {title}
  {Design automation for quantum architectures},}\ }\bibfield  {booktitle}
  {\emph {\bibinfo {booktitle} {Proceedings of Design, Automation \& Test in
  Europe Conference (DATE 2017)}},\ }\href {\doibase
  10.23919/date.2017.7927196} {\ ,\ \bibinfo {pages} {1312--1317} (\bibinfo
  {year} {2017})}\BibitemShut {NoStop}%
\bibitem [{\citenamefont {Merkel}\ \emph {et~al.}(2013)\citenamefont {Merkel},
  \citenamefont {Gambetta}, \citenamefont {Smolin}, \citenamefont {Poletto},
  \citenamefont {C{\'o}rcoles}, \citenamefont {Johnson}, \citenamefont {Ryan},\
  and\ \citenamefont {Steffen}}]{tomography_merkel2013selfconsistent}%
  \BibitemOpen
  \bibfield  {author} {\bibinfo {author} {\bibfnamefont {Seth~T}\ \bibnamefont
  {Merkel}}, \bibinfo {author} {\bibfnamefont {Jay~M}\ \bibnamefont
  {Gambetta}}, \bibinfo {author} {\bibfnamefont {John~A}\ \bibnamefont
  {Smolin}}, \bibinfo {author} {\bibfnamefont {Stefano}\ \bibnamefont
  {Poletto}}, \bibinfo {author} {\bibfnamefont {Antonio~D}\ \bibnamefont
  {C{\'o}rcoles}}, \bibinfo {author} {\bibfnamefont {Blake~R}\ \bibnamefont
  {Johnson}}, \bibinfo {author} {\bibfnamefont {Colm~A}\ \bibnamefont {Ryan}},
  \ and\ \bibinfo {author} {\bibfnamefont {Matthias}\ \bibnamefont {Steffen}},\
  }\bibfield  {title} {\enquote {\bibinfo {title} {Self-consistent quantum
  process tomography},}\ }\href {\doibase 10.1103/physreva.87.062119}
  {\bibfield  {journal} {\bibinfo  {journal} {Physical Review A}\ }\textbf
  {\bibinfo {volume} {87}},\ \bibinfo {pages} {062119} (\bibinfo {year}
  {2013})}\BibitemShut {NoStop}%
\bibitem [{\citenamefont {Blume-Kohout}\ \emph {et~al.}(2013)\citenamefont
  {Blume-Kohout}, \citenamefont {Gamble}, \citenamefont {Nielsen},
  \citenamefont {Mizrahi}, \citenamefont {Sterk},\ and\ \citenamefont
  {Maunz}}]{gstomography_blume2013robust}%
  \BibitemOpen
  \bibfield  {author} {\bibinfo {author} {\bibfnamefont {Robin}\ \bibnamefont
  {Blume-Kohout}}, \bibinfo {author} {\bibfnamefont {John~King}\ \bibnamefont
  {Gamble}}, \bibinfo {author} {\bibfnamefont {Erik}\ \bibnamefont {Nielsen}},
  \bibinfo {author} {\bibfnamefont {Jonathan}\ \bibnamefont {Mizrahi}},
  \bibinfo {author} {\bibfnamefont {Jonathan~D}\ \bibnamefont {Sterk}}, \ and\
  \bibinfo {author} {\bibfnamefont {Peter}\ \bibnamefont {Maunz}},\ }\bibfield
  {title} {\enquote {\bibinfo {title} {Robust, self-consistent, closed-form
  tomography of quantum logic gates on a trapped ion qubit},}\ }\href {\doibase
  10.1038/ncomms14485} {\bibfield  {journal} {\bibinfo  {journal} {arXiv
  preprint arXiv:1310.4492}\ } (\bibinfo {year} {2013}),\
  10.1038/ncomms14485},\ \Eprint {http://arxiv.org/abs/1310.4492} {1310.4492}
  \BibitemShut {NoStop}%
\bibitem [{\citenamefont
  {Greenbaum}(2015)}]{gstomography_greenbaum2015introduction}%
  \BibitemOpen
  \bibfield  {author} {\bibinfo {author} {\bibfnamefont {Daniel}\ \bibnamefont
  {Greenbaum}},\ }\bibfield  {title} {\enquote {\bibinfo {title} {Introduction
  to quantum gate set tomography},}\ }\href {\doibase 10.1063/pt.5.028530}
  {\bibfield  {journal} {\bibinfo  {journal} {arXiv preprint arXiv:1509.02921}\
  } (\bibinfo {year} {2015}),\ 10.1063/pt.5.028530},\ \Eprint
  {http://arxiv.org/abs/1509.02921} {1509.02921} \BibitemShut {NoStop}%
\bibitem [{\citenamefont {Emerson}\ \emph {et~al.}(2005)\citenamefont
  {Emerson}, \citenamefont {Alicki},\ and\ \citenamefont
  {\.{Z}yczkowski}}]{Emerson2005}%
  \BibitemOpen
  \bibfield  {author} {\bibinfo {author} {\bibfnamefont {Joseph}\ \bibnamefont
  {Emerson}}, \bibinfo {author} {\bibfnamefont {Robert}\ \bibnamefont
  {Alicki}}, \ and\ \bibinfo {author} {\bibfnamefont {Karol}\ \bibnamefont
  {\.{Z}yczkowski}},\ }\bibfield  {title} {\enquote {\bibinfo {title} {Scalable
  noise estimation with random unitary operators},}\ }\href {\doibase
  10.1088/1464-4266/7/10/021} {\bibfield  {journal} {\bibinfo  {journal} {J.
  Opt. B}\ }\textbf {\bibinfo {volume} {7}},\ \bibinfo {pages} {S347} (\bibinfo
  {year} {2005})},\ \Eprint {http://arxiv.org/abs/quant-ph/0503243}
  {quant-ph/0503243} \BibitemShut {NoStop}%
\bibitem [{\citenamefont {Knill}\ \emph {et~al.}(2008)\citenamefont {Knill},
  \citenamefont {Leibfried}, \citenamefont {Reichle}, \citenamefont {Britton},
  \citenamefont {Blakestad}, \citenamefont {Jost}, \citenamefont {Langer},
  \citenamefont {Ozeri}, \citenamefont {Seidelin},\ and\ \citenamefont
  {Wineland}}]{benchmark_knill2008randomized}%
  \BibitemOpen
  \bibfield  {author} {\bibinfo {author} {\bibfnamefont {Emanuel}\ \bibnamefont
  {Knill}}, \bibinfo {author} {\bibfnamefont {D}~\bibnamefont {Leibfried}},
  \bibinfo {author} {\bibfnamefont {R}~\bibnamefont {Reichle}}, \bibinfo
  {author} {\bibfnamefont {J}~\bibnamefont {Britton}}, \bibinfo {author}
  {\bibfnamefont {RB}~\bibnamefont {Blakestad}}, \bibinfo {author}
  {\bibfnamefont {John~D}\ \bibnamefont {Jost}}, \bibinfo {author}
  {\bibfnamefont {C}~\bibnamefont {Langer}}, \bibinfo {author} {\bibfnamefont
  {R}~\bibnamefont {Ozeri}}, \bibinfo {author} {\bibfnamefont {Signe}\
  \bibnamefont {Seidelin}}, \ and\ \bibinfo {author} {\bibfnamefont {David~J}\
  \bibnamefont {Wineland}},\ }\bibfield  {title} {\enquote {\bibinfo {title}
  {Randomized benchmarking of quantum gates},}\ }\href {\doibase
  10.1103/physreva.77.012307} {\bibfield  {journal} {\bibinfo  {journal}
  {Physical Review A}\ }\textbf {\bibinfo {volume} {77}},\ \bibinfo {pages}
  {012307} (\bibinfo {year} {2008})}\BibitemShut {NoStop}%
\bibitem [{\citenamefont {Kimmel}\ \emph {et~al.}(2014)\citenamefont {Kimmel},
  \citenamefont {da~Silva}, \citenamefont {Ryan}, \citenamefont {Johnson},\
  and\ \citenamefont {Ohki}}]{Kimmel2013}%
  \BibitemOpen
  \bibfield  {author} {\bibinfo {author} {\bibfnamefont {Shelby}\ \bibnamefont
  {Kimmel}}, \bibinfo {author} {\bibfnamefont {Marcus~P.}\ \bibnamefont
  {da~Silva}}, \bibinfo {author} {\bibfnamefont {Colm~A.}\ \bibnamefont
  {Ryan}}, \bibinfo {author} {\bibfnamefont {Blake~R.}\ \bibnamefont
  {Johnson}}, \ and\ \bibinfo {author} {\bibfnamefont {Thomas}\ \bibnamefont
  {Ohki}},\ }\bibfield  {title} {\enquote {\bibinfo {title} {Robust extraction
  of tomographic information via randomized benchmarking},}\ }\href {\doibase
  10.1103/PhysRevX.4.011050} {\bibfield  {journal} {\bibinfo  {journal} {Phys.
  Rev. X}\ }\textbf {\bibinfo {volume} {4}},\ \bibinfo {pages} {011050}
  (\bibinfo {year} {2014})},\ \Eprint {http://arxiv.org/abs/1306.2348}
  {arXiv:1306.2348} \BibitemShut {NoStop}%
\bibitem [{\citenamefont {Heeres}\ \emph
  {et~al.}(2017{\natexlab{b}})\citenamefont {Heeres}, \citenamefont {Reinhold},
  \citenamefont {Ofek}, \citenamefont {Frunzio}, \citenamefont {Jiang},
  \citenamefont {Devoret},\ and\ \citenamefont
  {Schoelkopf}}]{heeres2017implementing}%
  \BibitemOpen
  \bibfield  {author} {\bibinfo {author} {\bibfnamefont {Reinier~W}\
  \bibnamefont {Heeres}}, \bibinfo {author} {\bibfnamefont {Philip}\
  \bibnamefont {Reinhold}}, \bibinfo {author} {\bibfnamefont {Nissim}\
  \bibnamefont {Ofek}}, \bibinfo {author} {\bibfnamefont {Luigi}\ \bibnamefont
  {Frunzio}}, \bibinfo {author} {\bibfnamefont {Liang}\ \bibnamefont {Jiang}},
  \bibinfo {author} {\bibfnamefont {Michel~H}\ \bibnamefont {Devoret}}, \ and\
  \bibinfo {author} {\bibfnamefont {Robert~J}\ \bibnamefont {Schoelkopf}},\
  }\bibfield  {title} {\enquote {\bibinfo {title} {Implementing a universal
  gate set on a logical qubit encoded in an oscillator},}\ }\href {\doibase
  10.1038/s41467-017-00045-1} {\bibfield  {journal} {\bibinfo  {journal}
  {Nature communications}\ }\textbf {\bibinfo {volume} {8}},\ \bibinfo {pages}
  {94} (\bibinfo {year} {2017}{\natexlab{b}})}\BibitemShut {NoStop}%
\bibitem [{\citenamefont {Boixo}\ \emph {et~al.}(2018)\citenamefont {Boixo},
  \citenamefont {Isakov}, \citenamefont {Smelyanskiy}, \citenamefont {Babbush},
  \citenamefont {Ding}, \citenamefont {Jiang}, \citenamefont {Bremner},
  \citenamefont {Martinis},\ and\ \citenamefont
  {Neven}}]{supremacy_boixo2018characterizing}%
  \BibitemOpen
  \bibfield  {author} {\bibinfo {author} {\bibfnamefont {Sergio}\ \bibnamefont
  {Boixo}}, \bibinfo {author} {\bibfnamefont {Sergei~V}\ \bibnamefont
  {Isakov}}, \bibinfo {author} {\bibfnamefont {Vadim~N}\ \bibnamefont
  {Smelyanskiy}}, \bibinfo {author} {\bibfnamefont {Ryan}\ \bibnamefont
  {Babbush}}, \bibinfo {author} {\bibfnamefont {Nan}\ \bibnamefont {Ding}},
  \bibinfo {author} {\bibfnamefont {Zhang}\ \bibnamefont {Jiang}}, \bibinfo
  {author} {\bibfnamefont {Michael~J}\ \bibnamefont {Bremner}}, \bibinfo
  {author} {\bibfnamefont {John~M}\ \bibnamefont {Martinis}}, \ and\ \bibinfo
  {author} {\bibfnamefont {Hartmut}\ \bibnamefont {Neven}},\ }\bibfield
  {title} {\enquote {\bibinfo {title} {Characterizing quantum supremacy in
  near-term devices},}\ }\href {\doibase 10.1038/s41567-018-0124-x} {\bibfield
  {journal} {\bibinfo  {journal} {Nature Physics}\ ,\ \bibinfo {pages} {1}}
  (\bibinfo {year} {2018})},\ \Eprint {http://arxiv.org/abs/1608.00263}
  {1608.00263} \BibitemShut {NoStop}%
\bibitem [{\citenamefont {Neill}\ \emph {et~al.}(2018)\citenamefont {Neill},
  \citenamefont {Roushan}, \citenamefont {Kechedzhi}, \citenamefont {Boixo},
  \citenamefont {Isakov}, \citenamefont {Smelyanskiy}, \citenamefont {Megrant},
  \citenamefont {Chiaro}, \citenamefont {Dunsworth}, \citenamefont {Arya} \emph
  {et~al.}}]{supremacy_neill2018blueprint}%
  \BibitemOpen
  \bibfield  {author} {\bibinfo {author} {\bibfnamefont {C}~\bibnamefont
  {Neill}}, \bibinfo {author} {\bibfnamefont {P}~\bibnamefont {Roushan}},
  \bibinfo {author} {\bibfnamefont {K}~\bibnamefont {Kechedzhi}}, \bibinfo
  {author} {\bibfnamefont {S}~\bibnamefont {Boixo}}, \bibinfo {author}
  {\bibfnamefont {SV}~\bibnamefont {Isakov}}, \bibinfo {author} {\bibfnamefont
  {V}~\bibnamefont {Smelyanskiy}}, \bibinfo {author} {\bibfnamefont
  {A}~\bibnamefont {Megrant}}, \bibinfo {author} {\bibfnamefont
  {B}~\bibnamefont {Chiaro}}, \bibinfo {author} {\bibfnamefont {A}~\bibnamefont
  {Dunsworth}}, \bibinfo {author} {\bibfnamefont {K}~\bibnamefont {Arya}},
  \emph {et~al.},\ }\bibfield  {title} {\enquote {\bibinfo {title} {A blueprint
  for demonstrating quantum supremacy with superconducting qubits},}\ }\href
  {\doibase 10.1126/science.aao4309} {\bibfield  {journal} {\bibinfo  {journal}
  {Science}\ }\textbf {\bibinfo {volume} {360}},\ \bibinfo {pages} {195--199}
  (\bibinfo {year} {2018})}\BibitemShut {NoStop}%
\bibitem [{\citenamefont {Brif}\ \emph {et~al.}(2010)\citenamefont {Brif},
  \citenamefont {Chakrabarti},\ and\ \citenamefont
  {Rabitz}}]{grapeimprov_brif2010control}%
  \BibitemOpen
  \bibfield  {author} {\bibinfo {author} {\bibfnamefont {Constantin}\
  \bibnamefont {Brif}}, \bibinfo {author} {\bibfnamefont {Raj}\ \bibnamefont
  {Chakrabarti}}, \ and\ \bibinfo {author} {\bibfnamefont {Herschel}\
  \bibnamefont {Rabitz}},\ }\bibfield  {title} {\enquote {\bibinfo {title}
  {Control of quantum phenomena: past, present and future},}\ }\href {\doibase
  10.1088/1367-2630/12/7/075008} {\bibfield  {journal} {\bibinfo  {journal}
  {New Journal of Physics}\ }\textbf {\bibinfo {volume} {12}},\ \bibinfo
  {pages} {075008} (\bibinfo {year} {2010})}\BibitemShut {NoStop}%
\bibitem [{\citenamefont {Zahedinejad}\ \emph {et~al.}(2015)\citenamefont
  {Zahedinejad}, \citenamefont {Ghosh},\ and\ \citenamefont
  {Sanders}}]{grapeimprov_zahedinejad2015high}%
  \BibitemOpen
  \bibfield  {author} {\bibinfo {author} {\bibfnamefont {Ehsan}\ \bibnamefont
  {Zahedinejad}}, \bibinfo {author} {\bibfnamefont {Joydip}\ \bibnamefont
  {Ghosh}}, \ and\ \bibinfo {author} {\bibfnamefont {Barry~C}\ \bibnamefont
  {Sanders}},\ }\bibfield  {title} {\enquote {\bibinfo {title} {High-fidelity
  single-shot toffoli gate via quantum control},}\ }\href {\doibase
  10.1103/physrevlett.114.200502} {\bibfield  {journal} {\bibinfo  {journal}
  {Physical review letters}\ }\textbf {\bibinfo {volume} {114}},\ \bibinfo
  {pages} {200502} (\bibinfo {year} {2015})}\BibitemShut {NoStop}%
\bibitem [{\citenamefont {Palittapongarnpim}\ \emph {et~al.}(2016)\citenamefont
  {Palittapongarnpim}, \citenamefont {Wittek}, \citenamefont {Zahedinejad},
  \citenamefont {Vedaie},\ and\ \citenamefont
  {Sanders}}]{grapeimprov_palittapongarnpim2016learning}%
  \BibitemOpen
  \bibfield  {author} {\bibinfo {author} {\bibfnamefont {Pantita}\ \bibnamefont
  {Palittapongarnpim}}, \bibinfo {author} {\bibfnamefont {Peter}\ \bibnamefont
  {Wittek}}, \bibinfo {author} {\bibfnamefont {Ehsan}\ \bibnamefont
  {Zahedinejad}}, \bibinfo {author} {\bibfnamefont {Shakib}\ \bibnamefont
  {Vedaie}}, \ and\ \bibinfo {author} {\bibfnamefont {Barry~C}\ \bibnamefont
  {Sanders}},\ }\bibfield  {title} {\enquote {\bibinfo {title} {Learning in
  quantum control: High-dimensional global optimization for noisy quantum
  dynamics},}\ }\href {\doibase 10.1016/j.neucom.2016.12.087} {\bibfield
  {journal} {\bibinfo  {journal} {arXiv preprint arXiv:1607.03428}\ } (\bibinfo
  {year} {2016}),\ 10.1016/j.neucom.2016.12.087}\BibitemShut {NoStop}%
\bibitem [{\citenamefont {Chen}\ \emph {et~al.}(2014)\citenamefont {Chen},
  \citenamefont {Dong}, \citenamefont {Li}, \citenamefont {Chu},\ and\
  \citenamefont {Tarn}}]{rl_chen2014fidelity}%
  \BibitemOpen
  \bibfield  {author} {\bibinfo {author} {\bibfnamefont {Chunlin}\ \bibnamefont
  {Chen}}, \bibinfo {author} {\bibfnamefont {Daoyi}\ \bibnamefont {Dong}},
  \bibinfo {author} {\bibfnamefont {Han-Xiong}\ \bibnamefont {Li}}, \bibinfo
  {author} {\bibfnamefont {Jian}\ \bibnamefont {Chu}}, \ and\ \bibinfo {author}
  {\bibfnamefont {Tzyh-Jong}\ \bibnamefont {Tarn}},\ }\bibfield  {title}
  {\enquote {\bibinfo {title} {Fidelity-based probabilistic q-learning for
  control of quantum systems},}\ }\href {\doibase 10.1109/tnnls.2013.2283574}
  {\bibfield  {journal} {\bibinfo  {journal} {IEEE transactions on neural
  networks and learning systems}\ }\textbf {\bibinfo {volume} {25}},\ \bibinfo
  {pages} {920--933} (\bibinfo {year} {2014})}\BibitemShut {NoStop}%
\bibitem [{\citenamefont {Bukov}\ \emph {et~al.}(2017)\citenamefont {Bukov},
  \citenamefont {Day}, \citenamefont {Sels}, \citenamefont {Weinberg},
  \citenamefont {Polkovnikov},\ and\ \citenamefont
  {Mehta}}]{rl_bukov2017machine}%
  \BibitemOpen
  \bibfield  {author} {\bibinfo {author} {\bibfnamefont {Marin}\ \bibnamefont
  {Bukov}}, \bibinfo {author} {\bibfnamefont {Alexandre~GR}\ \bibnamefont
  {Day}}, \bibinfo {author} {\bibfnamefont {Dries}\ \bibnamefont {Sels}},
  \bibinfo {author} {\bibfnamefont {Phillip}\ \bibnamefont {Weinberg}},
  \bibinfo {author} {\bibfnamefont {Anatoli}\ \bibnamefont {Polkovnikov}}, \
  and\ \bibinfo {author} {\bibfnamefont {Pankaj}\ \bibnamefont {Mehta}},\
  }\bibfield  {title} {\enquote {\bibinfo {title} {Machine learning meets
  quantum state preparation. the phase diagram of quantum control},}\ }\href
  {\doibase 10.1103/physrevx.8.031086} {\bibfield  {journal} {\bibinfo
  {journal} {arXiv preprint arXiv:1705.00565}\ } (\bibinfo {year} {2017}),\
  10.1103/physrevx.8.031086}\BibitemShut {NoStop}%
\bibitem [{\citenamefont {Niu}\ \emph {et~al.}(2018)\citenamefont {Niu},
  \citenamefont {Boixo}, \citenamefont {Smelyanskiy},\ and\ \citenamefont
  {Neven}}]{rl_niu2018universal}%
  \BibitemOpen
  \bibfield  {author} {\bibinfo {author} {\bibfnamefont {Murphy~Yuezhen}\
  \bibnamefont {Niu}}, \bibinfo {author} {\bibfnamefont {Sergio}\ \bibnamefont
  {Boixo}}, \bibinfo {author} {\bibfnamefont {Vadim}\ \bibnamefont
  {Smelyanskiy}}, \ and\ \bibinfo {author} {\bibfnamefont {Hartmut}\
  \bibnamefont {Neven}},\ }\bibfield  {title} {\enquote {\bibinfo {title}
  {Universal quantum control through deep reinforcement learning},}\ }\href
  {\doibase 10.1038/543171a} {\bibfield  {journal} {\bibinfo  {journal} {arXiv
  preprint arXiv:1803.01857}\ } (\bibinfo {year} {2018}),\ 10.1038/543171a},\
  \Eprint {http://arxiv.org/abs/1803.01857} {1803.01857} \BibitemShut {NoStop}%
\bibitem [{\citenamefont {Egger}\ and\ \citenamefont
  {Wilhelm}(2014)}]{grapenm_egger2014adaptive}%
  \BibitemOpen
  \bibfield  {author} {\bibinfo {author} {\bibfnamefont {DJ}~\bibnamefont
  {Egger}}\ and\ \bibinfo {author} {\bibfnamefont {FK}~\bibnamefont
  {Wilhelm}},\ }\bibfield  {title} {\enquote {\bibinfo {title} {Adaptive hybrid
  optimal quantum control for imprecisely characterized systems},}\ }\href
  {\doibase 10.1103/physrevlett.112.240503} {\bibfield  {journal} {\bibinfo
  {journal} {Physical review letters}\ }\textbf {\bibinfo {volume} {112}},\
  \bibinfo {pages} {240503} (\bibinfo {year} {2014})}\BibitemShut {NoStop}%
\bibitem [{\citenamefont {Wu}\ \emph {et~al.}(2018)\citenamefont {Wu},
  \citenamefont {Chu}, \citenamefont {Owens},\ and\ \citenamefont
  {Rabitz}}]{grapenm_wu2018data}%
  \BibitemOpen
  \bibfield  {author} {\bibinfo {author} {\bibfnamefont {Re-Bing}\ \bibnamefont
  {Wu}}, \bibinfo {author} {\bibfnamefont {Bing}\ \bibnamefont {Chu}}, \bibinfo
  {author} {\bibfnamefont {David~H}\ \bibnamefont {Owens}}, \ and\ \bibinfo
  {author} {\bibfnamefont {Herschel}\ \bibnamefont {Rabitz}},\ }\bibfield
  {title} {\enquote {\bibinfo {title} {Data-driven gradient algorithm for
  high-precision quantum control},}\ }\href {\doibase
  10.1103/physreva.97.042122} {\bibfield  {journal} {\bibinfo  {journal}
  {Physical Review A}\ }\textbf {\bibinfo {volume} {97}},\ \bibinfo {pages}
  {042122} (\bibinfo {year} {2018})}\BibitemShut {NoStop}%
\bibitem [{\citenamefont {Kelly}\ \emph {et~al.}(2014)\citenamefont {Kelly},
  \citenamefont {Barends}, \citenamefont {Campbell}, \citenamefont {Chen},
  \citenamefont {Chen}, \citenamefont {Chiaro}, \citenamefont {Dunsworth},
  \citenamefont {Fowler}, \citenamefont {Hoi}, \citenamefont {Jeffrey} \emph
  {et~al.}}]{grapenm_kelly2014optimal}%
  \BibitemOpen
  \bibfield  {author} {\bibinfo {author} {\bibfnamefont {Julian}\ \bibnamefont
  {Kelly}}, \bibinfo {author} {\bibfnamefont {R}~\bibnamefont {Barends}},
  \bibinfo {author} {\bibfnamefont {B}~\bibnamefont {Campbell}}, \bibinfo
  {author} {\bibfnamefont {Y}~\bibnamefont {Chen}}, \bibinfo {author}
  {\bibfnamefont {Z}~\bibnamefont {Chen}}, \bibinfo {author} {\bibfnamefont
  {B}~\bibnamefont {Chiaro}}, \bibinfo {author} {\bibfnamefont {A}~\bibnamefont
  {Dunsworth}}, \bibinfo {author} {\bibfnamefont {Austin~G}\ \bibnamefont
  {Fowler}}, \bibinfo {author} {\bibfnamefont {I-C}\ \bibnamefont {Hoi}},
  \bibinfo {author} {\bibfnamefont {E}~\bibnamefont {Jeffrey}},  \emph
  {et~al.},\ }\bibfield  {title} {\enquote {\bibinfo {title} {Optimal quantum
  control using randomized benchmarking},}\ }\href {\doibase
  10.1103/physrevlett.112.240504} {\bibfield  {journal} {\bibinfo  {journal}
  {Physical review letters}\ }\textbf {\bibinfo {volume} {112}},\ \bibinfo
  {pages} {240504} (\bibinfo {year} {2014})}\BibitemShut {NoStop}%
\bibitem [{\citenamefont {Ferrie}\ and\ \citenamefont
  {Moussa}(2015)}]{grapespsa_ferrie2015robust}%
  \BibitemOpen
  \bibfield  {author} {\bibinfo {author} {\bibfnamefont {Christopher}\
  \bibnamefont {Ferrie}}\ and\ \bibinfo {author} {\bibfnamefont {Osama}\
  \bibnamefont {Moussa}},\ }\bibfield  {title} {\enquote {\bibinfo {title}
  {Robust and efficient in situ quantum control},}\ }\href {\doibase
  10.1103/physreva.91.052306} {\bibfield  {journal} {\bibinfo  {journal}
  {Physical Review A}\ }\textbf {\bibinfo {volume} {91}},\ \bibinfo {pages}
  {052306} (\bibinfo {year} {2015})}\BibitemShut {NoStop}%
\bibitem [{\citenamefont {Rol}\ \emph {et~al.}(2017)\citenamefont {Rol},
  \citenamefont {Bultink}, \citenamefont {O'Brien}, \citenamefont {De~Jong},
  \citenamefont {Theis}, \citenamefont {Fu}, \citenamefont {Luthi},
  \citenamefont {Vermeulen}, \citenamefont {de~Sterke}, \citenamefont {Bruno}
  \emph {et~al.}}]{expcalib_rol2017restless}%
  \BibitemOpen
  \bibfield  {author} {\bibinfo {author} {\bibfnamefont {MA}~\bibnamefont
  {Rol}}, \bibinfo {author} {\bibfnamefont {CC}~\bibnamefont {Bultink}},
  \bibinfo {author} {\bibfnamefont {TE}~\bibnamefont {O'Brien}}, \bibinfo
  {author} {\bibfnamefont {SR}~\bibnamefont {De~Jong}}, \bibinfo {author}
  {\bibfnamefont {LS}~\bibnamefont {Theis}}, \bibinfo {author} {\bibfnamefont
  {Xiang}\ \bibnamefont {Fu}}, \bibinfo {author} {\bibfnamefont
  {F}~\bibnamefont {Luthi}}, \bibinfo {author} {\bibfnamefont {RFL}\
  \bibnamefont {Vermeulen}}, \bibinfo {author} {\bibfnamefont {JC}~\bibnamefont
  {de~Sterke}}, \bibinfo {author} {\bibfnamefont {Alessandro}\ \bibnamefont
  {Bruno}},  \emph {et~al.},\ }\bibfield  {title} {\enquote {\bibinfo {title}
  {Restless tuneup of high-fidelity qubit gates},}\ }\href {\doibase
  10.1103/physrevapplied.7.041001} {\bibfield  {journal} {\bibinfo  {journal}
  {Physical Review Applied}\ }\textbf {\bibinfo {volume} {7}},\ \bibinfo
  {pages} {041001} (\bibinfo {year} {2017})},\ \Eprint
  {http://arxiv.org/abs/1611.04815} {1611.04815} \BibitemShut {NoStop}%
\bibitem [{\citenamefont {Gross}\ \emph {et~al.}(2010)\citenamefont {Gross},
  \citenamefont {Liu}, \citenamefont {Flammia}, \citenamefont {Becker},\ and\
  \citenamefont {Eisert}}]{Gross2010}%
  \BibitemOpen
  \bibfield  {author} {\bibinfo {author} {\bibfnamefont {David}\ \bibnamefont
  {Gross}}, \bibinfo {author} {\bibfnamefont {Yi-Kai}\ \bibnamefont {Liu}},
  \bibinfo {author} {\bibfnamefont {Steven~T.}\ \bibnamefont {Flammia}},
  \bibinfo {author} {\bibfnamefont {Stephen}\ \bibnamefont {Becker}}, \ and\
  \bibinfo {author} {\bibfnamefont {Jens}\ \bibnamefont {Eisert}},\ }\bibfield
  {title} {\enquote {\bibinfo {title} {Quantum state tomography via compressed
  sensing},}\ }\href {\doibase 10.1103/PhysRevLett.105.150401} {\bibfield
  {journal} {\bibinfo  {journal} {Phys. Rev. Lett.}\ }\textbf {\bibinfo
  {volume} {105}},\ \bibinfo {pages} {150401} (\bibinfo {year} {2010})},\
  \Eprint {http://arxiv.org/abs/0909.3304} {arXiv:0909.3304} \BibitemShut
  {NoStop}%
\bibitem [{\citenamefont {Flammia}\ \emph {et~al.}(2012)\citenamefont
  {Flammia}, \citenamefont {Gross}, \citenamefont {Liu},\ and\ \citenamefont
  {Eisert}}]{Flammia2012}%
  \BibitemOpen
  \bibfield  {author} {\bibinfo {author} {\bibfnamefont {S.~T.}\ \bibnamefont
  {Flammia}}, \bibinfo {author} {\bibfnamefont {D.}~\bibnamefont {Gross}},
  \bibinfo {author} {\bibfnamefont {Y.-K.}\ \bibnamefont {Liu}}, \ and\
  \bibinfo {author} {\bibfnamefont {J.}~\bibnamefont {Eisert}},\ }\bibfield
  {title} {\enquote {\bibinfo {title} {Quantum tomography via compressed
  sensing: Error bounds, sample complexity, and efficient estimators},}\ }\href
  {\doibase 10.1088/1367-2630/14/9/095022} {\bibfield  {journal} {\bibinfo
  {journal} {New J. Phys.}\ }\textbf {\bibinfo {volume} {14}},\ \bibinfo
  {pages} {095022} (\bibinfo {year} {2012})},\ \Eprint
  {http://arxiv.org/abs/1205.2300} {arXiv:1205.2300} \BibitemShut {NoStop}%
\bibitem [{\citenamefont {Flurin}\ \emph {et~al.}(2018)\citenamefont {Flurin},
  \citenamefont {Martin}, \citenamefont {Hacohen-Gourgy},\ and\ \citenamefont
  {Siddiqi}}]{flurin2018rnn}%
  \BibitemOpen
  \bibfield  {author} {\bibinfo {author} {\bibfnamefont {Emmanuel}\
  \bibnamefont {Flurin}}, \bibinfo {author} {\bibfnamefont {Leigh~S}\
  \bibnamefont {Martin}}, \bibinfo {author} {\bibfnamefont {Shay}\ \bibnamefont
  {Hacohen-Gourgy}}, \ and\ \bibinfo {author} {\bibfnamefont {Irfan}\
  \bibnamefont {Siddiqi}},\ }\bibfield  {title} {\enquote {\bibinfo {title}
  {Using a recurrent neural network to reconstruct quantum dynamics of a
  superconducting qubit from physical observations},}\ }\href@noop {}
  {\bibfield  {journal} {\bibinfo  {journal} {arXiv preprint arXiv:1811.12420}\
  } (\bibinfo {year} {2018})},\ \Eprint {http://arxiv.org/abs/1811.12420}
  {1811.12420} \BibitemShut {NoStop}%
\bibitem [{\citenamefont {Innes}(2018)}]{innes2018blackmagic}%
  \BibitemOpen
  \bibfield  {author} {\bibinfo {author} {\bibfnamefont {Michael}\ \bibnamefont
  {Innes}},\ }\bibfield  {title} {\enquote {\bibinfo {title} {Don't unroll
  adjoint: Differentiating ssa-form programs},}\ }\href@noop {} {\bibfield
  {journal} {\bibinfo  {journal} {arXiv preprint arXiv:1810.07951}\ } (\bibinfo
  {year} {2018})}\BibitemShut {NoStop}%
\bibitem [{\citenamefont {Abadi}\ \emph {et~al.}(2015)\citenamefont {Abadi},
  \citenamefont {Agarwal}, \citenamefont {Barham}, \citenamefont {Brevdo},
  \citenamefont {Chen}, \citenamefont {Citro}, \citenamefont {Corrado},
  \citenamefont {Davis}, \citenamefont {Dean}, \citenamefont {Devin},
  \citenamefont {Ghemawat}, \citenamefont {Goodfellow}, \citenamefont {Harp},
  \citenamefont {Irving}, \citenamefont {Isard}, \citenamefont {Jia},
  \citenamefont {Jozefowicz}, \citenamefont {Kaiser}, \citenamefont {Kudlur},
  \citenamefont {Levenberg}, \citenamefont {Man\'{e}}, \citenamefont {Monga},
  \citenamefont {Moore}, \citenamefont {Murray}, \citenamefont {Olah},
  \citenamefont {Schuster}, \citenamefont {Shlens}, \citenamefont {Steiner},
  \citenamefont {Sutskever}, \citenamefont {Talwar}, \citenamefont {Tucker},
  \citenamefont {Vanhoucke}, \citenamefont {Vasudevan}, \citenamefont
  {Vi\'{e}gas}, \citenamefont {Vinyals}, \citenamefont {Warden}, \citenamefont
  {Wattenberg}, \citenamefont {Wicke}, \citenamefont {Yu},\ and\ \citenamefont
  {Zheng}}]{diffprog_tensorflow2015-whitepaper}%
  \BibitemOpen
  \bibfield  {author} {\bibinfo {author} {\bibfnamefont {Mart\'{\i}n}\
  \bibnamefont {Abadi}}, \bibinfo {author} {\bibfnamefont {Ashish}\
  \bibnamefont {Agarwal}}, \bibinfo {author} {\bibfnamefont {Paul}\
  \bibnamefont {Barham}}, \bibinfo {author} {\bibfnamefont {Eugene}\
  \bibnamefont {Brevdo}}, \bibinfo {author} {\bibfnamefont {Zhifeng}\
  \bibnamefont {Chen}}, \bibinfo {author} {\bibfnamefont {Craig}\ \bibnamefont
  {Citro}}, \bibinfo {author} {\bibfnamefont {Greg~S.}\ \bibnamefont
  {Corrado}}, \bibinfo {author} {\bibfnamefont {Andy}\ \bibnamefont {Davis}},
  \bibinfo {author} {\bibfnamefont {Jeffrey}\ \bibnamefont {Dean}}, \bibinfo
  {author} {\bibfnamefont {Matthieu}\ \bibnamefont {Devin}}, \bibinfo {author}
  {\bibfnamefont {Sanjay}\ \bibnamefont {Ghemawat}}, \bibinfo {author}
  {\bibfnamefont {Ian}\ \bibnamefont {Goodfellow}}, \bibinfo {author}
  {\bibfnamefont {Andrew}\ \bibnamefont {Harp}}, \bibinfo {author}
  {\bibfnamefont {Geoffrey}\ \bibnamefont {Irving}}, \bibinfo {author}
  {\bibfnamefont {Michael}\ \bibnamefont {Isard}}, \bibinfo {author}
  {\bibfnamefont {Yangqing}\ \bibnamefont {Jia}}, \bibinfo {author}
  {\bibfnamefont {Rafal}\ \bibnamefont {Jozefowicz}}, \bibinfo {author}
  {\bibfnamefont {Lukasz}\ \bibnamefont {Kaiser}}, \bibinfo {author}
  {\bibfnamefont {Manjunath}\ \bibnamefont {Kudlur}}, \bibinfo {author}
  {\bibfnamefont {Josh}\ \bibnamefont {Levenberg}}, \bibinfo {author}
  {\bibfnamefont {Dandelion}\ \bibnamefont {Man\'{e}}}, \bibinfo {author}
  {\bibfnamefont {Rajat}\ \bibnamefont {Monga}}, \bibinfo {author}
  {\bibfnamefont {Sherry}\ \bibnamefont {Moore}}, \bibinfo {author}
  {\bibfnamefont {Derek}\ \bibnamefont {Murray}}, \bibinfo {author}
  {\bibfnamefont {Chris}\ \bibnamefont {Olah}}, \bibinfo {author}
  {\bibfnamefont {Mike}\ \bibnamefont {Schuster}}, \bibinfo {author}
  {\bibfnamefont {Jonathon}\ \bibnamefont {Shlens}}, \bibinfo {author}
  {\bibfnamefont {Benoit}\ \bibnamefont {Steiner}}, \bibinfo {author}
  {\bibfnamefont {Ilya}\ \bibnamefont {Sutskever}}, \bibinfo {author}
  {\bibfnamefont {Kunal}\ \bibnamefont {Talwar}}, \bibinfo {author}
  {\bibfnamefont {Paul}\ \bibnamefont {Tucker}}, \bibinfo {author}
  {\bibfnamefont {Vincent}\ \bibnamefont {Vanhoucke}}, \bibinfo {author}
  {\bibfnamefont {Vijay}\ \bibnamefont {Vasudevan}}, \bibinfo {author}
  {\bibfnamefont {Fernanda}\ \bibnamefont {Vi\'{e}gas}}, \bibinfo {author}
  {\bibfnamefont {Oriol}\ \bibnamefont {Vinyals}}, \bibinfo {author}
  {\bibfnamefont {Pete}\ \bibnamefont {Warden}}, \bibinfo {author}
  {\bibfnamefont {Martin}\ \bibnamefont {Wattenberg}}, \bibinfo {author}
  {\bibfnamefont {Martin}\ \bibnamefont {Wicke}}, \bibinfo {author}
  {\bibfnamefont {Yuan}\ \bibnamefont {Yu}}, \ and\ \bibinfo {author}
  {\bibfnamefont {Xiaoqiang}\ \bibnamefont {Zheng}},\ }\href {\doibase
  10.1145/3190508.3190551} {\enquote {\bibinfo {title} {{TensorFlow}:
  Large-scale machine learning on heterogeneous systems},}\ } (\bibinfo {year}
  {2015}),\ \bibinfo {note} {software available from
  tensorflow.org}\BibitemShut {NoStop}%
\bibitem [{\citenamefont {Cram{\'e}r}(2016)}]{cramer2016bound}%
  \BibitemOpen
  \bibfield  {author} {\bibinfo {author} {\bibfnamefont {Harald}\ \bibnamefont
  {Cram{\'e}r}},\ }\href {\doibase 10.1515/9781400883868} {\emph {\bibinfo
  {title} {Mathematical methods of statistics (PMS-9)}}},\ Vol.~\bibinfo
  {volume} {9}\ (\bibinfo  {publisher} {Princeton university press},\ \bibinfo
  {year} {2016})\BibitemShut {NoStop}%
\bibitem [{\citenamefont {Leung}\ \emph {et~al.}(2017)\citenamefont {Leung},
  \citenamefont {Abdelhafez}, \citenamefont {Koch},\ and\ \citenamefont
  {Schuster}}]{grape_leung2017speedup}%
  \BibitemOpen
  \bibfield  {author} {\bibinfo {author} {\bibfnamefont {Nelson}\ \bibnamefont
  {Leung}}, \bibinfo {author} {\bibfnamefont {Mohamed}\ \bibnamefont
  {Abdelhafez}}, \bibinfo {author} {\bibfnamefont {Jens}\ \bibnamefont {Koch}},
  \ and\ \bibinfo {author} {\bibfnamefont {David}\ \bibnamefont {Schuster}},\
  }\bibfield  {title} {\enquote {\bibinfo {title} {Speedup for quantum optimal
  control from automatic differentiation based on graphics processing units},}\
  }\href {\doibase 10.1103/physreva.95.042318} {\bibfield  {journal} {\bibinfo
  {journal} {Physical Review A}\ }\textbf {\bibinfo {volume} {95}},\ \bibinfo
  {pages} {042318} (\bibinfo {year} {2017})},\ \Eprint
  {http://arxiv.org/abs/1612.04929} {1612.04929} \BibitemShut {NoStop}%
\bibitem [{\citenamefont {Mavadia}\ \emph {et~al.}(2017)\citenamefont
  {Mavadia}, \citenamefont {Frey}, \citenamefont {Sastrawan}, \citenamefont
  {Dona},\ and\ \citenamefont {Biercuk}}]{mavadia2017prediction}%
  \BibitemOpen
  \bibfield  {author} {\bibinfo {author} {\bibfnamefont {Sandeep}\ \bibnamefont
  {Mavadia}}, \bibinfo {author} {\bibfnamefont {Virginia}\ \bibnamefont
  {Frey}}, \bibinfo {author} {\bibfnamefont {Jarrah}\ \bibnamefont
  {Sastrawan}}, \bibinfo {author} {\bibfnamefont {Stephen}\ \bibnamefont
  {Dona}}, \ and\ \bibinfo {author} {\bibfnamefont {Michael~J}\ \bibnamefont
  {Biercuk}},\ }\bibfield  {title} {\enquote {\bibinfo {title} {Prediction and
  real-time compensation of qubit decoherence via machine learning},}\ }\href
  {\doibase 10.1038/ncomms14106} {\bibfield  {journal} {\bibinfo  {journal}
  {Nature communications}\ }\textbf {\bibinfo {volume} {8}},\ \bibinfo {pages}
  {14106} (\bibinfo {year} {2017})}\BibitemShut {NoStop}%
\bibitem [{\citenamefont {Pezz{\`e}}\ \emph {et~al.}(2017)\citenamefont
  {Pezz{\`e}}, \citenamefont {Ciampini}, \citenamefont {Spagnolo},
  \citenamefont {Humphreys}, \citenamefont {Datta}, \citenamefont {Walmsley},
  \citenamefont {Barbieri}, \citenamefont {Sciarrino},\ and\ \citenamefont
  {Smerzi}}]{expdesign_pezze2017optimal}%
  \BibitemOpen
  \bibfield  {author} {\bibinfo {author} {\bibfnamefont {Luca}\ \bibnamefont
  {Pezz{\`e}}}, \bibinfo {author} {\bibfnamefont {Mario~A}\ \bibnamefont
  {Ciampini}}, \bibinfo {author} {\bibfnamefont {Nicol{\`o}}\ \bibnamefont
  {Spagnolo}}, \bibinfo {author} {\bibfnamefont {Peter~C}\ \bibnamefont
  {Humphreys}}, \bibinfo {author} {\bibfnamefont {Animesh}\ \bibnamefont
  {Datta}}, \bibinfo {author} {\bibfnamefont {Ian~A}\ \bibnamefont {Walmsley}},
  \bibinfo {author} {\bibfnamefont {Marco}\ \bibnamefont {Barbieri}}, \bibinfo
  {author} {\bibfnamefont {Fabio}\ \bibnamefont {Sciarrino}}, \ and\ \bibinfo
  {author} {\bibfnamefont {Augusto}\ \bibnamefont {Smerzi}},\ }\bibfield
  {title} {\enquote {\bibinfo {title} {Optimal measurements for simultaneous
  quantum estimation of multiple phases},}\ }\href {\doibase
  10.1103/PhysRevLett.119.130504} {\bibfield  {journal} {\bibinfo  {journal}
  {Physical review letters}\ }\textbf {\bibinfo {volume} {119}},\ \bibinfo
  {pages} {130504} (\bibinfo {year} {2017})}\BibitemShut {NoStop}%
\bibitem [{\citenamefont {Matsumoto}(2002)}]{crbnoncom_matsumoto2002new}%
  \BibitemOpen
  \bibfield  {author} {\bibinfo {author} {\bibfnamefont {Keiji}\ \bibnamefont
  {Matsumoto}},\ }\bibfield  {title} {\enquote {\bibinfo {title} {A new
  approach to the cram{\'e}r-rao-type bound of the pure-state model},}\ }\href
  {\doibase 10.1088/0305-4470/35/13/307} {\bibfield  {journal} {\bibinfo
  {journal} {Journal of Physics A: Mathematical and General}\ }\textbf
  {\bibinfo {volume} {35}},\ \bibinfo {pages} {3111} (\bibinfo {year}
  {2002})}\BibitemShut {NoStop}%
\bibitem [{\citenamefont {Yang}\ \emph {et~al.}(2018)\citenamefont {Yang},
  \citenamefont {Pang}, \citenamefont {Zhou},\ and\ \citenamefont
  {Jordan}}]{crbnoncom_yang2018optimal}%
  \BibitemOpen
  \bibfield  {author} {\bibinfo {author} {\bibfnamefont {Jing}\ \bibnamefont
  {Yang}}, \bibinfo {author} {\bibfnamefont {Shengshi}\ \bibnamefont {Pang}},
  \bibinfo {author} {\bibfnamefont {Yiyu}\ \bibnamefont {Zhou}}, \ and\
  \bibinfo {author} {\bibfnamefont {Andrew~N}\ \bibnamefont {Jordan}},\
  }\bibfield  {title} {\enquote {\bibinfo {title} {Optimal measurements for
  quantum multi-parameter estimation with general states},}\ }\href@noop {}
  {\bibfield  {journal} {\bibinfo  {journal} {arXiv preprint arXiv:1806.07337}\
  } (\bibinfo {year} {2018})},\ \Eprint {http://arxiv.org/abs/1806.07337}
  {1806.07337} \BibitemShut {NoStop}%
\bibitem [{\citenamefont {Nesterov}(1983)}]{opt_nesterov1983method}%
  \BibitemOpen
  \bibfield  {author} {\bibinfo {author} {\bibfnamefont {Yurii}\ \bibnamefont
  {Nesterov}},\ }\bibfield  {title} {\enquote {\bibinfo {title} {A method of
  solving a convex programming problem with convergence rate {$O(1/k^2)$}},}\
  }in\ \href {\doibase 10.1137/1.9781611970791} {\emph {\bibinfo {booktitle}
  {Soviet Mathematics Doklady}}},\ Vol.~\bibinfo {volume} {27}\ (\bibinfo
  {year} {1983})\ pp.\ \bibinfo {pages} {372--376}\BibitemShut {NoStop}%
\bibitem [{\citenamefont {Kingma}\ and\ \citenamefont
  {Ba}(2014)}]{opt_kingma2014adam}%
  \BibitemOpen
  \bibfield  {author} {\bibinfo {author} {\bibfnamefont {Diederik~P}\
  \bibnamefont {Kingma}}\ and\ \bibinfo {author} {\bibfnamefont {Jimmy}\
  \bibnamefont {Ba}},\ }\bibfield  {title} {\enquote {\bibinfo {title} {Adam: A
  method for stochastic optimization},}\ }\href {\doibase 10.1063/pt.5.028530}
  {\bibfield  {journal} {\bibinfo  {journal} {arXiv preprint arXiv:1412.6980}\
  } (\bibinfo {year} {2014}),\ 10.1063/pt.5.028530},\ \Eprint
  {http://arxiv.org/abs/1412.6980} {1412.6980} \BibitemShut {NoStop}%
\bibitem [{\citenamefont {MacKay}(1992)}]{optbayes_mackay1992practical}%
  \BibitemOpen
  \bibfield  {author} {\bibinfo {author} {\bibfnamefont {David~JC}\
  \bibnamefont {MacKay}},\ }\bibfield  {title} {\enquote {\bibinfo {title} {A
  practical bayesian framework for backpropagation networks},}\ }\href
  {\doibase 10.1162/neco.1992.4.3.448} {\bibfield  {journal} {\bibinfo
  {journal} {Neural computation}\ }\textbf {\bibinfo {volume} {4}},\ \bibinfo
  {pages} {448--472} (\bibinfo {year} {1992})}\BibitemShut {NoStop}%
\bibitem [{\citenamefont {Pratt}(1993)}]{transfer_pratt1993discriminability}%
  \BibitemOpen
  \bibfield  {author} {\bibinfo {author} {\bibfnamefont {Lorien~Y}\
  \bibnamefont {Pratt}},\ }\bibfield  {title} {\enquote {\bibinfo {title}
  {Discriminability-based transfer between neural networks},}\ }in\ \href
  {\doibase 10.1080/095400996116866} {\emph {\bibinfo {booktitle} {Advances in
  neural information processing systems}}}\ (\bibinfo {year} {1993})\ pp.\
  \bibinfo {pages} {204--211}\BibitemShut {NoStop}%
\bibitem [{\citenamefont {Tsang}(2009)}]{weak_tsang2009time}%
  \BibitemOpen
  \bibfield  {author} {\bibinfo {author} {\bibfnamefont {Mankei}\ \bibnamefont
  {Tsang}},\ }\bibfield  {title} {\enquote {\bibinfo {title} {Time-symmetric
  quantum theory of smoothing},}\ }\href {\doibase
  10.1103/PhysRevLett.102.250403} {\bibfield  {journal} {\bibinfo  {journal}
  {Physical Review Letters}\ }\textbf {\bibinfo {volume} {102}},\ \bibinfo
  {pages} {250403} (\bibinfo {year} {2009})}\BibitemShut {NoStop}%
\bibitem [{\citenamefont {Cortez}\ \emph {et~al.}(2017)\citenamefont {Cortez},
  \citenamefont {Chantasri}, \citenamefont {Garc{\'\i}a-Pintos}, \citenamefont
  {Dressel},\ and\ \citenamefont {Jordan}}]{weak_cortez2017rapid}%
  \BibitemOpen
  \bibfield  {author} {\bibinfo {author} {\bibfnamefont {Luis}\ \bibnamefont
  {Cortez}}, \bibinfo {author} {\bibfnamefont {Areeya}\ \bibnamefont
  {Chantasri}}, \bibinfo {author} {\bibfnamefont {Luis~Pedro}\ \bibnamefont
  {Garc{\'\i}a-Pintos}}, \bibinfo {author} {\bibfnamefont {Justin}\
  \bibnamefont {Dressel}}, \ and\ \bibinfo {author} {\bibfnamefont {Andrew~N}\
  \bibnamefont {Jordan}},\ }\bibfield  {title} {\enquote {\bibinfo {title}
  {Rapid estimation of drifting parameters in continuously measured quantum
  systems},}\ }\href@noop {} {\bibfield  {journal} {\bibinfo  {journal}
  {Physical Review A}\ }\textbf {\bibinfo {volume} {95}},\ \bibinfo {pages}
  {012314} (\bibinfo {year} {2017})}\BibitemShut {NoStop}%
\bibitem [{\citenamefont {Kueng}\ \emph {et~al.}(2016)\citenamefont {Kueng},
  \citenamefont {Long}, \citenamefont {Doherty},\ and\ \citenamefont
  {Flammia}}]{Kueng2016}%
  \BibitemOpen
  \bibfield  {author} {\bibinfo {author} {\bibfnamefont {Richard}\ \bibnamefont
  {Kueng}}, \bibinfo {author} {\bibfnamefont {David~M.}\ \bibnamefont {Long}},
  \bibinfo {author} {\bibfnamefont {Andrew~C.}\ \bibnamefont {Doherty}}, \ and\
  \bibinfo {author} {\bibfnamefont {Steven~T.}\ \bibnamefont {Flammia}},\
  }\bibfield  {title} {\enquote {\bibinfo {title} {Comparing experiments to the
  fault-tolerance threshold},}\ }\href {\doibase
  10.1103/PhysRevLett.117.170502} {\bibfield  {journal} {\bibinfo  {journal}
  {Phys. Rev. Lett.}\ }\textbf {\bibinfo {volume} {117}},\ \bibinfo {pages}
  {170502} (\bibinfo {year} {2016})},\ \Eprint
  {http://arxiv.org/abs/1510.05653} {arXiv:1510.05653} \BibitemShut {NoStop}%
\bibitem [{\citenamefont {Tibshirani}(1996)}]{Tibshirani1996}%
  \BibitemOpen
  \bibfield  {author} {\bibinfo {author} {\bibfnamefont {Robert}\ \bibnamefont
  {Tibshirani}},\ }\bibfield  {title} {\enquote {\bibinfo {title} {Regression
  shrinkage and selection via the lasso},}\ }\href {\doibase 10.2307/2346178}
  {\bibfield  {journal} {\bibinfo  {journal} {Journal of the Royal Statistical
  Society: Series B (Methodological)}\ }\textbf {\bibinfo {volume} {58}},\
  \bibinfo {pages} {267--288} (\bibinfo {year} {1996})}\BibitemShut {NoStop}%
\bibitem [{\citenamefont {Akaike}(1974)}]{akaike1974new}%
  \BibitemOpen
  \bibfield  {author} {\bibinfo {author} {\bibfnamefont {Hirotugu}\
  \bibnamefont {Akaike}},\ }\bibfield  {title} {\enquote {\bibinfo {title} {A
  new look at the statistical model identification},}\ }\href {\doibase
  10.1007/978-1-4612-1694-0_16} {\bibfield  {journal} {\bibinfo  {journal}
  {IEEE transactions on automatic control}\ }\textbf {\bibinfo {volume} {19}},\
  \bibinfo {pages} {716--723} (\bibinfo {year} {1974})}\BibitemShut {NoStop}%
\end{thebibliography}%

\newpage{}\clearpage{}

\appendix

\section*{Supplementary Materials for Stochastic Estimation of Dynamical Variables}

The software we provide can be used both for running our method of
stochastic Hamiltonian estimation and in reverse for performing gradient-based
control (\`{a} la GRAPE). We admit both constant-in-time drive pulses and time-dependent
drive pulses. For Hamiltonian dynamics we use an integrator that directly
computes the evolution operator through diagonalization of the Hamiltonian.
For Lindbladian dynamics we use an RK4 integrator. All integrators
are implemented as fully differentiable operations, i.e.\ their gradients
with respect to any parameters (control pulse parameters, parameters
of the Hamiltonian or the Lindbladian) are computable analytically,
to be used for parameter estimation (e.g.\ our stochastic Hamiltonian
estimation method), experimental design (e.g.\ maximizing Fisher information),
or control. Later in this appendix we describe the implementation
in more details.

For most of the numerical tests of our method we use a simulated hardware
with $Q=3$ qubits governed by a Hamiltonian containing Pauli drives
($\sigma_{x}$, $\sigma_{y}$, and $\sigma_{z}$) for each of the
three qubits (marked in many of the figures as $X1$ through $Z3$)
as well as nearest neighbor exchange interactions $\sigma_{i}^{+}\sigma_{j}^{-}+h.c.$
(marked in plots as pairs $12$, $23$, and $31$, depending on which
pair of qubits they correspond to). For exact numerical values (which
are on the order of 1, generated randomly), consult the referenced source
code. We have $D=12$ drives for each of the $M=12$ components of
the Hamiltonian. Moreover, there is no mixing between the drives (i.e.\
the true $\bm{\alpha}$ is diagonal). The $\sigma_{z}$ and the nearest
neighbor contributions to the Hamiltonian are present even in the
absence of drives, i.e.\ the corresponding $\bm{\beta}$ coefficients
are non-zero; the rest of the $\bm{\beta}$ vector ($\sigma_{x}$
and $\sigma_{y}$) is zero.

The source code for this project is available at \href{https://www.github.com/Krastanov/hamiltonian_estimation}{github.com/Krastanov/hamiltonian\_{}estimation}.

\section{\label{sec:validation-choice}The
Choice of Validation Function: How to Evaluate the Quality of Parameter
Estimation}

\begin{figure*}
\includegraphics[height=4cm]{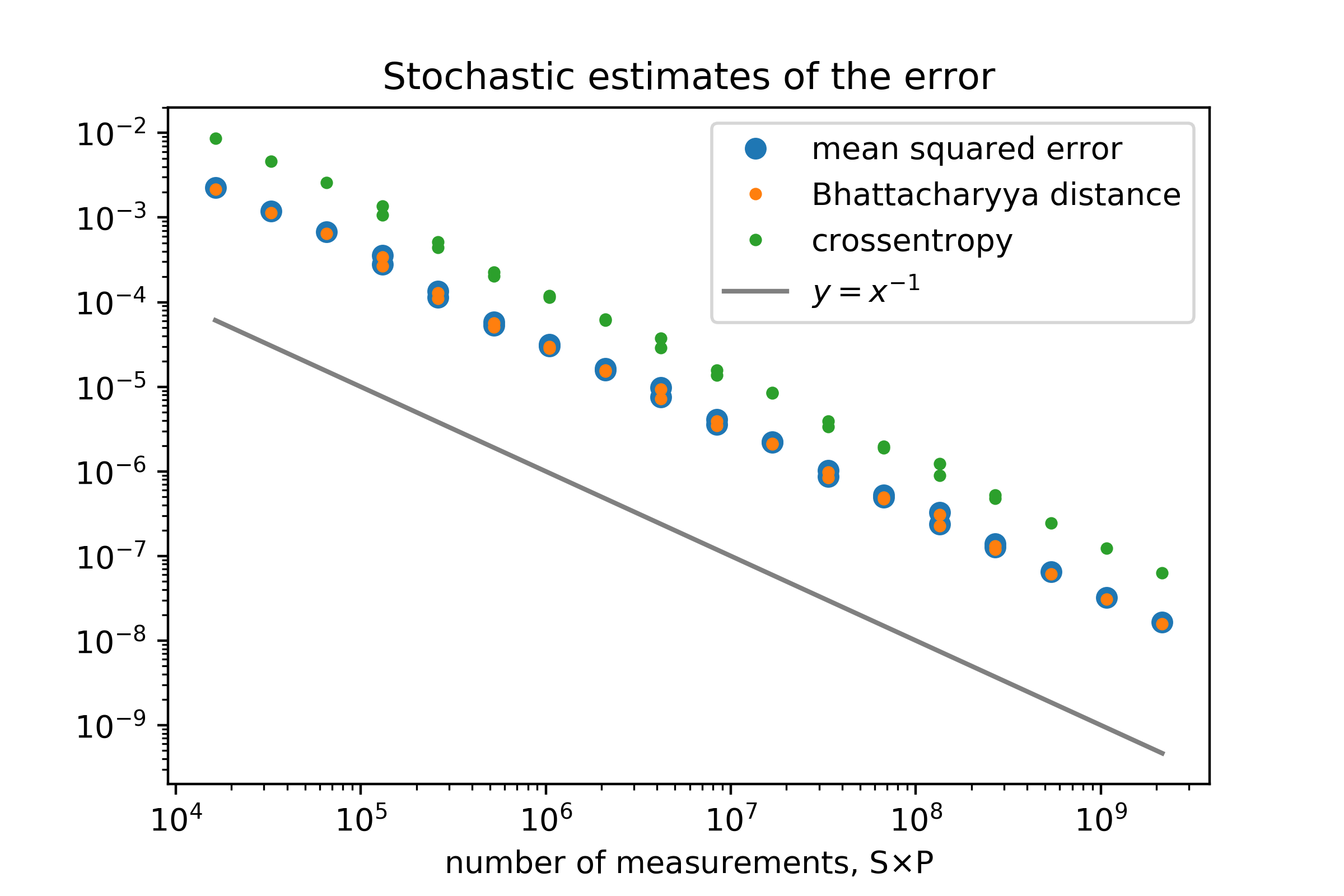}(a)\includegraphics[height=4cm]{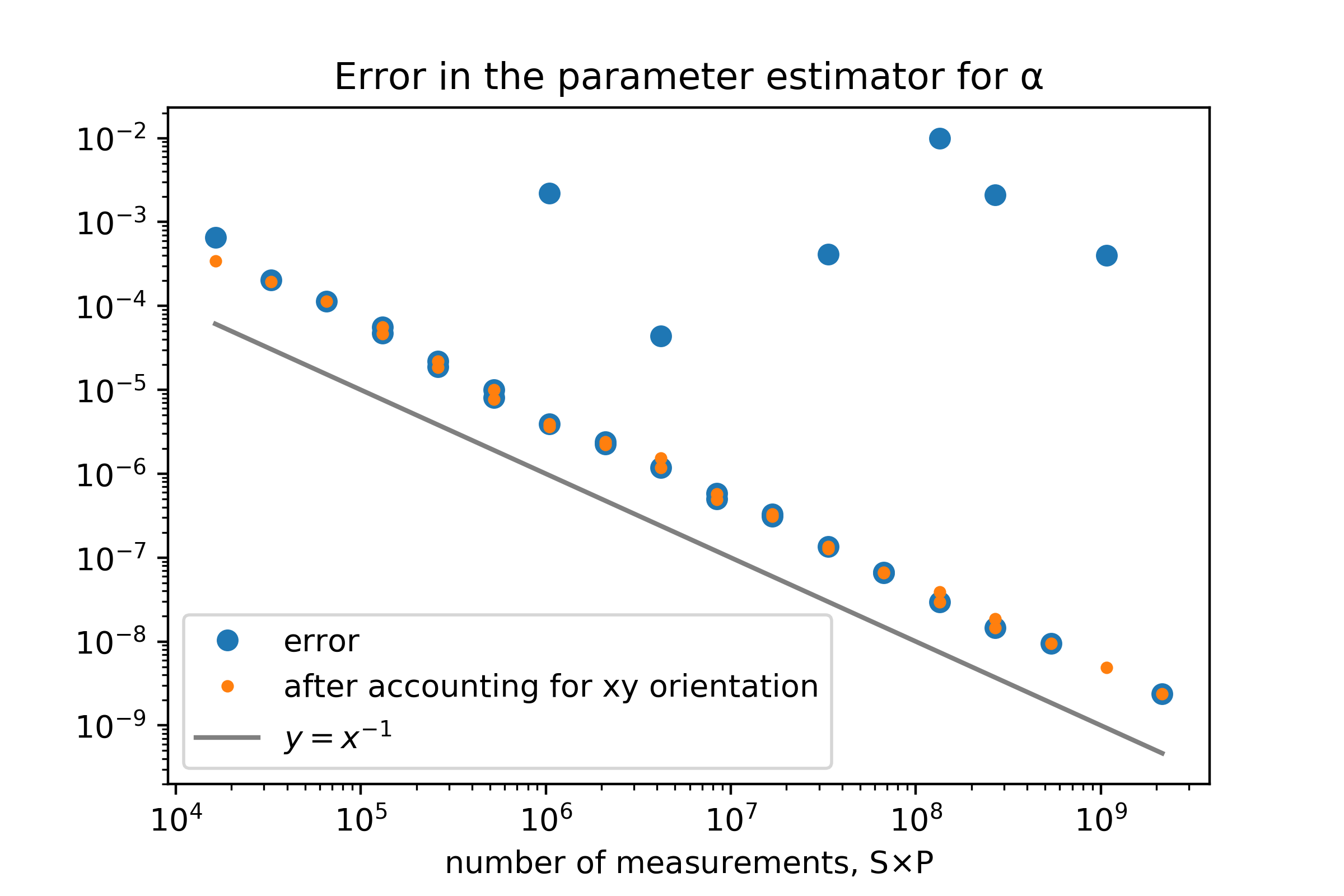}(b)\includegraphics[height=4cm]{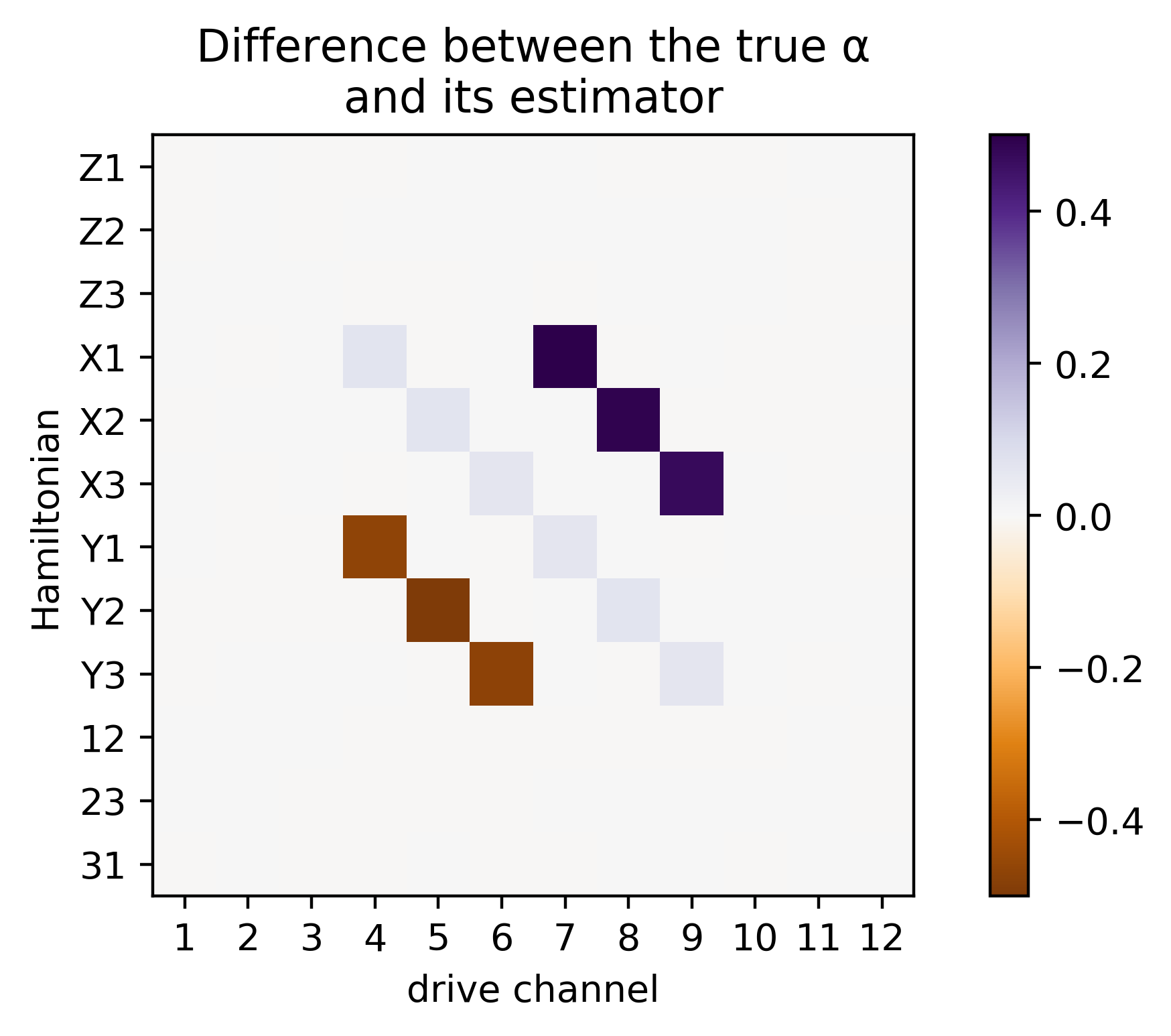}(c)

\caption{\label{fig:distances}(a)
The simple validation function we use is a good proxy for more principled
measures of distance between the measured and predicted population
distributions. This assures that the predicted probability distribution
of states closely resembles the distribution of states that actually
occurs on the hardware. The distance measures in the plot are ``mean
squared error'' $\left\langle \left(\hat{\bm{p}}-\tilde{\bm{p}}\right)^{2}\right\rangle $,
the cross entropy surplus $\left\langle \hat{\bm{p}}.\log\left(\hat{\bm{p}}\right)-\hat{\bm{p}}.\log\left(\tilde{\bm{p}}\right)\right\rangle $,
and the Bhattacharya distance $\left\langle -\log\left(\sqrt{\hat{\bm{p}}.\tilde{\bm{p}}}\right)\right\rangle $.
The x axis is the number of measurements taken $S\times P$. (b) The exact error in the estimator. As mentioned
in the main text, we are estimating arrays $\bm{\alpha}$ and $\bm{\beta}$
such that $\tilde{H}(\bm{\alpha},\bm{\beta};\bm{d})=\stackrel[k=1]{M}{\sum}a_{k}A_{k}$,
where $a_{k}=\stackrel[l=1]{M}{\sum}\alpha_{kl}d_{l}+\beta_{k}$.
We plot the mean squared difference between the estimate for components
of $\bm{\alpha}$ and their true values. Notably, there are a number
of outliers for which the estimator for $\bm{\alpha}$ is particularly
bad. However, when we calculate the value of the gauge degree of freedom
and perform the corresponding global rotation, the error in the estimator
drops to the expected error floor. The majority of test cases do not
show errors even without accounting for the global rotation due to
the regularization we have imposed on our parameters; it causes the
simplest/sparsest parameters to be chosen by the estimator. In (c)
we show the worst of the outliers and we can see that the large error
in $\bm{\alpha}$ is due to a gauge degree of freedom (a simultaneous
rotation around all three z axes) that does not affect the measurements
(represented graphically is the difference between the matrix $\bm{\alpha}$
and the estimate we have obtained for it).}
\end{figure*}

Our parameter
estimation procedure is inherently stochastic, which is both a blessing
and a curse. It is what permits us to use the entirety of the measurements
performed on the hardware and reach the information-theoretic limits
of precision. For numerical simplicity we choose a simple cost function---the mean squared error between measured and predicted populations---however 
we do observe consistent behavior independent of the particular
choice of stochastic validation function (see Fig.~\ref{fig:distances}a).

However, there
might be a ``gauge'' degree of freedom in the way we have parameterized
the Hamiltonian, which does not affect the actual dynamics of the
system. As such, there might be parameters whose value will neither
affect the measurement data we gather, nor will be of any consequence
when designing control pulses down the road. Our choice of ``indirect''
validation function permits us to disregard this degree of freedom,
as the validation function is sensitive only to the prediction of
our model, not to the particular parameterization we have used in
our model.

However, it
would be instructive to observe how some of these ``unimportant''
parameters behave in our estimator. For instance, the test system
used in much of this manuscript has a Hamiltonian of the form 
\begin{align*}
H(\bm{d})= & (\varepsilon_{1}+\delta_{1})\sigma_{z1}+(\varepsilon_{2}+\delta_{2})\sigma_{z2}+(\varepsilon_{3}+\delta_{3})\sigma_{z3}\\
+ & \delta_{4}\sigma_{x1}+\delta_{5}\sigma_{x2}+\delta_{6}\sigma_{x3}+\delta_{7}\sigma_{y1}+\delta_{8}\sigma_{y2}+\delta_{9}\sigma_{y3}\\
+ & (\eta_{1}+\delta_{10})\left(\sigma_{1}^{+}\sigma_{2}^{-}{\scriptstyle +h.c.}\right)\\
+ & (\eta_{2}+\delta_{11})\left(\sigma_{2}^{+}\sigma_{3}^{-}{\scriptstyle +h.c.}\right)\\
+ & (\eta_{3}+\delta_{12})\left(\sigma_{3}^{+}\sigma_{1}^{-}{\scriptstyle +h.c.}\right),\\
 & \text{where }\delta_{i}=\kappa_{i}d_{i}
\end{align*}
 where $\kappa_{i}$ denote the coupling strengths for each drive
pulse component, and $\varepsilon_{i}$ and $\eta_{i}$ are the strengths
of Hamiltonian components that are present even in the absence of
drives. Our estimator tries to model this Hamiltonian as $\tilde{H}(\bm{\alpha},\bm{\beta};\bm{d})=\stackrel[k=1]{M}{\sum}a_{k}A_{k}$,
where $a_{k}=\stackrel[l=1]{M}{\sum}\alpha_{kl}d_{l}+\beta_{k}$ and
the set of possible $A_{k}$ is $\{\sigma_{z1},\sigma_{z2},\sigma_{z3},\sigma_{x1},\sigma_{x2},\sigma_{x3},\sigma_{y1},\sigma_{y2},\sigma_{y3},\sigma_{1}^{+}\sigma_{2}^{-}+h.c.,\sigma_{2}^{+}\sigma_{3}^{-}+h.c.,\sigma_{3}^{+}\sigma_{1}^{-}+h.c.,\}$.
We can see that the ``true'' $\bm{\beta}$ is zero for the subset
of operators $\{\sigma_{x1},\sigma_{x2},\sigma_{x3},\sigma_{y1},\sigma_{y2},\sigma_{y3}\}$,
and the ``true'' $\bm{\alpha}$ is a diagonal matrix. However, in
the measurement data used by the estimator, there is nothing that
defines the orientation of the x axis in the xy plane. If we rotate
all the qubits by the same angle around the z axis, the physics of
the system (and the measured data) will not change. In other words,
for this particular Hamiltonian, the change of xy basis generator
$\sigma_{z1}\otimes\sigma_{z2}\otimes\sigma_{z3}$ commutes
with the zero drive Hamiltonian $H(\bm{0})$, hence non-diagonal $\bm{\alpha}$
such as the one in Fig.~\ref{fig:distances}c would predict the same evolution
as the diagonal one.

In the case
of a microwave superconducting circuit implementation, 
this freedom in the parameter would be due the fact that the 
master oscillator used in mixing the control signals has an arbitrary 
initial phase (this phase defines
our choice of x and y axis in the rotating frame for all qubits).

More generally,
as long as the random pulses we use in our stochastic cost function
are representative of the pulses that will be used to control the
hardware, such ``gauge'' degrees of freedom are inconsequential:
if they do not affect our cost function, they will not affect the
result of a computation running on the hardware either.

\section{The Choice of Cost Function, Estimator Efficiency, and the Cram\'{e}r--Rao
Bound}

\begin{figure}
\includegraphics[width=7cm]{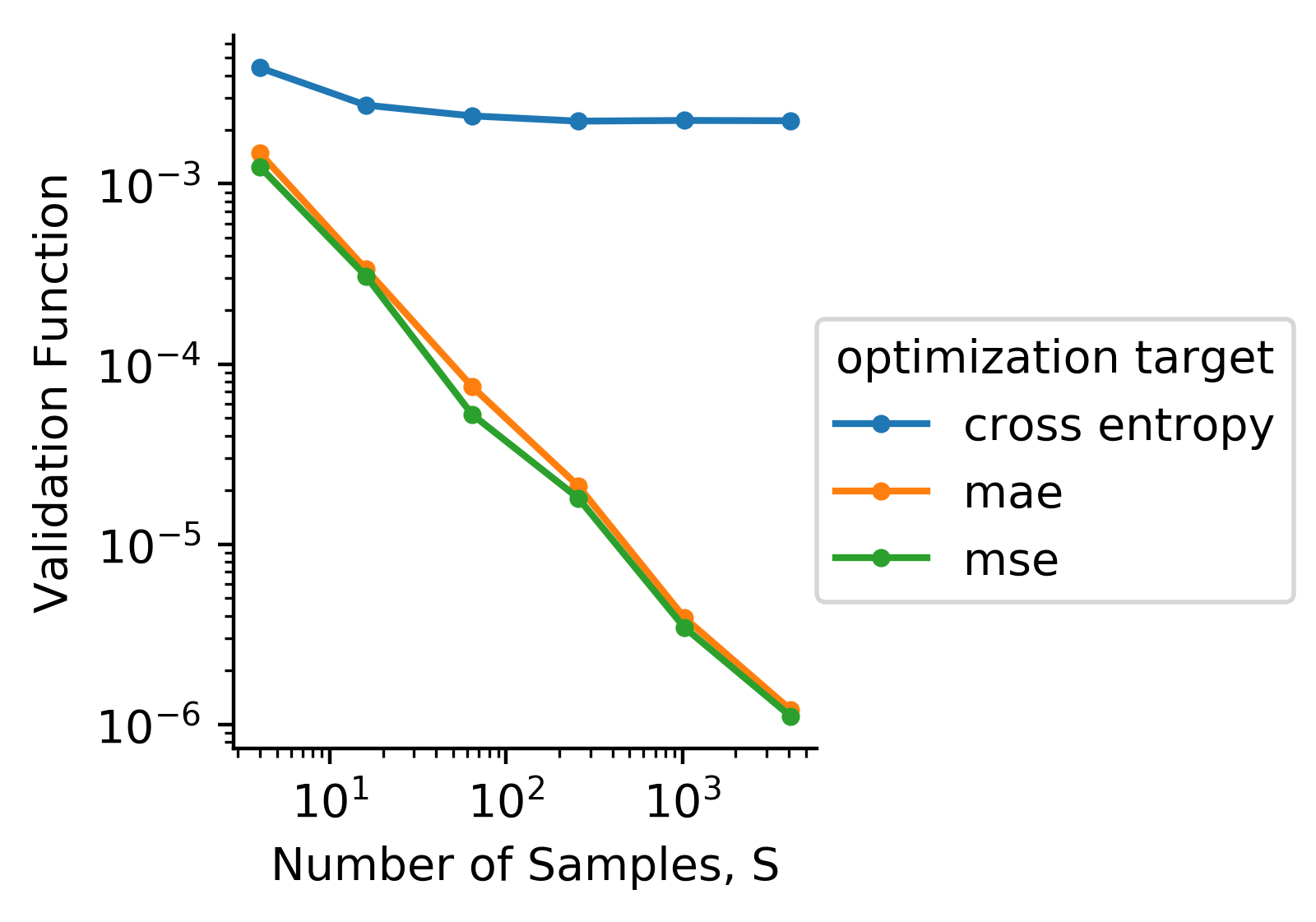}

\caption{\label{fig:costfunctions}This
figure shows the optimized validation function (employing always a
mean square error as a distance measure) for different choices of distance
measure in the cost function. We can see that for our particular test
system, using mean squared error (mse) or mean absolute error (mae)
work equally well. However, if we use cross entropy as our distance
measure, our estimator fails to converge. Although cross entropy is
known to be the best choice when convergence is guaranteed (because
it provides a maximum likelihood estimator), in many real settings,
like ours, it causes numerical issues, and other distance functions
need to be considered.}
\end{figure}

In the main text we invoked the Cram\'{e}r--Rao bound in order
to explain the $\frac{1}{P\times S}$ fidelity scaling that we obtain in the case
of unbiased estimation (before considering the effects of SPAM). That
bound is a general information inequality that expresses a lower bound
on the variance. Besides verifying numerically that we follow the
bound (as done in the main text), it would be instructive to derive
the variance of our particular estimator and compare it to this bound.
This would help inform and defend our choice of $\text{dist}(\hat{\bm{p}},\tilde{\bm{p}})$,
which until now has been based on computational convenience.

Let $\bm{G}(\bm{\omega})$ be the gradient of $C(\bm{\omega})$, and
$\bm{H}(\bm{\omega})$ be the Hessian matrix. For our estimator we
have $\bm{G}(\bm{\hat{\omega}})=0$ where $\bm{\hat{\omega}}$ is
our estimate for the true value $\bm{\omega_{0}}$. Expanding $\bm{G}(\bm{\omega})$
around $\bm{\omega_{0}}$ we get 
\[
\bm{G}(\bm{\omega})\approx\bm{G}(\bm{\omega_{0}})+\bm{H}(\bm{\omega_{0}}).(\bm{\omega}-\bm{\omega_{0}}),
\]
 which leads us to 
\[
\bm{\hat{\omega}}\approx\bm{\omega_{0}}-\bm{H}(\bm{\omega_{0}})^{-1}.\bm{G}(\bm{\omega_{0}}).
\]

This permits us to estimate variances and covariances for the parameters
\begin{align*}
\mathbb{E}\left(\left(\hat{\omega}_{i}-\omega_{0i}\right)\left(\hat{\omega}_{j}-\omega_{0j}\right)\right)\approx & \bm{\mathcal{I}_{C}}(\bm{\omega_{0}})_{ij}^{-1},\\
\text{where }\bm{\mathcal{I}_{C}}(\bm{\omega_{0}}) = & BA^{-1}B\\
A_{ij}= & \mathbb{E}\left.\left(\frac{\partial C}{\partial\omega_{i}}\frac{\partial C}{\partial\omega_{j}}\right)\right|_{\bm{\omega}=\bm{\omega_{0}}}\\
B_{ij}= & \mathbb{E}\left.\left(-\frac{\partial^{2}C}{\partial\omega_{i}\partial\omega_{j}}\right)\right|_{\bm{\omega}=\bm{\omega_{0}}}.
\end{align*}

If we pick a maximal likelihood estimator (i.e.\ if the distance function
is the cross entropy $\text{dist}(\hat{\bm{p}},\tilde{\bm{p}})=\tilde{\bm{p}}.\log(\hat{\bm{p}})$),
then we have $A=B=\bm{\mathcal{I}_{F}}$ and $\bm{\mathcal{I}_{C}}=\bm{\mathcal{I}_{F}}$,
where $\bm{\mathcal{I}_{F}}$ is the Fisher information matrix, therefore
proving that we would saturate the Cram\'{e}r--Rao bound and
have a fully efficient estimator. In practice, one would need to consider
the numerical stability of gradient descent as well: the estimator
would be run with a number of different distance functions, to see
empirically which one proves most reliable numerically. This led us
to use the mean squared error in our examples, but other choices might
be more performant in other settings (Fig.~\ref{fig:costfunctions}).

For completeness we also derive the explicit form of the Fisher information
matrix used in arguments in the main text. First we denote the log-likelihood
function 
\[
l(\omega)=\log\left(\prod_{i=1}^{P}\left(S!\prod_{k=1}^{2^{Q}}\frac{\tilde{p}_{ik}(\bm{\omega})^{\hat{p}_{ik}S}}{(\hat{p}_{ik}S)!}\right)\right),
\]
 where $\tilde{p}_{ik}(\bm{\omega})$ is the predicted population in the $k$-th
state for the $i$-th pulse in the training set $\{\bm{d}_{1},\dots,\bm{d}_{P}\}$
and $\hat{p}_{ik}$ is the measured population for that state and pulse
(by taking $S$ samples). This leads us to an expression for the Fisher
information 
\begin{align*}
\bm{{\mathcal{I}_{F}}}_{ij}= & - \mathbb{E}\left(\frac{\partial^{2}l(\omega)}{\partial\omega_{i}\partial\omega_{j}}\right)\\
= & -\mathbb{E}\left(\sum_{l=1}^{P}\sum_{k=1}^{2^{Q}}S\hat{p}_{lk}\left(\frac{1}{\tilde{p}_{lk}}\frac{\partial^{2}\tilde{p}_{lk}}{\partial\omega_{i}\partial\omega_{j}}-\frac{1}{\tilde{p}_{lk}^{2}}\frac{\partial \tilde{p}_{lk}}{\partial\omega_{i}}\frac{\partial \tilde{p}_{lk}}{\partial\omega_{j}}\right)\right)\\
= & S\mathbb{E}\left(\sum_{l=1}^{P}\sum_{k=1}^{2^{Q}}\hat{p}_{lk}\left(\frac{1}{\tilde{p}_{lk}^{2}}\frac{\partial \tilde{p}_{lk}}{\partial\omega_{i}}\frac{\partial \tilde{p}_{lk}}{\partial\omega_{j}}\right)\right)\\
\approx & PS\left\langle \sum_{k=1}^{2^{Q}}\frac{1}{\tilde{p}_{k}}\frac{\partial \tilde{p}_{k}}{\partial\omega_{i}}\frac{\partial \tilde{p}_{k}}{\partial\omega_{j}}\right\rangle ,
\end{align*}
 where $\langle\dots\rangle$ denotes average over all pulses in the
training set. Also, $\sum_{k=1}^{2^{Q}}\frac{1}{\tilde{p}_{k}}\frac{\partial \tilde{p}_{k}}{\partial\omega_{i}}\frac{\partial \tilde{p}_{k}}{\partial\omega_{j}}$
happens to be the Fisher information for a single measurement of a
single pulse, i.e.\ the Fisher information is additive.

\section{Convergence, Overfitting, and Model Errors}

In the previous discussion we neglected issues of convergence, overfitting, and model errors. 
We use two complementary tools to fight these problems.
Firstly, overfitting or being stuck in a valley of the cost function can both be 
avoided with an annealed regularization cost applied to the parameters. 

The specific annealed and regularized cost function that we use 
as a function of iteration $k$ is of the form
\begin{equation*}
	C(\bm{\omega}) + \lambda_k \| \bm{\omega}\|_1
\end{equation*}
where $C(\bm{\omega})$ is given by Eq.~\ref{eq:cost} and the $\lambda_k \ge 0$ are annealed as $k$ increases, meaning that our estimator is essentially a LASSO-type estimator~\cite{Tibshirani1996}.

From a
purely practical point of view this avoids the initial steps of the gradient
descent going in a wildly unphysical direction of pathological values for the
parameters. More importantly, minimizing the 1-norm of the parameters provides
for a sparser realistic parametrization. For particularly difficult programs we
observed that empirical tricks akin to MacKay's regularization annealing
schedule~\citep{optbayes_mackay1992practical}, where the variance in the
regularization is forced to follow the variance in the cost enables reliable
convergence.

\begin{figure}
\includegraphics[width=7cm]{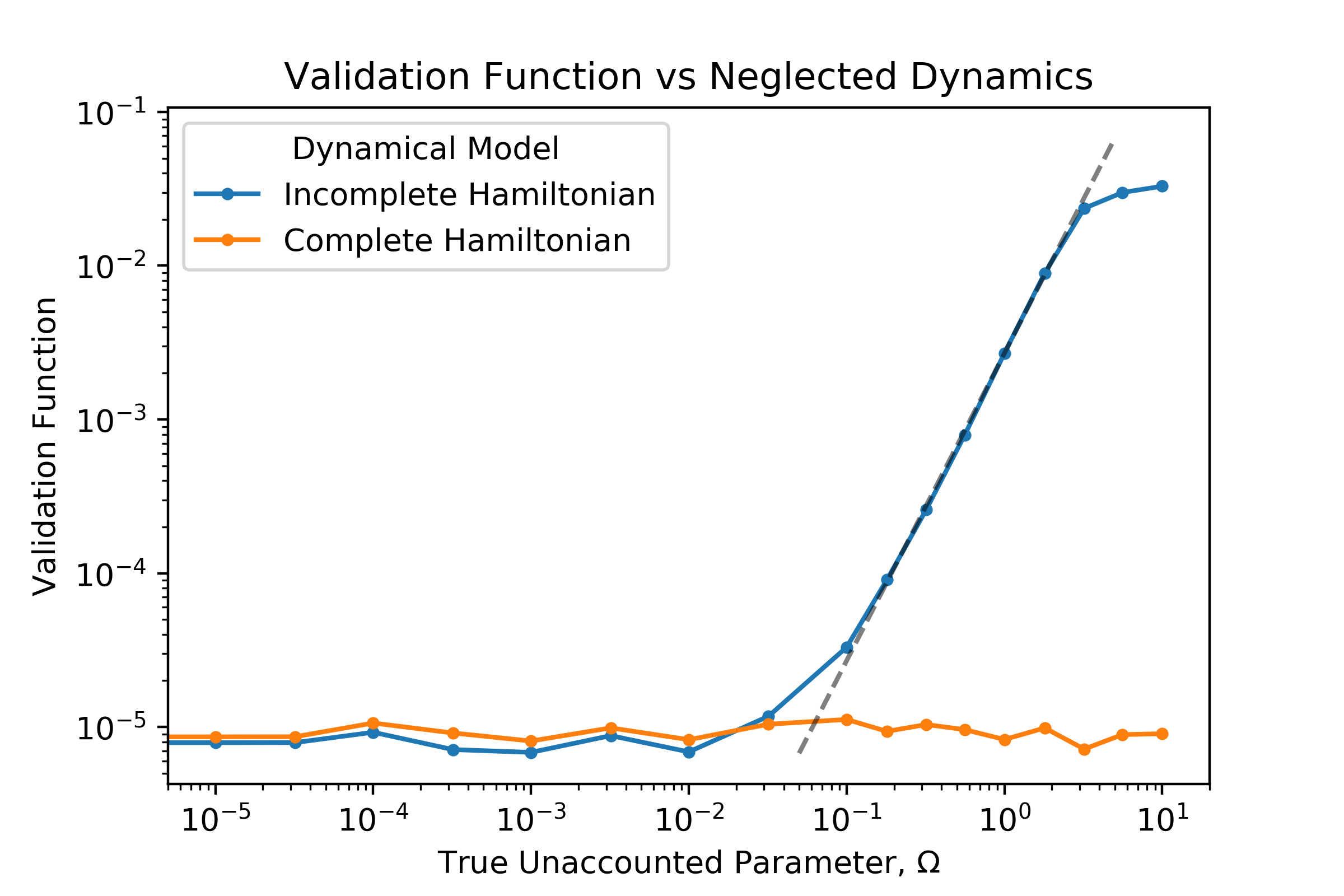}

\caption{\label{fig:incomplete}Similar
comparison to the one done in Fig.~\ref{fig:lindblad} from the main
text, however here we are concerned only with unitary dynamics. The
test system for this plot is the test system we have considered in
the rest of the manuscript, but with the nearest neighbor interaction
between qubit 1 and 2 constant (undriven), and set to be the $\Omega\sigma_{1}^{+}\sigma_{2}^{-}+h.c.$,
where $\Omega$ is a real parameter. The ``complete model'' is the
model we have used in the rest of the manuscript, and it is capable
of expressing this dynamics. The ``incomplete model'' has the $\sigma_{1}^{+}\sigma_{2}^{-}+h.c.$
term deleted (i.e.\ $\Omega$ implicitly set to 0) and can not represent
the exact dynamics of the system. As such, $\Omega$ becomes a parameter
describing how ``incomplete'' the incomplete model is, similarly
to $\Gamma$ in the case of non-unitary dynamics. As expected, we
see that for large $\Omega$, the incomplete model reaches a performance
floor. The gray line represents the $y\propto x^{2}$ power law, the
same one as in the case of non-unitary dynamics.}
\end{figure}

The second tool involves extending the dynamics permitted in the
model. Throughout the main text and in the following sections we discuss the
biasing effect of various types of model errors (e.g.\ intrinsic SPAM not being
included in the simplest models, or the effect of neglected parasitic couplings
or other coherent errors in Fig.~\ref{fig:incomplete}, or incoherent noise that can
not be expressed in a Hamiltonian formalism). A quick fix solution we discussed
is making the estimator more sensitive to parameter errors than to model errors
by, e.g., using stronger pulses that accentuate the parameter errors. However,
the more powerful solution, as long as it is computationally feasible, is to
extend the model to include these otherwise neglected dynamics. Both of these
approaches are described in the following sections.

Lastly, it is important to note the interaction between the regularization
approach and the model extension approach. Permitting many more degrees of
freedom in the model also leads to a higher risk of overfitting and convergence
issues, hence leading to the need for annealed regularization. For particularly
difficult problems one can even envision extending STEADY to a ``hyper parameter''
estimator where a discrete optimization algorithm evaluates a family of models
by the Akaike information criterion~\citep{akaike1974new}.

\section{Intrinsic SPAM}

As described in the main text, we used the following model for the
intrinsic state preparation error 
\[
\rho_{\text{init}}=(1-Qs)|0\rangle\langle0|+s\sum_{q=1}^{Q}|0\dots01_{q}0\dots0\rangle\langle0\dots01_{q}0\dots0|,
\]
and the following model for the measurement error: $\bm{p}_{\text{meas}}=\mathcal{S}\boldsymbol{p}$,
where 
\[
\mathcal{S}=(1-Qs)I+\mathcal{S}^{\prime}
\]
 and the only non-zero components of $\mathcal{S^{\prime}}$ are $\mathcal{S^{\prime}}_{ij}=s$
where the Hamming distance between the binary representations of $i$
and $j$ is 1.

To find the bias $\bm{b}$ in our estimator $\hat{\bm{\omega}}$ we will need to take a more careful look at the
minimum of $C_{s}(\bm{\omega})$ ($C_{s}(\bm{\omega})$ is what $C(\bm{\omega})$
becomes in the presence of SPAM errors as described below). By definition
of $\bm{b}$, that minimum will be at $\bm{\omega_{0}}+\bm{b}$. For
simplicity let us consider only measurement errors, which would cause
the vector $\hat{\bm{p}}_{i}$ to become $\hat{\bm{\pi}}_{i}=\mathcal{S}\hat{\bm{p}}_{i}$,
where $\mathcal{S}$ is an almost-diagonal stochastic matrix, with
off-diagonal components on the order of $\mathcal{O}\left(s\right)$,
i.e.\ $\mathcal{S}=I+s\mathcal{S}^{\prime}+\mathcal{O}(s^{2})$, where
$\mathcal{S}^{\prime}$ contains constant components on the order
of unity. $\mathcal{S}$ describes the chance that a measurement of
a given state is wrongly reported as another state. While the state
preparation errors are more complicated to express, because they happen
before the dynamical evolution of the state, linearity guarantees
that a similar treatment would work for an appropriately chosen ensemble
of states. This leaves us with 
\begin{align*}
C_{s}(\bm{\omega})&= \frac{1}{P}\sum_{i=1}^{P}\text{dist}\left(\mathcal{S}\hat{\bm{p}}_{i},\tilde{\bm{p}}_{i}(\bm{\omega})\right)\\
 & =\frac{1}{P}\sum_{i=1}^{P}\text{dist}\left(\hat{\bm{p}}_{i}+s\mathcal{S}^{\prime}\hat{\bm{p}}_{i}+\mathcal{O}(s^{2}),\tilde{\bm{p}}_{i}(\bm{\omega})\right)\\
 & =C(\bm{\omega})+\frac{1}{P}\sum_{i=1}^{P}s\mathcal{S}^{\prime}\hat{\bm{p}}_{i}.\left(\hat{\bm{p}}_{i}-\tilde{\bm{p}}_{i}(\bm{\omega})\right)\\
 & \hphantom{=} +\mathcal{O}(s^{2}).
\end{align*}
 Evaluated at $\bm{\omega_{0}}+\bm{b}$ it results in 
\begin{align*}
C_{s}(\bm{\bm{\omega_{0}}}+\bm{\bm{b}}) = & C(\bm{\omega}+\bm{\bm{b}})\\
 & +\frac{1}{P}\sum_{i=1}^{P}s\mathcal{S}^{\prime}\hat{\bm{p}}_{i}.\left(\hat{\bm{p}}_{i}-\tilde{\bm{p}}_{i}(\bm{\omega})\right)\\
 & +\delta s^{2}+\text{higher orders in \ensuremath{s} or \ensuremath{|\bm{b}|}},
\end{align*}
 where $\delta$ is a positive real number. Moreover, we have $C_{s}(\bm{\omega_{0}}+\bm{b})-C(\bm{\omega_{0}})\sim\alpha s^{2}+\mathcal{O}(s^{3})$
(because $s\mapsto\underset{\bm{\omega}}{\min}\left(C_{s}(\bm{\omega})\right)$
has its minimum at $s=0$) and $C(\bm{\omega_{0}}+\bm{b})-C(\bm{\omega_{0}})\sim\beta|\bm{b}|^{2}+\mathcal{O}(|\bm{b}|^{3})$
(because $\bm{b}\mapsto\underset{\bm{\omega}}{\min}\left(C(\bm{\omega}+\bm{b})\right)$
has its minimum at $\bm{b}=0$), where $\alpha$ and $\beta$
are positive real numbers. Therefore 
\begin{align*}
 C_{s}(\bm{\omega_{0}}+\bm{b})& -C(\bm{\omega_{0}}+\bm{b})=\\
=\, & \alpha s^{2}-\beta|\bm{b}|^{2}+\text{higher orders in \ensuremath{s} or \ensuremath{|\bm{b}|}}\\
=\, & s\frac{1}{P}\sum_{i=1}^{P}\left(\mathcal{S}^{\prime}\hat{\bm{p}}_{i}\right).\left(\hat{\bm{p}}_{i}-\tilde{\bm{p}}_{i}(\bm{\omega}_0+\bm{b})\right)\\
 & +\delta s^{2}+\text{higher orders in \ensuremath{s} or \ensuremath{|\bm{b}|}}\\
=\, & s\frac{1}{P}\sum_{i=1}^{P}\left(\mathcal{S}^{\prime}\hat{\bm{p}}_{i}\right).\left(\hat{\bm{p}}_{i}-\bm{p_{i}}+\bm{\mathcal{O}}\left(|\bm{b}|\right)\right)\\
 & +\delta s^{2}+\text{higher orders in \ensuremath{s} or \ensuremath{|\bm{b}|}}.
\end{align*}
 Given that we are interested in the regime where the bias overwhelms
the statistical error, we can take the limit $S\to\infty$ which results
to leading order in $\left(\alpha-\delta\right)s^{2}-\beta|\bm{b}|^{2}-\gamma s|\bm{b}|=0$,
where $\gamma$ is a positive real number. This leads to $|\bm{b}|\propto s$
and the observed error floor of $V\propto|\bm{b}|^{2}\propto s^{2}$
(see Fig.~\ref{fig:spam}d).

This error floor is unsurmountable by simply increasing the number
of acquired measurements. However, if we redo the expansion while
keeping record of $T$, we are left with $|\bm{b}|\propto\frac{s}{T}$.
This immediately suggests a way to decrease the bias of our estimator:
simply use longer (or more powerful) control pulses. Fig.~\ref{fig:spam}c
demonstrates the improvements due to this approach. 

\section{Effects of
Non-unitary Decay on Information Content}

As we have seen
in the rest of the manuscript, having an incomplete model, one that
is incapable of expressing the entire dynamics, would cause bias in
our estimator. This was observed both in the case of intrinsic SPAM,
and in the case of Lindbladian dynamics. This issue can be addressed
in some cases by making the estimator more sensitive to estimator
errors (by using longer pulses). When this fails one can instead extend
the model to include the missing dynamics. Both of these approaches
were discussed in the main text.

However, the
good performance of the Lindbladian model estimator from Fig.~\ref{fig:lindblad}
can be counterintuitive. Taken to the extreme, strong decay would
cause all of the information about the unitary evolution to leak out
to the environment before the measurement. This extreme example seems
to imply that the performance of the estimator should drop at extremely
high decay parameters, which we do not observe. Similarly to the discussion
in \secref{validation-choice}, this stems from our choice of validation
function; that is, we evaluate the quality of the predictions we can make
about our system, and we do not evaluate directly how precise each parameter
is estimated. Hence, when we are interested in the quality of predictions,
we do not need to worry about parameters that do not affect the dynamics
strongly. In the case of strong decays, all other parameters become
unimportant, and this is why we do not see a drop in performance. 
The same phenomenon that makes parameters hard to estimate also makes
them inconsequential to the dynamics of the system.

This can be
observed in the Fisher information. Consider for simplicity a two
level system. The excited state Born probability is $p(\omega)$ in
the case of unitary evolution. In the presence of decay, a first order
approximation for that same probability is $p(\omega)\mathrm{e}^{-\Gamma T}$.
The Fisher information with respect to $\omega$ for a single measurement
is then $\bm{{\mathcal{I}_{F}}}_{\omega}=\frac{\mathrm{e}^{-\Gamma T}}{p}p^{\prime2}(\omega)+\frac{\mathrm{e}^{-2\Gamma T}}{1-p\mathrm{e}^{-\Gamma T}}p^{\prime2}(\omega)=p^{\prime2}(\omega)\frac{\mathrm{e}^{-\Gamma T}}{p\left(1-p\mathrm{e}^{-\Gamma T}\right)}$,
which in the limiting cases is $\bm{{\mathcal{I}_{F}}}_{\omega}\underset{\Gamma\rightarrow\infty}{\sim}\frac{p^{\prime2}(\omega)\mathrm{e}^{-\Gamma T}}{p(\omega)}$
and $\bm{{\mathcal{I}_{F}}}_{\omega}\underset{\Gamma\rightarrow0}{\sim}\frac{p^{\prime2}(\omega)}{p(\omega)}+\mathcal{O}(\Gamma T)$.
Similarly, the Fisher information with respect to $\Gamma$ is $\bm{{\mathcal{I}_{F}}}_{\Gamma}=p^{2}T^{2}\mathrm{e}^{-2\Gamma T}\left(\frac{1}{p\mathrm{e}^{-\Gamma T}}+\frac{1}{1-p\mathrm{e}^{-\Gamma T}}\right)$,
with limiting behavior $\bm{{\mathcal{I}_{F}}}_{\Gamma}\underset{\Gamma\rightarrow\infty}{\sim}p^{2}T^{2}\mathrm{e}^{-\Gamma T}$
and $\bm{{\mathcal{I}_{F}}}_{\Gamma}\underset{\Gamma\rightarrow0}{\sim}\frac{p}{1-p}T^{2}+\mathcal{O}(\Gamma T)$.

\begin{figure}
\includegraphics[width=7cm]{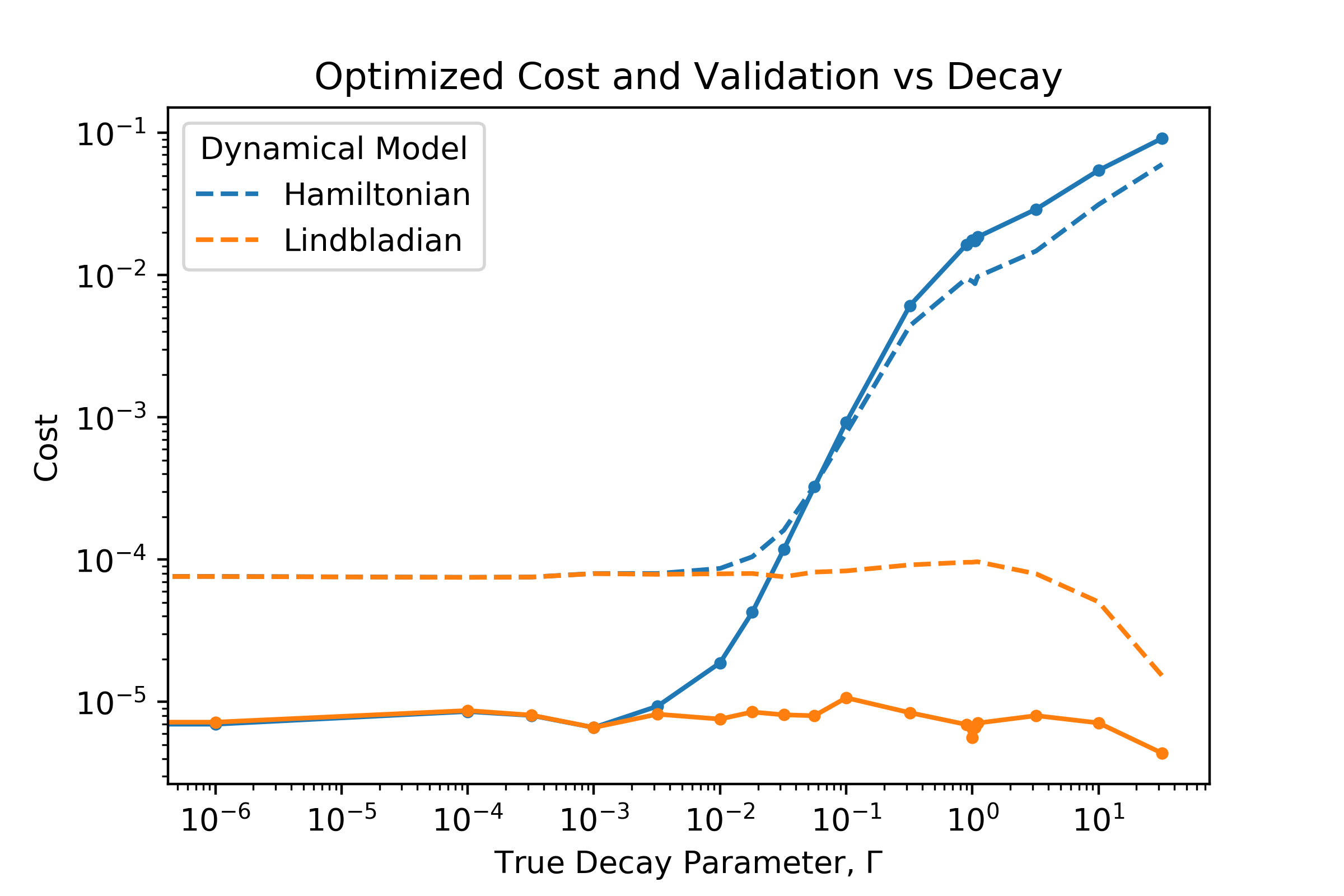}

\caption{\label{fig:lindblad_cost}Same
as Fig.~\ref{fig:lindblad} from the main text, but it also includes
the optimized cost function (dashed line) besides the validation function
(solid line). At very high value for the decay parameter, one can
observe the asymptotic $\mathrm{e}^{-\Gamma T}$ behavior (both the cost and
validation functions become much lower). To reiterate, this is due
to the fact that strong decays make the system ``uninteresting''
and trivial to characterize, not due to a particular advantage of
any characterization method one could deploy. The fact that the incomplete
Hamiltonian model, incapable of describing the decay, performs poorly
was already discussed in the main text. }
\end{figure}

However, the
validation function is to first order $V\propto\left(p(\omega_{0})\mathrm{e}^{-\Gamma_{0}T}-p(\omega)\mathrm{e}^{-\Gamma T}\right)^{2}$,
where $\omega_{0}$ and $\Gamma_{0}$ are the true values of the parameters.
Hence $V\propto \mathrm{e}^{-2\Gamma_{0}T}\left(p^{\prime}(\omega_{0})(\omega_{0}-\omega)-Tp(\omega_{0})(\Gamma_{0}-\Gamma)\right)^{2}\propto \mathrm{e}^{-2\Gamma_{0}T}\left(p^{\prime2}(\omega_{0})\sigma_{\omega}^{2}+T^{2}p^{2}(\omega_{0})\sigma_{\Gamma}^{2}\right)$,
where $\sigma_{\omega}$ and $\sigma_{\Gamma}$ are the variances
in the estimators for each of the parameters (we neglect correlations).
Due to the Cram\'{e}r--Rao bound both variances scale as the
inverse of the corresponding Fisher information, which leaves us with
$V\underset{\Gamma\rightarrow\infty}{\sim}p(\omega_{0})\mathrm{e}^{-\Gamma T}$
(see Fig.~\ref{fig:lindblad_cost}). Therefore, as the decay rate
goes higher and leaves us with less and less available information
per measurement, it also causes less diversity in the final measurements
(high probability that the final state is the ground state), leading
to small (``good'') values for the validation function. In practice,
this effect becomes noticeable only at impractically high values for
$\Gamma$: while we do precisely predict in such cases how the state
decays, this is not of particular use for devising control protocols
for the system. 
Generally a system with strong decays is not particularly
useful as a quantum hardware, unless we use ``engineered dissipation''
control schemes or control schemes employing virtual states like
STIRAP~\citep{ucontrol_unanyan1998robust},
which are beyond the scope of this manuscript.

\section{Variance of Parameter Estimator in Linear Least Squares}

Here we give a concrete analytical example of cases where the cost
function has a floor dictated by $P$, while the validation function
has a floor dictated by $S\times P$. The description here is generic; it does not refer to a model of quantum dynamics.

The model we are learning is $f(\omega;x)$ parameterized by $\omega$.
We denote by $(y_{i},x_{i})$ the pairs of (population estimate, pulse) that we are
learning from (we have $P$ such pairs). For each pair we used $S$
repetitions in order to estimate $y_{i}$, the population, for the
given $x_{i}$, the pulse.

The true population is $y_{i}^{\circ}=y_{i}-e_{i}$ (given that we
are performing multinomial sampling, we approximate $e_{i}$ as drawn
from a normal distribution with $\sigma_{0}=\frac{\sqrt{p}}{\sqrt{S}}$).

We perform parameter estimation by minimizing least squares $C(\omega)=\frac{1}{P}\stackrel[1]{P}{\sum}\left(y_{i}-f(\omega,x_{i})\right)^{2}=\frac{1}{P}\stackrel[1]{P}{\sum}\left(y_{i}^{\circ}+e_{i}-f(\omega,x_{i})\right)^{2}$
and the validation function is $V(\omega)=\frac{1}{P}\stackrel[1]{P}{\sum}\left(y_{i}^{\circ}-f(\omega,x_{i})\right)^{2}$.

For the purpose of this example, consider linear least squares: the
model we are fitting to has the parameters $\omega=a,b$, and $f(a,b;x)=a+bx$.
$a$ and $b$ denote the parameter values which minimize $C$. However,
the data are governed by the ``true'' model $y^{\circ}=f(a^{\circ},b^{\circ};x)=a^{\circ}+b^{\circ}x$.

The minimum of $C(\omega)$ is reached at $b=\frac{\hat{\sigma}_{xy}}{\hat{\sigma}_{xx}}$
and $a=\bar{y}-b\bar{x}$, where:
\begin{itemize}
\item a bar denotes the typical \emph{estimator} of an average value;
\item $a=\bar{y}-b\bar{x}=\bar{y}^{\circ}+\bar{e}-b\bar{x}$;
\item $\hat{\sigma}_{xx}$ denotes the \emph{estimator} of the variance of $x$;
\item $\hat{\sigma}_{xy}$ is the \emph{estimator} of the covariance of $x$ and $y$. 
\end{itemize}
The estimated covariance can be calculated as, 
\begin{align*}
\hat{\sigma}_{xy} & =\frac{1}{P}\sum_{i=1}^P(x_{i}-\bar{x})(y_{i}-\bar{y})\\
 & =\frac{1}{P}\sum_{i=1}^P(x_{i}-\bar{x})(y_{i}^{\circ}-\bar{y}^{\circ}+e_{i}-\bar{e})\\
 & =\frac{1}{P}\sum_{i=1}^P(x_{i}-\bar{x})b^{\circ}(x_{i}-\bar{x})+\frac{1}{P}\sum_{i=1}(x_{i}-\bar{x})(e_{i}-\bar{e})\\
 & =b^{\circ}\sigma_{x}^{2}+\hat{\sigma}_{xe}.
\end{align*}
Substituting the expressions for $y_i^\circ$ and $a$ in $V$ and factoring out
the $(x_{i}-\bar{x})$ term leaves us with:
\begin{align*}
V_{\text{opt}} & =\frac{1}{P}\sum_{i=1}^P\left(y_{i}^{\circ}-a-bx_{i}\right)^{2}\\
 & =\frac{1}{P}\sum_{i=1}^P\left(b^{\circ}x_{i}-b^{\circ}\bar{x}-\bar{e}+b\bar{x}-bx_{i}\right)^{2}\\
 & =\frac{1}{P}\sum_{i=1}^P\left(\frac{\hat{\sigma}_{xe}(x_{i}-\bar{x})}{\sigma_{x}^{2}}-\bar{e}\right)^{2}\\
 & \approx\frac{1}{P}\sum_{i=1}^P\bar{e}^{2}
\end{align*}

Hence the expectation value of $V$ is $\frac{p}{PS}$.

\section{Software Implementation}

The estimators we implement are all fully differentiable programs,
that can run ``batches'' of measurement data. They can run on
both CPUs and GPUs, as well as any other optimized tensor processing
units supported by Tensorflow, like Google's TPU chips. The code is
documented inline and extremely short (<50 lines of code per estimator).
We implement the following estimators (all of them can be run either
as stochastic parameter estimators where measurements are provided
as training data, or as optimal control optimizers where the parameters
are fixed but the control pulses are optimized for a given target).
\begin{itemize}
\item 
    \texttt{StateProbabilitiesPaulied}
solves Schroedinger's equation for 
\[
\begin{array}{c}
\tilde{H}(\bm{\alpha},\bm{\beta};\bm{d})=\stackrel[k=1]{M}{\sum}a_{k}A_{k},\\
\text{where }a_{k}=\stackrel[l=1]{D}{\sum}\alpha_{kl}d_{l}+\beta_{k},
\end{array}
\]
and where $\{A_{1},\dots,A_{M}\}$ are fixed in advance. Control pulses
for this solver are constant in time (with implied duration of $T=1$)
provided as arrays of shape $P\times D$. The time-evolution operator
is calculated through diagonalization of the Hamiltonian.
\item 
    \texttt{StateProbabilities}
 solves Schroedinger's equation for 
\[
\tilde{H}_{ij}(\bm{\sigma},\bm{h};\bm{d})=h_{ij}+\stackrel[k=1]{D}{\sum}\sigma_{ijk}d_{k},
\]
 with explicitly ensured Hermiticity (see the source code for details).
Control pulses for this solver are constant in time (with implied
duration of $T=1$) provided as arrays of shape $P\times D$. The
time-evolution operator is calculated through diagonalization of the
Hamiltonian.
\item Non-linear drives can be straightforwardly added to the above solvers
by adding higher-order terms like $\sigma_{ijkl}d_{k}d_{l}$, etc.
\item 
    \texttt{StateProbabilitiesTimeDep}
 solves Schroedinger's equation in either of the previous two forms,
but for time dependent control pulses $d_{k}$. Control pulses are
piece-wise constant of implied total duration $T=1$ and a predetermined
number of timesteps $\Theta$. The control pulses are provided as
arrays of shape $P\times\Theta\times D$. The time evolution is computed
through evaluating the $\Theta$ consecutive time-evolution operators.
Each operator is computed either, as above, through diagonalization,
or through faster, but less precise, Taylor expansion of the matrix
exponential (in Horner's form).
\item 
    \texttt{StateProbabilitiesTimeDepLindblad}
 solves Lindblad's Master equation for time dependent control pulses
$d_{k}$. The Hamiltonian part is provided in either of the two forms
discussed above. The control pulses are provided in the same format
as above. The non-unitary evolution is modeled by fixed predetermined
collapse operators $\{L_{1},\dots,L_{C}\}$, however the strengths
of each collapse operator $\{c_{1},\dots,c_{C}\}$ is a parameter
optimized by the estimator. The full equation being modeled is 
\begin{align*}
\dot{\rho}=&-i[\tilde{H}(\bm{\omega};\bm{d}),\rho]\\
&+\sum_{i=1}^C c_{i}\left(L_{i}\rho L_{i}^{\dagger}-L_{i}^{\dagger}L_{i}\rho/2-\rho L_{i}^{\dagger}L_{i}/2\right).
\end{align*}
The integrator uses Euler's method.
\item 
    \texttt{StateProbabilitiesTimeDepLindbladRK4}
 works as above but it uses the RK4 method for numerical integration.
\item Unlike gradients, Hessians do not permit fast backpropagation methods
for their calculation (and hence are not well supported by differentiable
programming frameworks yet). This led us to writing a simple numerical
procedure for estimating the Hessians needed for the calculation of
the Fisher information in 
        \texttt{AvgFIStateProbabilitiesPaulied.}
It is a proof-of-concept that works only on the first of the estimators
described above, but it is the tool that permitted us to run experimental-design
optimizations as described in the main text. It is implemented in
a separate Jupyter notebook.
\end{itemize}
\begin{figure*}
\includegraphics[width=2\columnwidth]{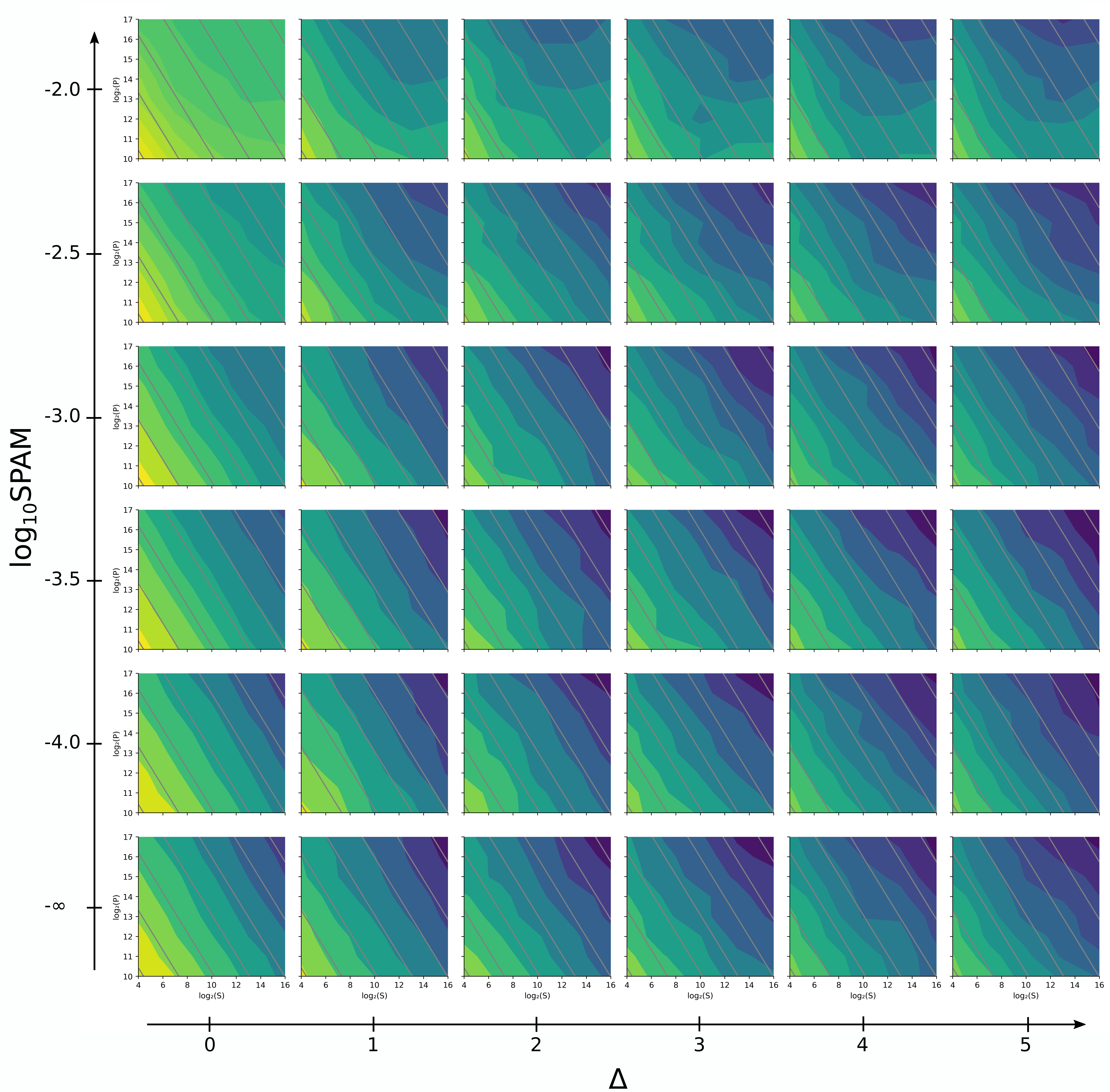}
\caption{The complete set of simulations of SPAM errors and counter-SPAM measures
using random control pulses. Going top-to-bottom, SPAM increases.
Going left to right, the length of the random control pulses increases.}
\end{figure*}

\begin{figure*}
\includegraphics[height=4cm]{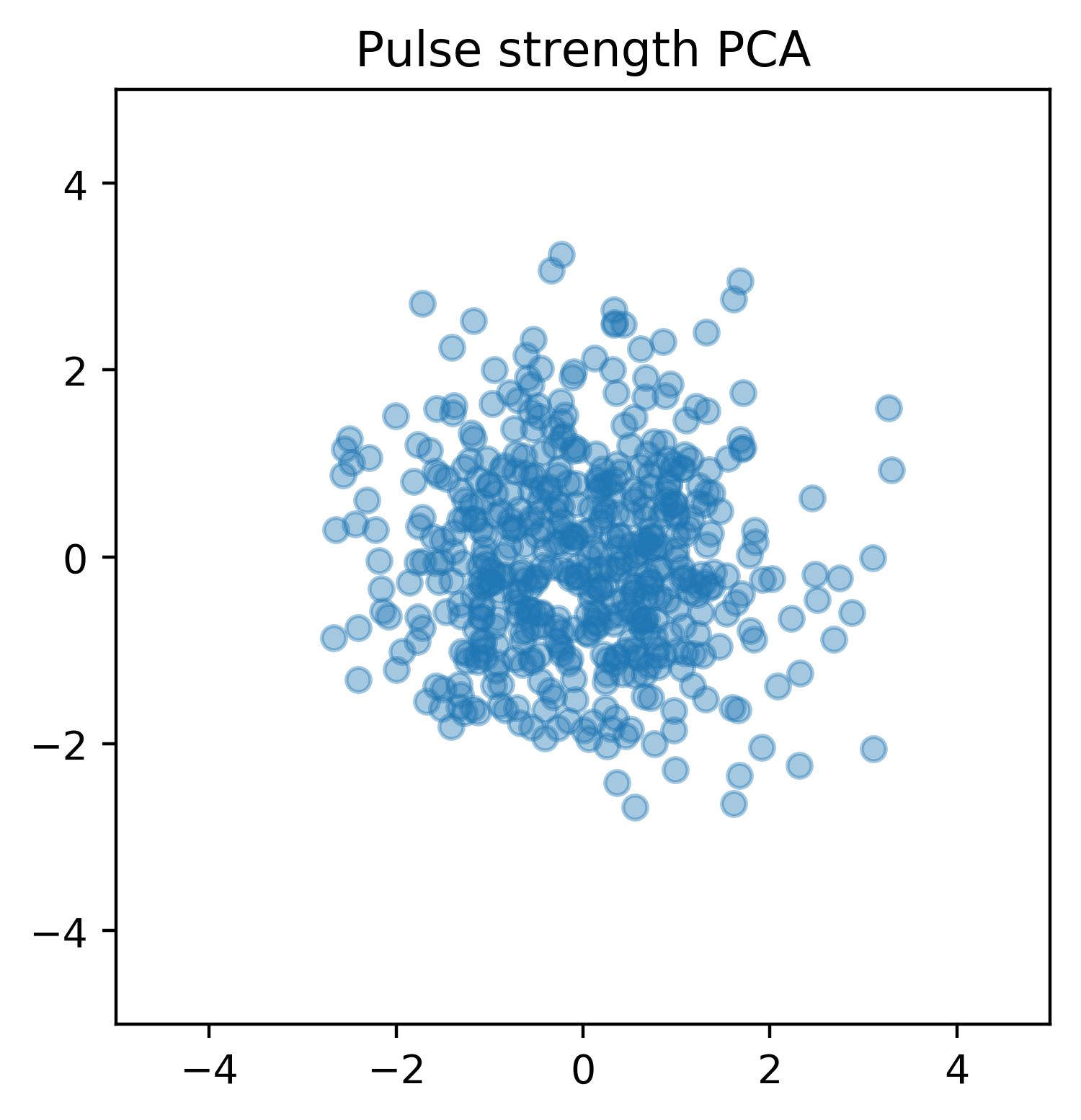}\includegraphics[height=4cm]{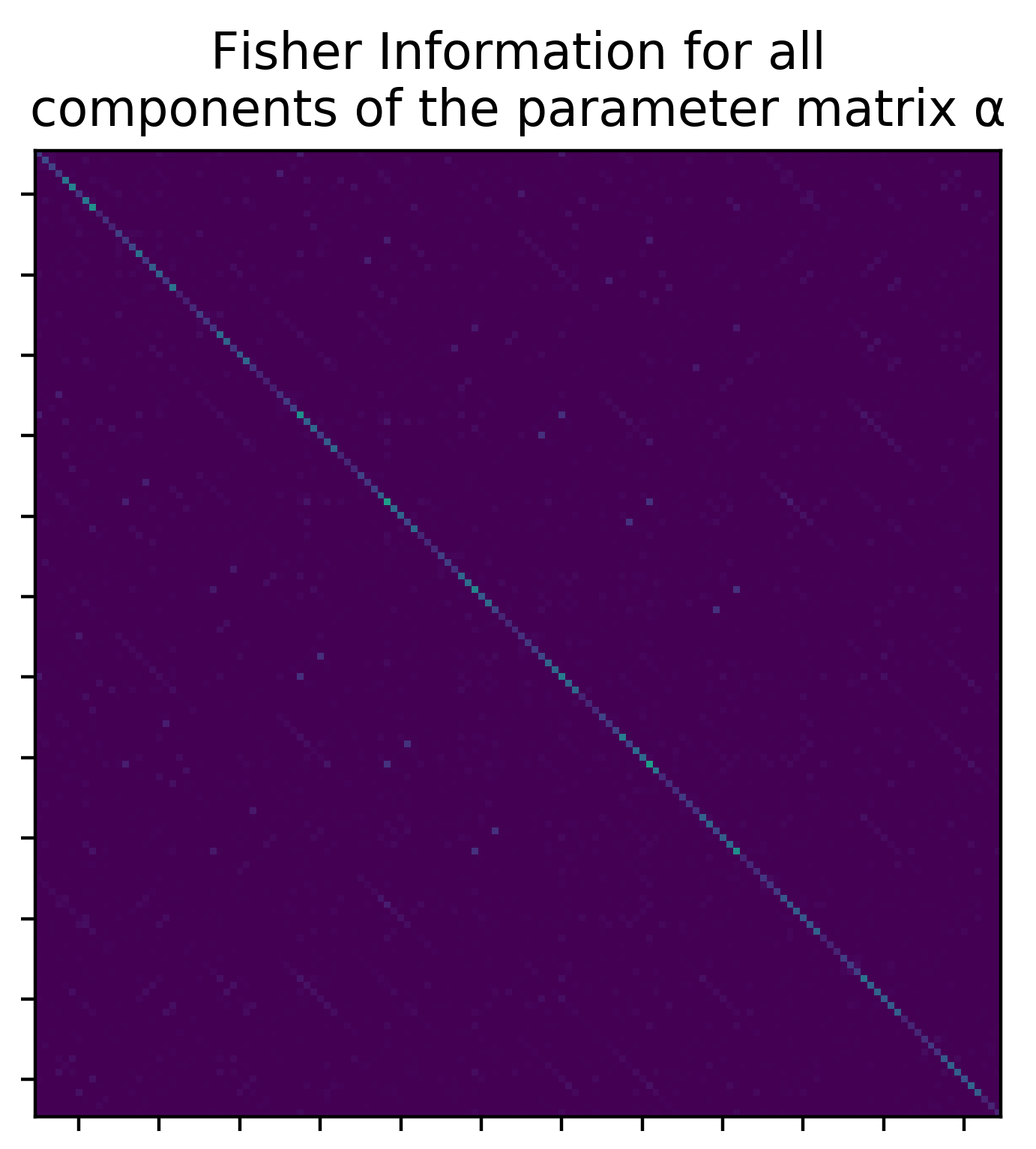}\includegraphics[height=4cm]{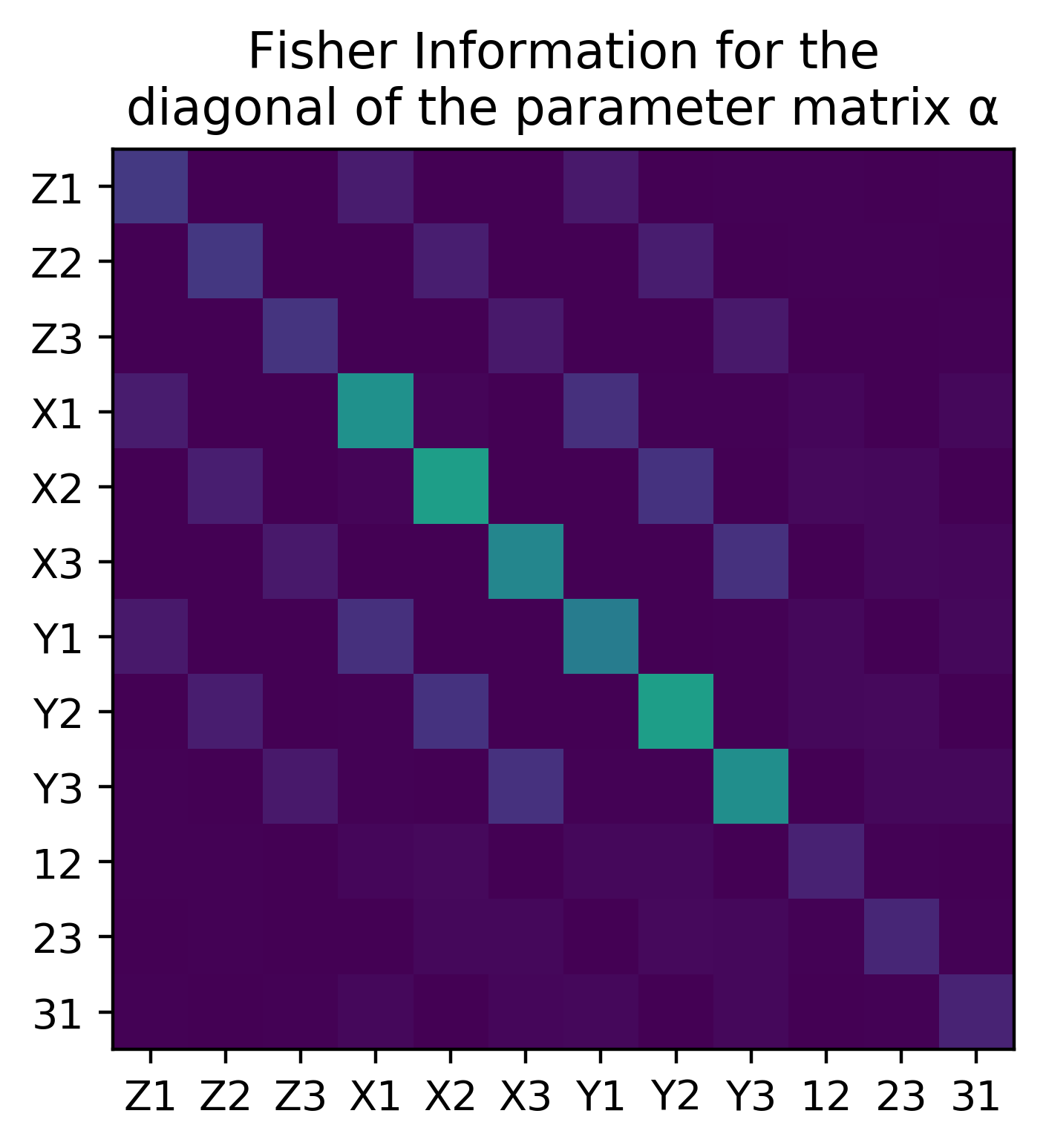}\includegraphics[height=4cm]{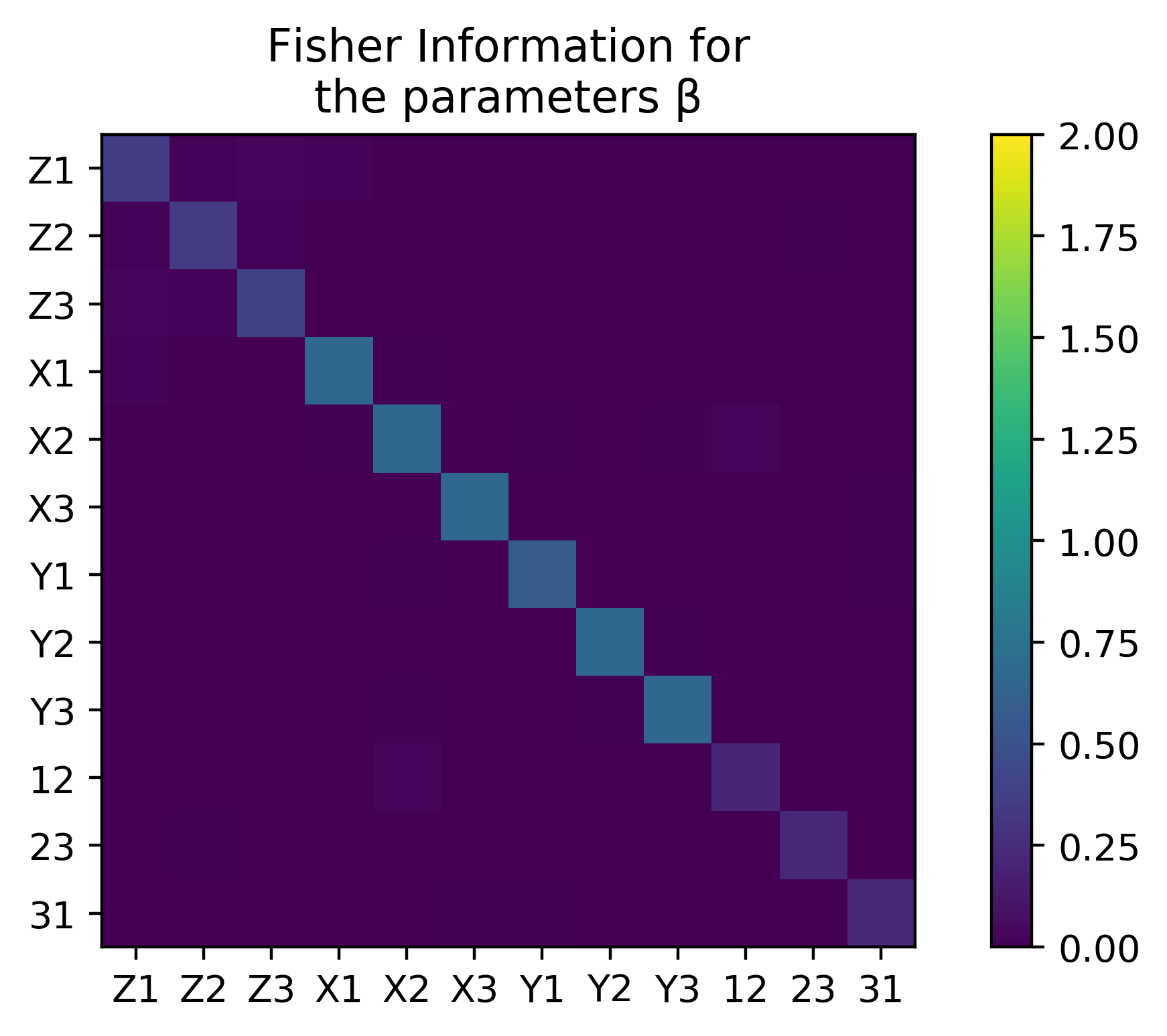}

\includegraphics[height=4cm]{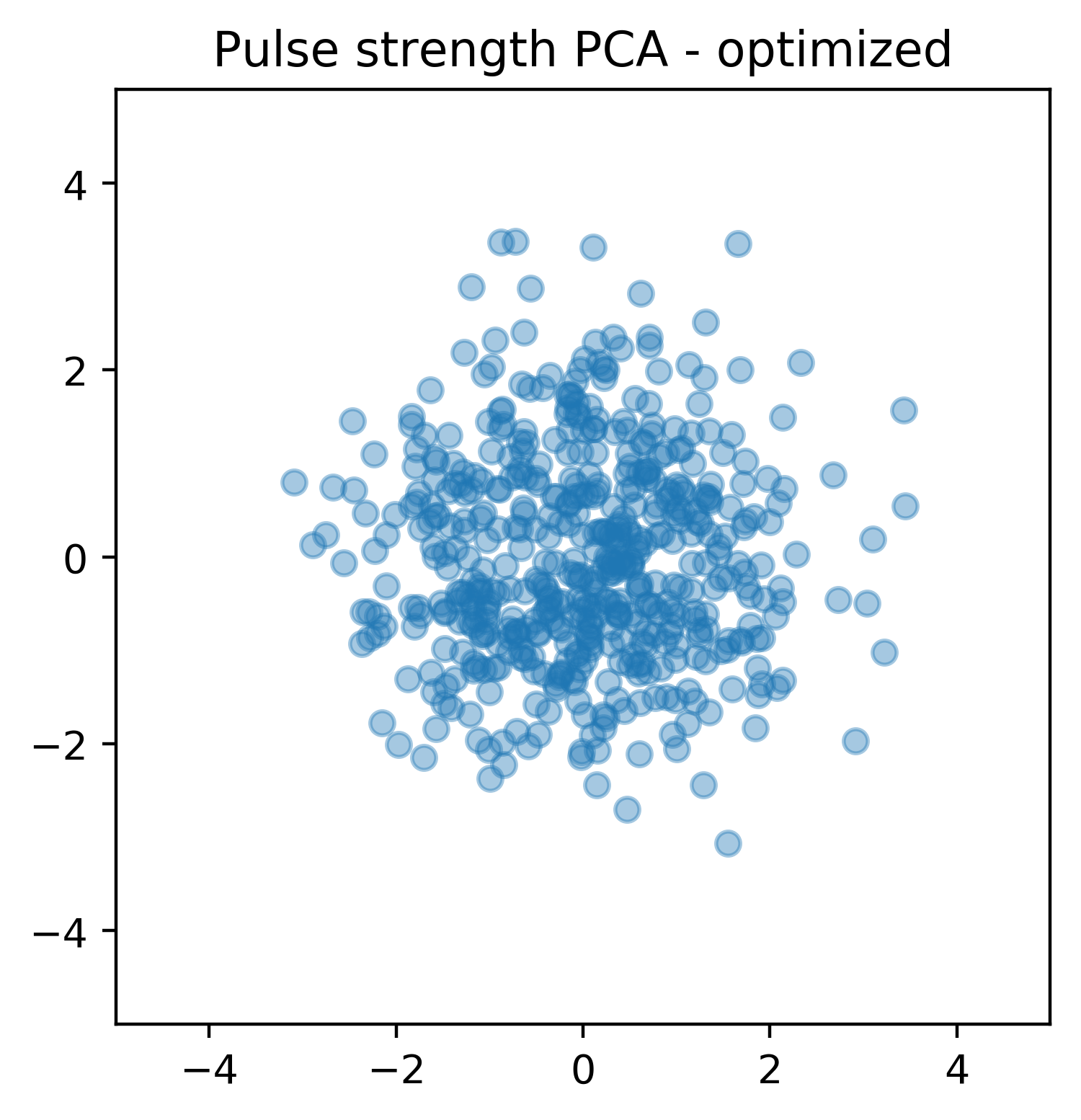}\includegraphics[height=4cm]{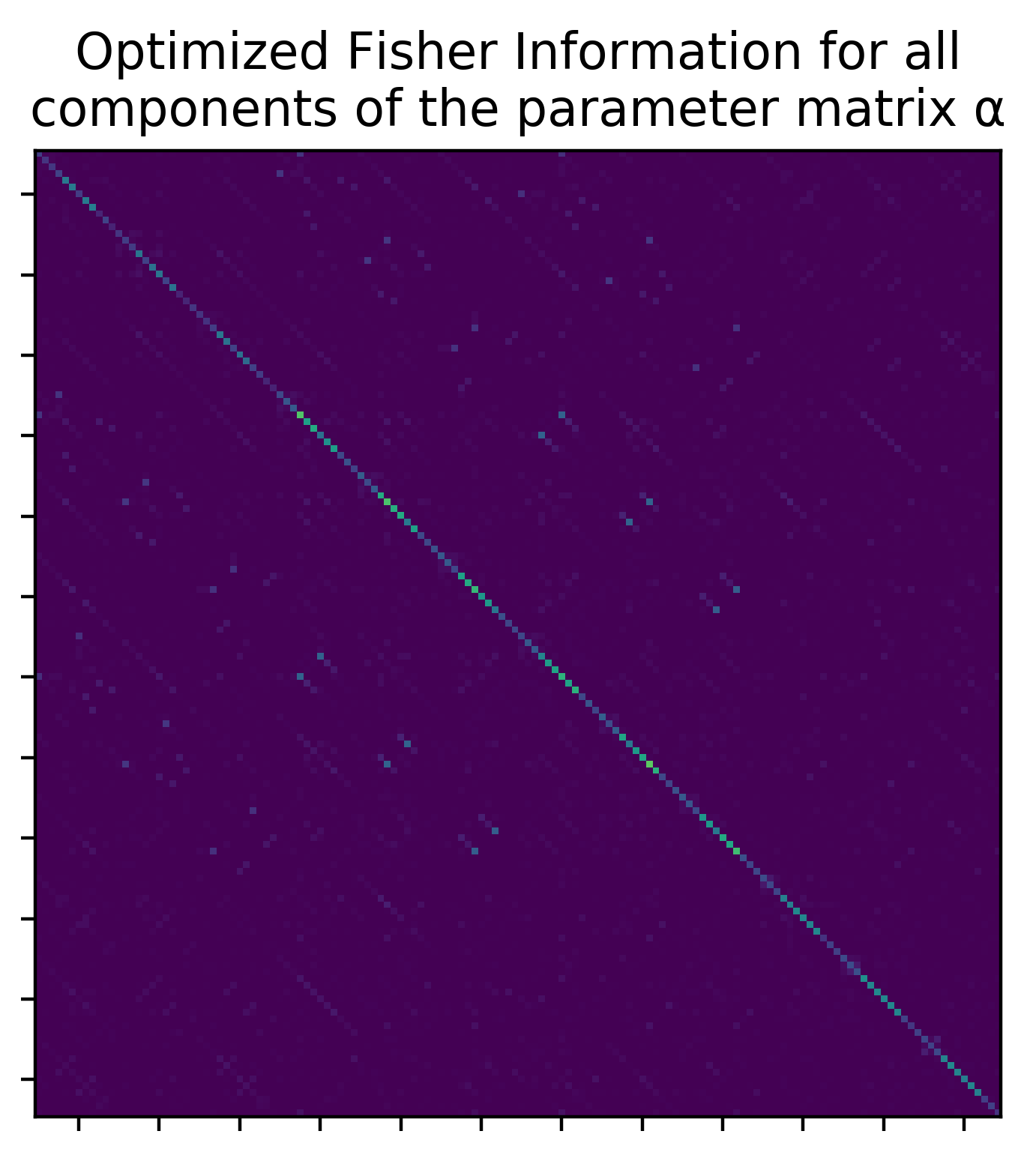}\includegraphics[height=4cm]{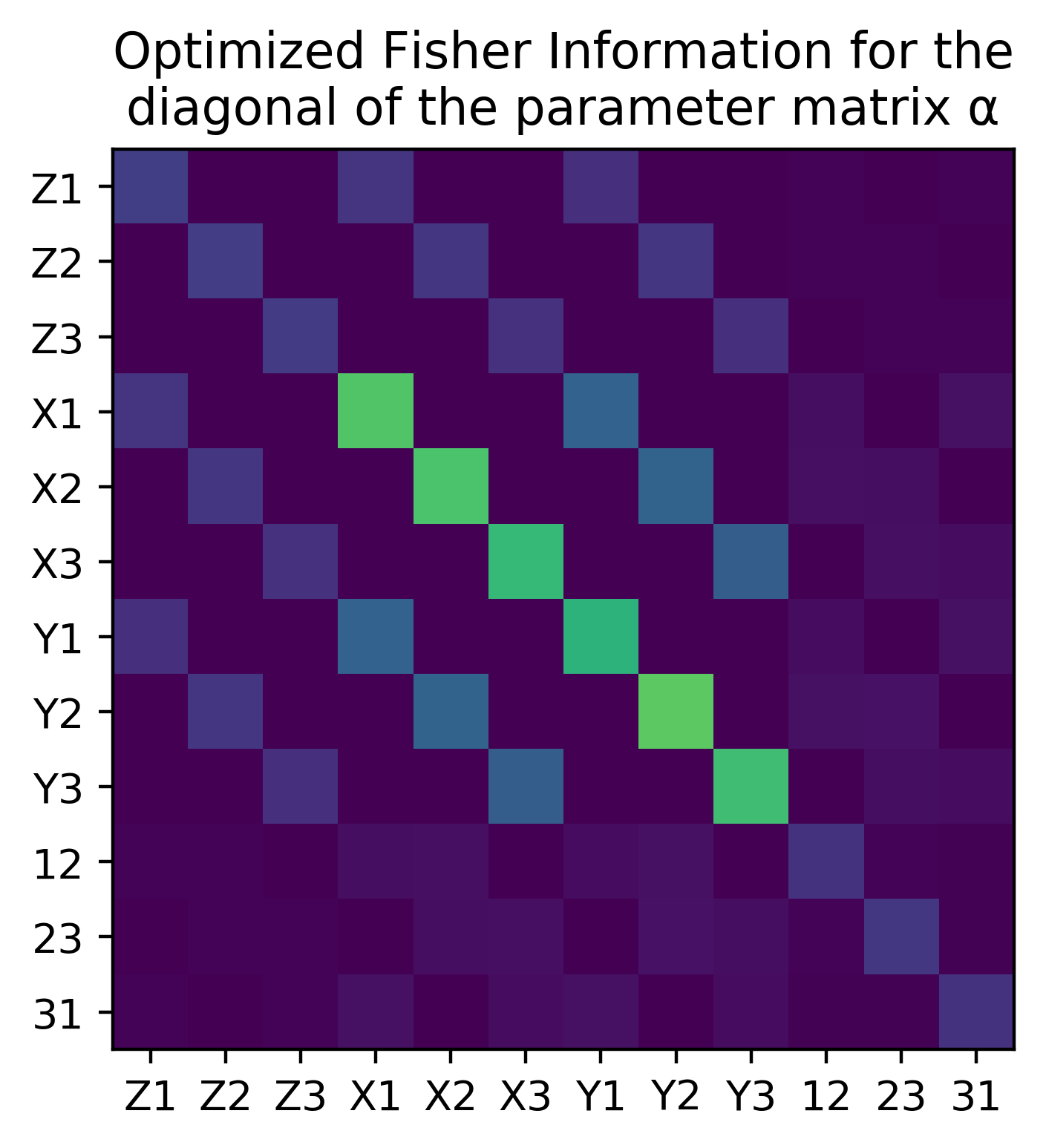}\includegraphics[height=4cm]{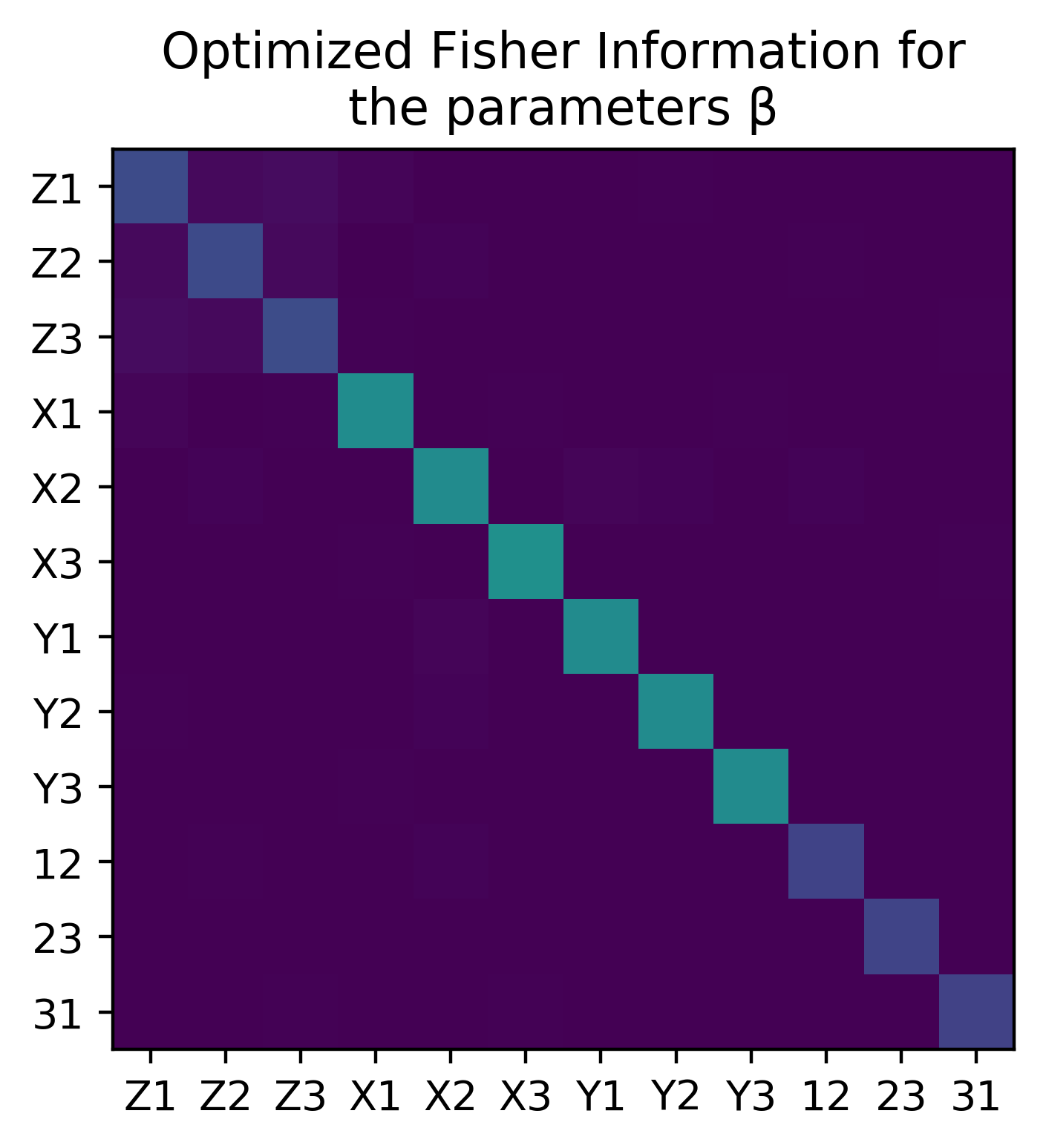}~~~~~~~~

\includegraphics[height=4cm]{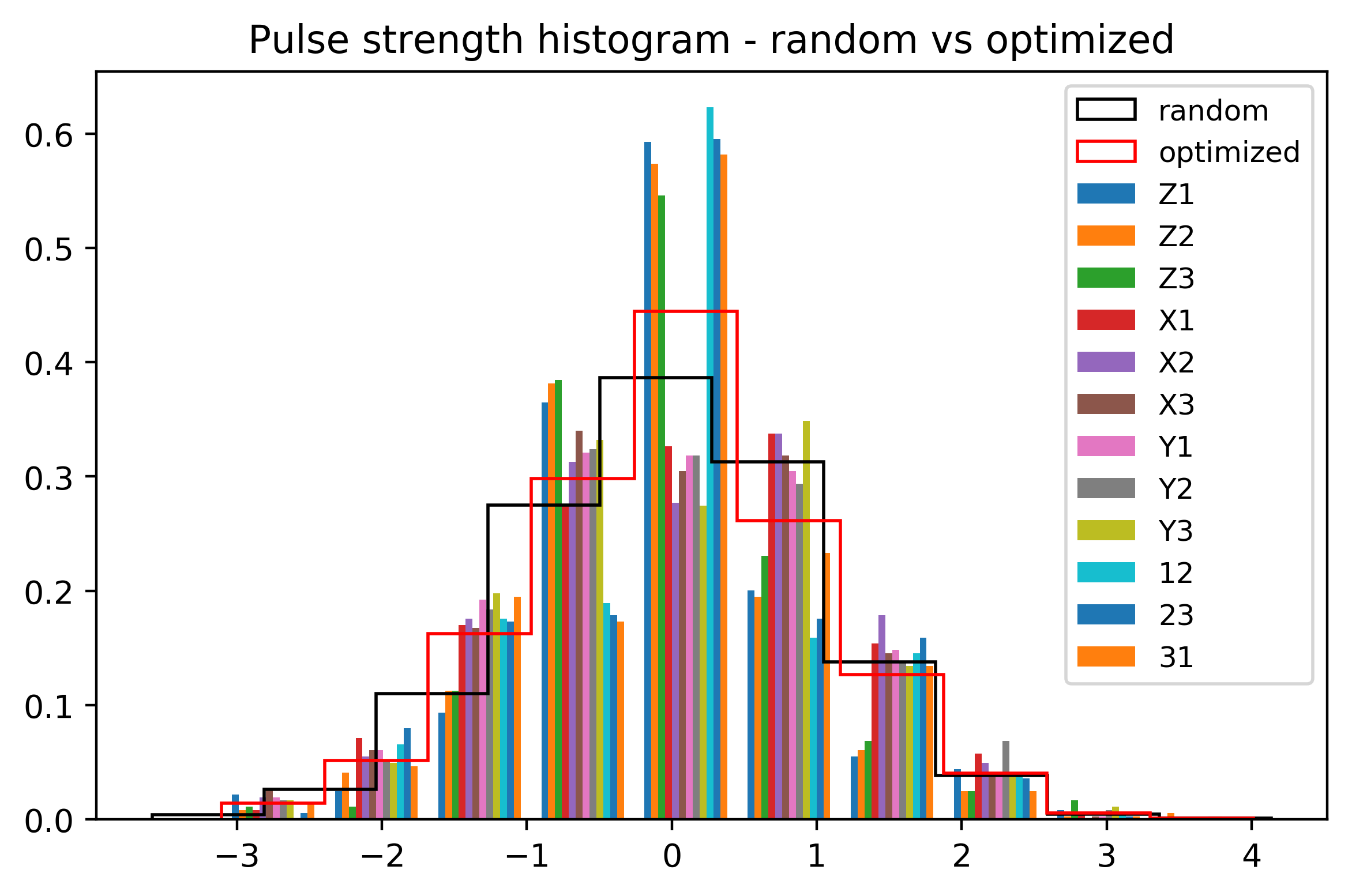}\includegraphics[height=4cm]{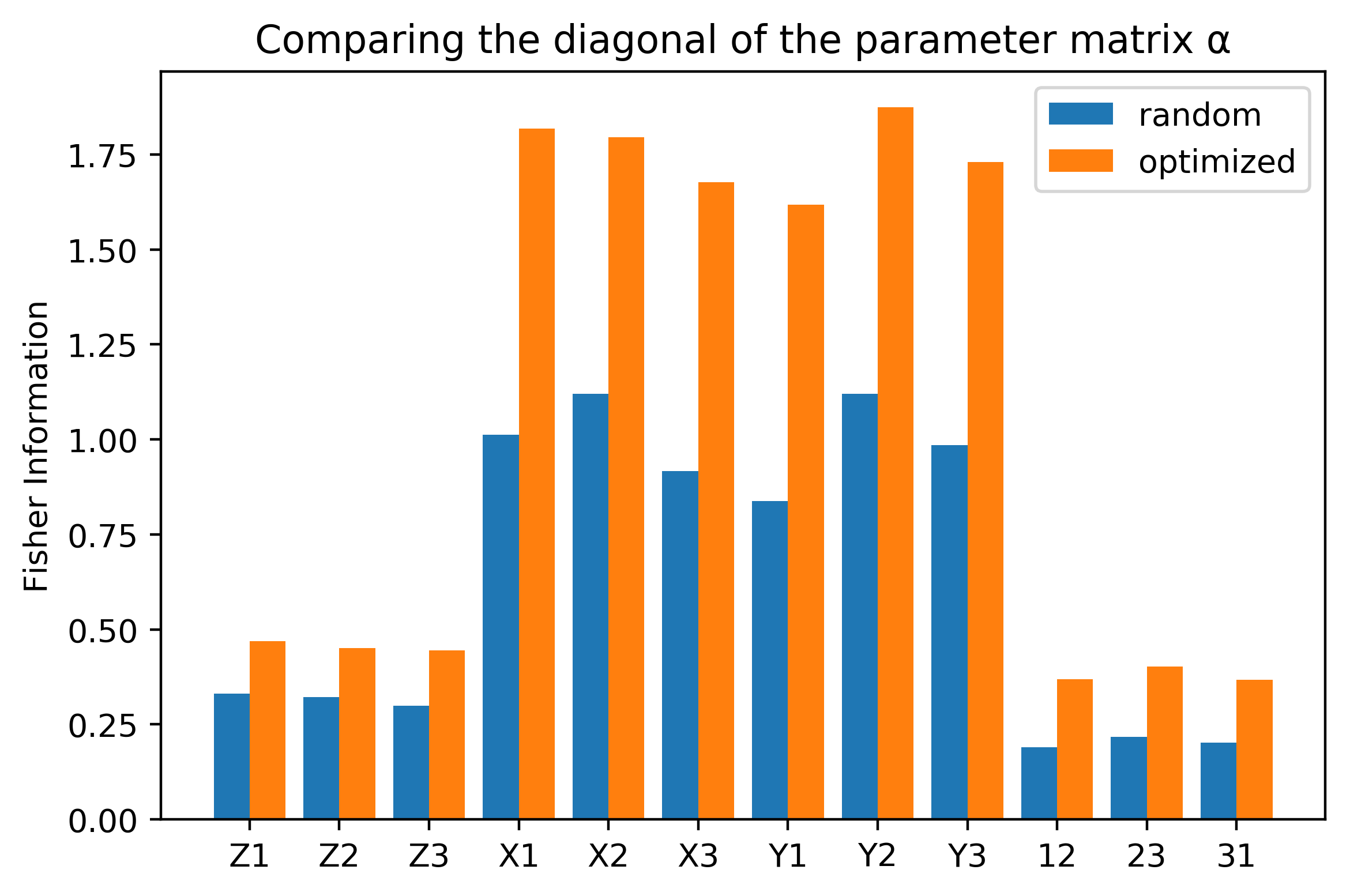}\includegraphics[height=4cm]{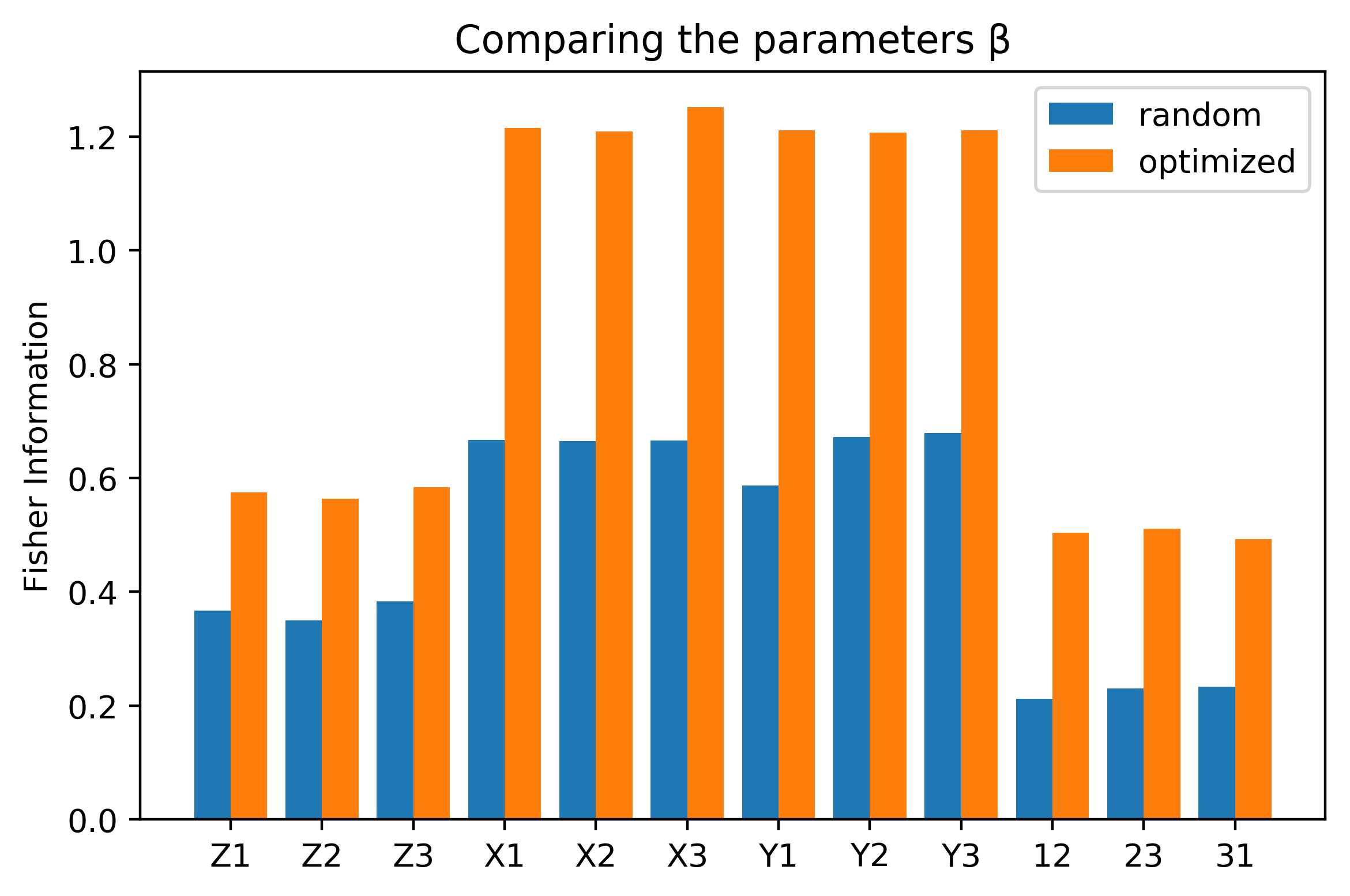}

\caption{Random versus optimized pulses in terms of Fisher information. Including:
Principle component analysis of the pulses to ensure there is no preferential
axis; a histogram of the pulses before and after optimization; a detailed
depiction of various components of the Fisher information matrix.}
\end{figure*}

\end{document}